\documentclass[runningheads,fleqn]{svmult}

\usepackage{makeidx}   
\usepackage{graphicx}  
\usepackage{subeqnar}  
\usepackage{multicol}  
\usepackage{physmult}  
\makeindex             

\newcommand{\Cuncs}{$\kappa$-(BEDT-TTF)$_2$Cu(NCS)$_2$}
\newcommand{\cuncs}{$\kappa$-(ET)$_2$Cu(NCS)$_2$}
\newcommand{\cuncsh}{$\kappa$-(H$_8$-ET)$_2$Cu(NCS)$_2$}
\newcommand{\cuncsd}{$\kappa$-(D$_8$-ET)$_2$Cu(NCS)$_2$}

\newcommand{\brom}{$\kappa$-(ET)$_2$Cu[N(CN)$_2$]Br}
\newcommand{\bromh}{$\kappa$-(H$_8$-ET)$_2$Cu[N(CN)$_2$]Br}
\newcommand{\bromd}{$\kappa$-(D$_8$-ET)$_2$Cu[N(CN)$_2$]Br}

\newcommand{\cl}{$\kappa$-(ET)$_2$Cu[N(CN)$_2$]Cl}
\newcommand{\sfcho}{$\beta''$-(ET)$_2$SF$_5$CH$_2$CF$_2$SO$_3$}
\newcommand{\tc}{$T_c$}

\newcommand{\tst}{$T^{\ast}$}
\newcommand{\tg}{$T_g$}

\newcommand{\etzx}{(ET)$_2$X}

\newcommand{\et}{[(CH$_2$)$_2$]}

\newcommand{\xs}{$\chi_{\rm Spin}$}
\newcommand{\ces}{$C_{\rm es}$}

\begin{document}
\title*{Organic Superconductors}
\toctitle{Organic Superconductors}
\titlerunning{Organic Superconductors}
\author{Michael Lang
\and Jens M\"uller}
\authorrunning{Michael Lang \and Jens M\"uller}

\maketitle

\section{Introduction}
Over the past twenty years the research on organic conductors has developed into a most
active branch of modern condensed-matter physics. The main difference of molecular
conductors compared to conventional metals is that the former are made up of building
blocks constructed from carbon atoms and their combinations with other elements such as
sulfur, selenium or oxygen. As a result, when forming a crystal, these molecular units
preserve to a large extent their specific features such as the molecular orbitals, the
ionization energy and the intramolecular vibrational modes. Molecular materials thus have
the potential of providing a flexible building-block system where the physical properties
can be tuned by small modifications in the arrangements and bridging of these functional units. \\
In contrast to the molecular crystals formed by weakly van der Waals-bond neutral
entities, organic conductors consist of open-shell molecular units which are the result
of a partial oxidation and reduction of the donor and acceptor molecules in the
crystal-growth process. It is the unpaired electron residing in the $\pi$-molecular
orbital ($\pi$-hole) of the donor unit which is responsible for the electronic properties
of these charge-transfer salts. Due to a $\pi$-orbital overlap between adjacent
molecules, the $\pi$-holes can delocalize throughout the crystal
giving rise to metallic conductivity.\\
The unique chemistry of carbon, which, on the one hand, provides a rich basis of potential
organic donor molecules, and the manifold possibilities of combining them with
charge-compensating acceptor units, on the other, have enabled the synthesis of an
enormous number of conducting organic charge-transfer salts. The geometry of the building
blocks and the way they are packed together in the crystal determine the effective
dimensionality of the electronic structure of the compound. The planar shape of the TMTSF
(tetramethyltetraselenafulvalene) molecule - a derivative of the prototype TTF
(tetrathiafulvalene) - permits an infinite stacking of these units in the crystal
structure. As a consequence, a significant intermolecular overlap of the $\pi$-orbitals
occurs only along the stacking axis giving rise to a quasi-one-dimensional conduction
band. In contrast, packing of the larger BEDT-TTF (bisethylenedithio-tetrathiafulvalene)
molecules often results in a quasi-two-dimensional electronic structure.\\

These low-dimensional organic conductors constitute a peculiar class of materials which
has gained strong interest among scientists due to the wealth of interesting phenomena
that have been observed over the last two decades or so. Above all it was the discovery
of superconductivity in pressurized (TMTSF)$_2$PF$_6$ by J\'{e}rome et al.\ in 1979
\cite{Jerome 80} which created much excitement among both solid-state chemists and
physicists. Indeed, this finding was a result of interdisciplinary efforts in
synthesizing and characterizing organic conductors - activities which started up in the
early 1950's by the pioneering work of Akamatu et al.\ \cite{Akamatu 54} and which
received a fresh impetus in 1964 thanks to the work of Little \index{Little's model}
\cite{Little 64}. He proposed that in a suitably designed one-dimensional conductor
embedded in a polarizable medium, superconductivity at high temperatures should be
possible. Although it soon became clear that the nature of the superconducting state in
the (TMTSF)$_2$PF$_6$ salt is of a fundamentally different type from what had been
discussed in \index{Little's model} Little's model, this finding set the stage for the
discovery of a large number of conducting charge-transfer salts including up to the
latest count, more than $80$ superconductors. Many of these systems superconduct even at
ambient pressure with the highest $T_c$ value of about $12$\,K received in the
quasi-two-dimensional BEDT-TTF-based systems. Besides superconductivity, the organic
charge-transfer salts reveal a variety of other interesting collective phenomena
including spin-Peierls and density-wave states as well as phases with localized charges
and commensurate-type antiferromagnetic order. These states have been found to depend
most sensitively on factors such as the acceptor ions,
the magnetic field or external pressure.\\

One of the key features underlying the above phenomena is the low dimensionality of the
materials. The confinement of the carrier motion to one or two spatial dimensions
together with the low charge-carrier concentration enhance the effect of the interactions
between the electrons. Another important feature specific to these molecular systems is a
considerably strong coupling of the charge carriers to the lattice degrees of freedom.
For molecular crystals, this coupling includes both the interactions with high-frequency
intramolecular modes and the low-lying intermolecular vibrations. The present organic
metals thus represent ideal model systems for exploring the interplay of strong
electron-electron and electron-phonon interactions in reduced dimensions. In particular,
the close proximity of superconductivity to a magnetically-ordered state encountered not
only in the organic materials but also in other strongly-correlated electron systems such
as the heavy-fermion metals and high-temperature superconductors continues to challenge
our understanding of superconductivity. While for the latter systems evidence is growing
that, indeed, the magnetism and superconductivity are intimately related to each other,
the nature of the superconducting state for the present materials is far from
being understood.\\
The discussion of the superconducting-state properties for the present organic materials
is complicated by the existence of contradictory experimental evidences. In some cases,
this controversy even encompasses results on the same quantity when the measurements have
been carried out by different groups. While about one half of the data seem to support an
anisotropic superconducting state, with a $d$-wave order parameter being the most favored
one, the other half are consistent with an order parameter which
 is finite everywhere on the Fermi surface.\\
Closely related to the issue of superconductivity in the proximity to antiferromagnetic
order is the nature of the state above $T_c$. For the quasi-two-dimensional organic
conductors, for example, unusual metallic properties have been observed posing the
question whether this phenomenon is related to the equally unusual behavior seen in the
other class of quasi-two-dimensional
superconductors, the high-$T_c$ cuprates \cite{McKenzie 97}.\\
This article will give an overview on the normal- and superconducting-state properties of
organic superconductors. There are a number of review articles on this subject
\cite{Ishiguro 98,Jerome 02,Bulaevskii 88,Farges 94,Lang 96,Singleton 02} - most of them
focus on either the quasi-one-dimensional or two-dimensional materials. The intention of
the present review is therefore to provide a discussion which covers aspects common to
both families on the same footing. Instead of reviewing the whole diversity of behaviors
found among the various compounds with all their structural and chemical modifications,
we will mainly focus on selected compounds of both families. These are the most
extensively studied and best characterized (TMTSF)$_2$X and (BEDT-TTF)$_2$X slats whose
properties are representative for a wide class of materials. Special attention is paid to
the more recent developments including the controversial discussions of some aspects, in
particular the discussion on the nature and symmetry of the superconducting state.

\clearpage

\section{Characteristics of organic charge-transfer conductors}\label{Kapitel2}
\subsection{Molecular building blocks}
The prerequisites of forming conducting molecular solids are essentially (i) the creation
of unpaired electrons and (ii) their delocalization throughout the crystal. For the
organic superconductors discussed in this article, condition (i) is satisfied by a
partial transfer of charge between the two constituent parts of a \index{charge transfer}
charge-transfer complex: an organic electron-donor molecule D is combined with an - in
most cases inorganic - electron-acceptor complex X according to the reaction $[{\rm D}_m]
+ [{\rm X}_n]$ $\rightarrow$ $[{\rm D}_m]^{+ \delta} + [{\rm X}_n]^{- \delta}$, where $m$
and $n$ are integers. Since the $\pi$-electrons \index{$\pi$-electrons, -orbital} of the
D molecule with orbits extending perpendicularly to the planes of the molecules have low
binding energies - much lower than those of \index{$\sigma$-electrons, -orbital} the
$\sigma$-electrons - they can easily be excited which can be seen in the charge-transfer
process. Accordingly, the so created charge carriers have a $\pi$-electron (hole)
character. In most cases, the negatively charged anions $[{\rm X}_n]^{- \delta}$ adopt a
closed-shell configuration and thus do not contribute to the electrical conductivity.
When the crystal is formed, a delocalization of the charge carriers (ii) may be obtained
by a dense packing \index{packing} of the donor molecules. As a result, the
$\pi$-orbitals of the partially filled outer molecular shells overlap and an electronic
\index{band structure} band structure is formed.\\ The conductivity of solids depends on
both the number of free carriers and their mobility in the crystal. In organic
charge-transfer \index{charge transfer} salts, the carrier concentration \index{carrier
concentration} is determined by the electronegativity of the donor, the electron affinity
of the acceptor molecule and the chemical bonds. A high mobility is obtained by a
relatively large bandwidth \index{bandwidth} which is the result of a considerably strong
overlap of the $\pi$-orbitals \index{$\pi$-electrons, -orbital}
from adjacent molecules.\\
Organic superconductors have been derived from a variety of different organic
electron-donor molecules, where most of them are derivatives of the archetype TTF
\index{TTF} molecule. Its combination with the electron acceptor TCNQ \index{TCNQ}
(tetra\-cyano\-quino\-di\-methane) in 1973 led to the synthesis of the first
quasi-one-dimensional (quasi-1D) organic conductor TTF-TCNQ \cite{Coleman 73,Ferraris 73}.
This material and its de\-ri\-va\-tives served for a long time as model systems for
exploring the physical properties of quasi-1D conductors.

\begin{figure}[t]
\includegraphics[width=\textwidth]{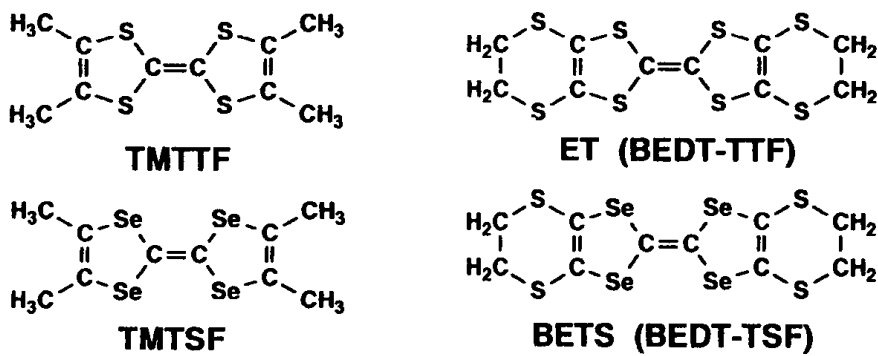}
\caption[]{Principal structures of donor molecules that furnish organic
superconductors.\\}\label{Donor}
\end{figure}

In the vast majority of D$_m$X$_n$ salts, the donor-acceptor molecular ratio $m:n$ is
fixed to $2:1$, i.e., two donor molecules transfer \index{charge transfer} one electron to
the acceptor X. Organic and organometallic chemistry have provided an enormous number of
donor molecules which may serve as building blocks for organic conductors and
superconductors. The main examples are the TTF \index{TTF} derivatives \index{TMTSF}
TMTSF, \index{TMTTF} TMTTF\,\footnote{TMTTF stands for tetramethyltetrathiafulvalene},
\index{BEDT-TTF} BEDT-TTF (commonly abbreviated as ET) \index{ET} and
BEDT-TSF\,\footnote{BEDT-TSF stands for bisethylenedithio-tetraselenafulvalene}
\index{BEDT-TSF} (or simply \index{BETS} BETS), see Fig.~\ref{Donor}. The TMTSF
\index{TMTSF} differs
from TMTTF \index{TMTTF} in that the four sulfur atoms of the latter have been replaced by
selenium. In the same way, BEDT-TSF \index{BEDT-TSF} is derived from \index{BEDT-TTF}
BEDT-TTF. It is interesting to note that organic superconductors have been synthesized
not only by using these symmetric molecules, but also on the basis of asymmetric
complexes. These are the DMET (dimethylethylenedithio-diselenedithiafulvalene)
\cite{Papavassiliou 88,Kini 89}, which is a hybride of TMTSF and \index{BEDT-TTF}
BEDT-TTF, MDT-TTF (methylenedithio-TTF) \cite{Papavassiliou 88,Kini 89} and BEDO-TTF
(bis\-ethylene\-di\-thioxy-TTF) \cite{Beno
90}, see \cite{Ishiguro 98} for an over\-view.\\
The TMTSF \index{TMTSF} molecule provides the basis for the so-called \index{Bechgaard
salts} Bechgaard salts (TMTSF)$_2$X which form with a variety of inorganic monovalent
acceptor molecules X. Indeed, it was (TMTSF)$_2$PF$_6$ where in 1979 superconductivity
had been observed for the first time in an organic compound \cite{Jerome 80}. At
ambient-pressure conditions, the material was found to undergo a metal-to-insulator
transition around 12\,K which had been identified as a spin-density-wave (SDW)
\index{density wave} ordering \cite{Jerome 02}. By the application of hydrostatic
pressure of 12\,kbar the SDW instability can be suppressed and superconductivity forms
below $T_c=0.9$\,K \cite{Bechgaard 80,Jerome 80}. A replacement of PF$_6$ by ClO$_4$ has
resulted in the first and - till now - only member of the quasi-1D salts which becomes
superconducting at ambient pressure \cite{Bechgaard 81}.\\

Apparently, the strong tendency of the Bechgaards salts to undergo a metal-insulator
transition, inherent to quasi-1D electron systems, counteracts with their ability to
become superconducting. In order to achieve ambient-pressure superconductivity in these
organic complexes with possibly even higher transition temperatures it is necessary to
increase their dimensionality. According to this strategy, in 1982 the first metallic
compound based on the new electron donor molecule \index{BEDT-TTF} BEDT-TTF
\cite{Montgomery 94}, cf. Fig.~\ref{Donor}, was synthesized by Saito et al.\ \cite{Saito
82}. The underlying idea was to enhance the overlap between $\pi$-orbitals
\index{$\pi$-electrons, -orbital} of adjacent molecules by enlarging the $\pi$-electron
system on each molecule. This has been accomplished by adding rings of carbon and sulfur
atoms at the outer ends of the TTF \index{TTF} skeleton. In contrast to the Bechgaard
salts \index{Bechgaard salts} where the donor molecules form infinite stacks, steric
effects specific to the \index{BEDT-TTF} BEDT-TTF molecules prevent such an infinite
face-to-face stacking in the (BEDT-TTF)$_2$X salts. As a consequence, the side-by-side
overlap between $\pi$-orbitals of adjacent molecules becomes stronger and, in some cases,
comparable to the face-to-face interaction resulting in a quasi-2D electronic structure
of the BEDT-TTF-based salts. The combination of BEDT-TTF with the monovalent anion X =
Cu(NCS)$_2$ achieved in 1988 led to the discovery of the first ambient-pressure
superconductor in this class of materials - the second generation of organic
superconductors - with transition temperatures in the range of 10\,K \cite{Oshima 88}.

The quasi-1D and -2D systems tend to grow in needle- and plate-like shapes, with the
latter usually showing well developed smooth and shiny metallic surfaces. The synthesis
of organic molecular conductors can be divided into two steps: (1) the synthesis of the
neutral donor and acceptor molecules and (2) the oxidation and reduction of the donors
and acceptors to radical cations and radical anions. For step (1) several independent
methods have been discussed and used even for selected donor and acceptor molecules, see
e.g.\ \cite{Montgomery 94}. Conversely, to perform step (2), the redox process, the
electrocrystallization technique has proved to be the method of choice for synthesizing
high-quality crystals. For details on the chemical synthesis and crystal growth
\index{crystal growth} techniques, the reader is referred to \cite{Williams 92,Montgomery
94,Ishiguro 98} and references cited therein.

\subsection{Structural aspects}\label{Structural aspects}
\begin{figure}[]
\sidecaption
\includegraphics[width=.5\textwidth]{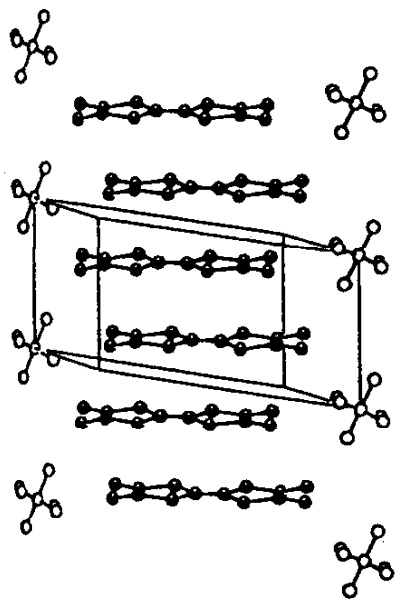}
\caption[]{Crystal structure of (TMTSF)$_2$PF$_6$ (side-viewed tilted). The axis with the
highest conductivity is the vertical $a'$-axis, the one with the lowest conductivity the
horizontal $c$-axis. $a'$ is the projection of the $a$-axis onto the direction
perpendicular to the $bc$-plane.\\}\label{PF6} 
\end{figure}
Figure~\ref{PF6} shows exemplarily the crystal structure of (TMTSF)$_2$PF$_6$ viewed
almost perpendicular to the stacking axis ($a$-axis) - the direction of highest
conductivity. All members of the (TM)$_2$X family, where TM stands for TMTSF and TMTTF,
are isostructural with triclinic symmetry. The conducting stacks are separated by the
anions which are located at inversion centers of the lattice. The donor molecules which
are nearly planar and almost perpendicular to the chain axis are arranged in a
zigzag-type manner along the chains with two slightly different intrachain distances
corresponding to a weak \index{dimerization} dimerization. The fairly close ${\rm Se}
\cdots {\rm Se}$ distance of 3.87\,$\rm{\AA}$ along the $b$-axis being smaller than the
sum of the van der Waals radii \index{van der Waals radius} of 4\,$\rm{\AA}$ results in a
weak interchain overlap \cite{Jerome 02} and thus a weakly 2D electronic character. An
important structural aspect of the (TM)$_2$X compounds is related to the symmetry of the
anion X. Where these anions are centrosymmetric such as the octahedral X = PF$_6$,
AsF$_6$, SbF$_6$ or TaF$_6$ complexes, their orientation is fixed in the structure. In
contrast, (TM)$_2$X salts formed with non-centrosymmetric tetrahedral anions
\index{non-centrosymmetric anion} such as ClO$_4$ or FeO$_4$ undergo a structural
transition from a disordered \index{disorder} high-temperature state to an ordered
low-temperature phase \cite{Moret 86}.\\

\begin{figure}[t]
\includegraphics[width=\textwidth]{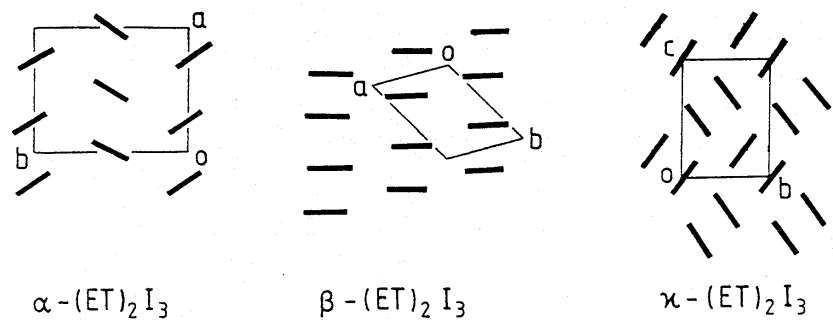}
\caption[]{Schematic packing \index{packing} motifs of the ET molecules in the $\alpha$-,
$\beta$- and $\kappa$-phases of (ET)$_2$I$_3$ viewed along the long axis of the ET
molecules.}\label{alphabetakappa}
\end{figure}

While all members of the (TM)$_2$X family share the same crystal structure, the rather
loose intra-stack coupling of the (BEDT-TTF)$_2$X salts gives rise to a variety of
polymorphic phases (packing motifs) \index{packing} which are distinguished by Greek
characters; the most important amongst them are the $\alpha$-, $\beta$-, and
$\kappa$-phases, see e.g.\ \cite{Mori 98u99} for a comprehensive review of the structural
properties of these salts. In some cases, as for instance realized in the compound with
the linear anion X = I$_3$, various structural modifications exist even for the same
anion, cf. Fig.~\ref{alphabetakappa}. The $\alpha$-type structure consists of stacks
arranged in a herring-bone pattern. The ET molecules are connected via ${\rm S} \cdots
{\rm S}$ contacts being shorter than the sum of the van-der-Waals radii \index{van der
Waals radius} of $3.6$\,\AA. The $\beta$-type packing \index{packing} is reminiscent of
the stacking arrangement found in the \index{Bechgaard salts} Bechgaard salts. However,
the smaller inter-stack distances in
$\beta$-(ET)$_2$I$_3$ lead to a more two-dimensional electronic structure. The
$\kappa$-phase is unique in that it does not consist of interacting stacks but rather of
interacting dimers \index{dimerization} formed by two face-to-face aligned ET \index{ET}
molecules. Adjacent dimers are arranged almost orthogonal to each other so that the {\em
intra}- and {\em inter}-dimer interactions are of the same size. This results in a
quasi-2D electronic structure with a small in-plane anisotropy. The $\kappa$-type
compounds with polymere-like anions are of particular interest with respect to their
superconducting properties as they exhibit the highest transition temperatures.\\ In
forming the crystal, apart from the ethylene [(CH$_2$)$_2$] groups at the outer ends of
the molecules, the charged ET molecules C$_6$S$_8$[(CH$_2$)$_2$]$_2$ untwist at their
centre and become planar.
\begin{figure}[b]
\sidecaption
\includegraphics[width=.5\textwidth]{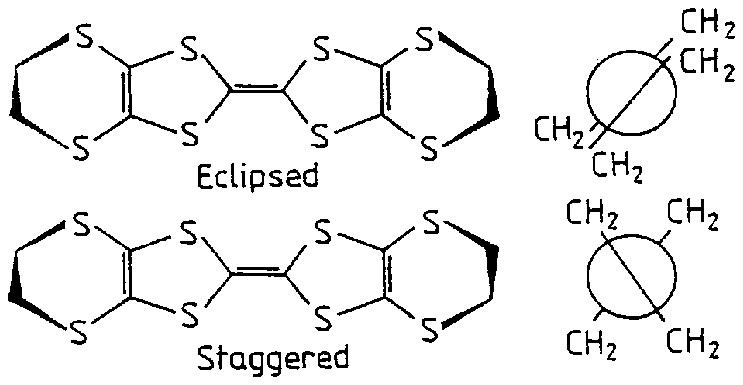}
\caption[]{Schematic view of the relative orientations of the ethylene endgroups
[(CH$_2$)$_2$] of the ET molecule. Right side shows the view along the long axis of the
molecule.}\label{Ethylen}
\end{figure}
As shown schematically in Fig.~\ref{Ethylen}, the relative orientation of the outer C$-$C
bonds can either be parallel (eclipsed) \index{eclipsed conformation} or canted
\index{staggered conformation} (staggered). At high temperatures, the ethylene endgroups
\index{ethylene endgroups} become disordered \index{disorder} due to the strong thermal
vibrations. Upon cooling to low temperatures, the endgroups \index{ethylene endgroups}
adopt one of the two possible \index{ethylene conformation} conformations, depending on
the anion and the crystal structure. As will be discussed in section~\ref{glassy
phenomena} for the $\kappa$-(ET)$_2$X salts, disorder \index{disorder} in the
conformation of the [(CH$_2$)$_2$] groups can have a severe influence on the electronic
properties in these compounds, in particular the superconductivity.\\ The planar
C$_6$S$_8$ skeleton of the ET molecules \index{ET} permits a rather dense packing
\index{packing} with a variety of possible packing \index{packing} arrangements. As a
result, the interdimer \index{dimerization} interaction becomes comparable to that within
the dimers giving rise to a quasi-2D electronic structure. Besides the intermolecular
${\rm S} \cdots {\rm S}$ contacts, i.e.\ the donor-donor interaction, the donor-acceptor
couplings also play an important role for the physical properties of these multilayer
systems. The latter interaction is provided by electrostatic forces as a consequence of
the charged molecules and the hydrogen bonds \index{H-bonding} joining between the carbon
atoms at the donor site and the sulfur, carbon or nitrogen atoms being located at the
acceptor site. The relative strength of these different interactions, the conformational
degrees of freedom of the ethylene groups along with the flexibility of the molecular
framework give rise to a variety of different ET complexes \cite{Ishiguro 98,Mori 98u99}.
\begin{figure}[]
\center
\includegraphics[width=\textwidth]{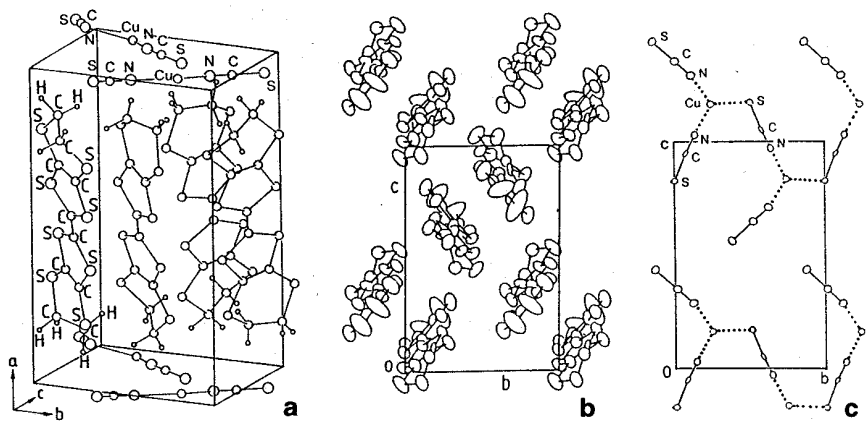}
\caption[]{(\textbf{a}) Crystal structure of \Cuncs. The arrangement of the ET molecules
(\textbf{b}) and the Cu(NCS)$_2$ anions (\textbf{c}) when viewed along the $a^{\ast}$
direction, i.e.\ perpendicular to the conducting planes. The $a$-axis is slightly tilted
from the $a^{\ast}$-axis which is normal to the conducting
$bc$-plane.}\label{KriststrCuNCS}
\end{figure}

\begin{figure}[t]
\sidecaption
\includegraphics[width=.5\textwidth]{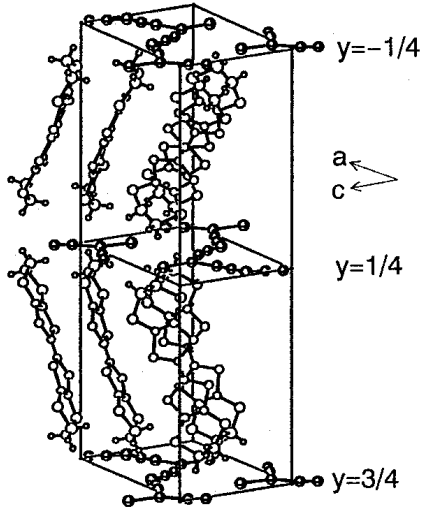}
\caption[]{Crystal structure of $\kappa$-(BEDT-TTF)$_2$Cu[N(CN)$_2$]Z, with Z = Br and
Cl. Here the direction perpendicular to the conducting plane is the crystallographic
$b$-axis. The anion layers are parallel to the $ac$-plane at ${\rm y} = -1/4$, $1/4$ and
$3/4$. The polymeric-like anion chains are running along the $a$
direction.\\}\label{KristruBrCl}
\end{figure}

Despite their complex crystal structure with rather low symmetry (cf.\
Table~\ref{Strukturtabelle}) it is convenient to think of the \etzx\ compounds as layered
systems consisting of conducting sheets formed by the ET molecules which are intersected
by more or less thick insulating anion layers. Prime examples are the $\kappa$-phase
\etzx\ salts with X=Cu(NCS)$_2$, Cu[N(CN)$_2$]Br and Cu[N(CN)$_2$]Cl which are the most
intensively studied and best characterized members of this class of materials. These
compounds are of particular interest not only because of their relatively high
superconducting transition temperatures but also owing to certain similarities in their
normal- and superconducting-state properties with those of the high-temperature cuprate
superconductors \cite{Lang 96,McKenzie 97}. Figures~\ref{KriststrCuNCS} and
\ref{KristruBrCl} display the crystal structures of \cuncs\ and
$\kappa$-(ET)$_2$Cu[N(CN)$_2$]Z. In both cases the layered structure consists of
conducting planes with the characteristic $\kappa$-type arrangement of the ET molecules
separated by insulating anion layers. While the crystal structure of \cuncs\ has
monoclinic symmetry with two dimers, i.e.\ two formular units per unit cell, the
$\kappa$-(ET)$_2$Cu[N(CN)$_2$]Z salts are orthorhombic with a unit cell containing four
dimers, see Table~\ref{Strukturtabelle}. Due to the particular arrangement of their
polymeric anions, these crystals lack a center of inversion symmetry.\\ Subtle changes in
the intermolecular spacing or relative orientation of the ET molecules as e.g.\ induced
by either external pressure or anion substitution may significantly alter the
$\pi$-electron overlap \index{$\pi$-electrons, -orbital} between adjacent molecules. This
can have a severe influence on the electronic properties as demonstrated for the
$\kappa$-(ET)$_2$Cu[N(CN)$_2$]Z system for various Z: while the compound with Z=Br is a
\begin{table}[b]
\caption{Room-temperature crystallographic data of some \etzx\ superconductors including
the space group SG, lattice parameters $a$,$b$ and $c$, unit-cell volume $V$, number of
formular units per unit cell $z$ as well as $T_c$ values. In the case of quasi-2D
(ET)$_2$X salts the lattice parameter perpendicular to the conducting planes is
underlined.}
\renewcommand{\arraystretch}{1.5}
\begin{tabular}{lccccccc}
 & SG \ & \ $a\,({\rm \AA})$ \ & \ $b\,({\rm \AA})$ \ & \ $c\,({\rm \AA})$ \ & \ $V\,({\rm
\AA}^3)$ \ & \ z\ \ & \ $T_c\,({\rm K})$ \\ \hline \hline (TMTSF)$_2$PF$_6$ & $P\bar{1}$ &
7.297 & 7.711 & 13.522 & 713.14 & 1 & 1.1 (6.5\,kbar) \\ \hline
$\kappa$-(ET)$_2$Cu(NCS)$_2$
& $P2_1$ & $\underline{16.248}$ & 8.440 & 13.124 & 1688 & 2 & 10.4 \\
$\kappa$-(ET)$_2$Cu[N(CN)$_2$]Br & $Pnma$ & 12.949 & $\underline{30.016}$ & 8.539 & 3317 &
4 & 11.2 \\ $\kappa$-(ET)$_2$Cu[N(CN)$_2$]Cl & $Pnma$ & 12.977 & $\underline{29.977}$ &
8.480 & 3299 & 4 & 12.8 (300\,bar)
\\ $\beta''$-(ET)$_2$SF$_5$CH$_2$CF$_2$SO$_3$ & $P\bar{1}$ & 9.260 & 11.635 & $\underline{17.572}$ & 1836 & 2 &
5.3 \\ $\alpha$-(ET)$_2$NH$_4$Hg(SCN)$_4$ & $P\bar{1}$ & 10.091 & $\underline{20.595}$ & 9.963 & 2008 & 2 & 1.1 \\
$\kappa$-(ET)$_2$I$_3$ & $P{2_{1}}/c$ & $\underline{16.387}$ & 8.466 & 12.832 & 1688 & 2 &
3.5 \\ \hline $\lambda$-(BETS)$_2$GaCl$_4$ & $P\bar{1}$ & 16.141 & 18.58 & 6.594 & 1774 &
2 & 6 \\  \label{Strukturtabelle}
\end{tabular}
\end{table}
supercondutor with $T_c = 11.2$\,K \cite{Kini 90}, replacement of Br by the slightly
smaller Cl results in an antiferromagnetic insulating ground state. On the other hand,
the application of hydrostatic pressure of only about $300$\,bar drives the latter system
to a superconductor with a $T_c$ of 12.8\,K \cite{Williams 90,Wang 91,Sushko 93},
the highest transition temperature found among this class of materials so far.\\

A new class of materials which has recently gained considerable interest is based on the
donor molecule BETS \index{BETS} and its combination with the discrete anions MX$_4$ (M=
Fe, Ga, In; X= Cl, Br). Two structural modifications have been found. These are the
orthorhombic $\kappa$-type structure (Pnma) which results in plate-like crystals and the
triclinic ($P\bar{1}$) $\lambda$-type variant which grow in a needle-like manner
\cite{Kobayashi 93,Kobayashi 93a,Montgomery 92}. The $\lambda$-(BETS)$_2$GaCl$_4$ salt is
a superconductor with $T_c = 6$\,K \cite{Kobayashi 97}. Upon substituting Ga by Fe in
$\lambda$-(BETS)$_2$Fe$_x$Ga$_{1-x}$Cl$_4$ superconductivity becomes continuously
suppressed with increasing $x$ \cite{Uji 02} and, for $x \geq 0.5$, replaced by an
antiferromagnetic insulating ground state.

Some structural data for a selection of organic superconductors are summarized in
Table~\ref{Strukturtabelle}.

\section{Normal-state properties}\label{Kapitel3}
\subsection{Electronic structure}
As for ordinary metals, the electronic properties of organic charge-trans\-fer salts are
determined by the quasiparticles at the Fermi surface (FS).\,\footnote{This implies the
applicability of the Fermi-liquid \index{Fermi liquid} concept which is questionable for
the most anisotropic (TM)$_2$X salts, see e.g.\ \cite{Jerome 94}.} The energy-band
structures \index{band structure} for both the quasi-1D (TM)$_2$X and the quasi-2D \etzx\
salts have been calculated employing a tight-binding scheme with a few simplifications
\cite{Link 3a}. The calculations are based on the assumption that the {\em
intra}molecular interactions are much stronger than the interactions {\em between}
adjacent molecules reducing the complexity of the problem enormously. In a first step,
$\sigma$- and $\pi$-molecular \index{$\sigma$-electrons, -orbital} orbitals
\index{$\pi$-electrons, -orbital} are constructed using linear combinations of atomic $s$-
and $p$-orbitals of the constituent atoms. In the molecular-orbital (MO) approximation,
the electrons (holes) are considered to be spread over the whole molecule and only those
electrons (holes) near the Fermi \index{Fermi surface} surface in the \underline{h}ighest
\underline{o}ccupied (HOMO) \index{highest occupied molecular orbital (HOMO)} and
\underline{l}owest \underline{u}noccupied \underline{m}olecular \underline{o}rbitals
(LUMO) \index{lowest unoccupied molecular orbital (LUMO)} are taken into account. Due to
the overlap between molecular orbitals of adjacent molecules, the corresponding
$\pi$-electrons (holes) \index{$\pi$-electrons, -orbital} are delocalized. Using available
structural data, the overlap integrals \index{overlap integral} and transfer energies can
be obtained from quantum chemistry. These are input parameters for a standard
tight-binding calculation based on molecular orbitals obtained by the extended H\"uckel
approximation (EHA) from which the band structure \index{band structure} and Fermi
\index{Fermi surface} surfaces are derived \cite{Mori 84,Whangbo 90,Geiser 91,Mori 98u99}.

\begin{figure}[t]
\sidecaption
\includegraphics[width=.65\textwidth]{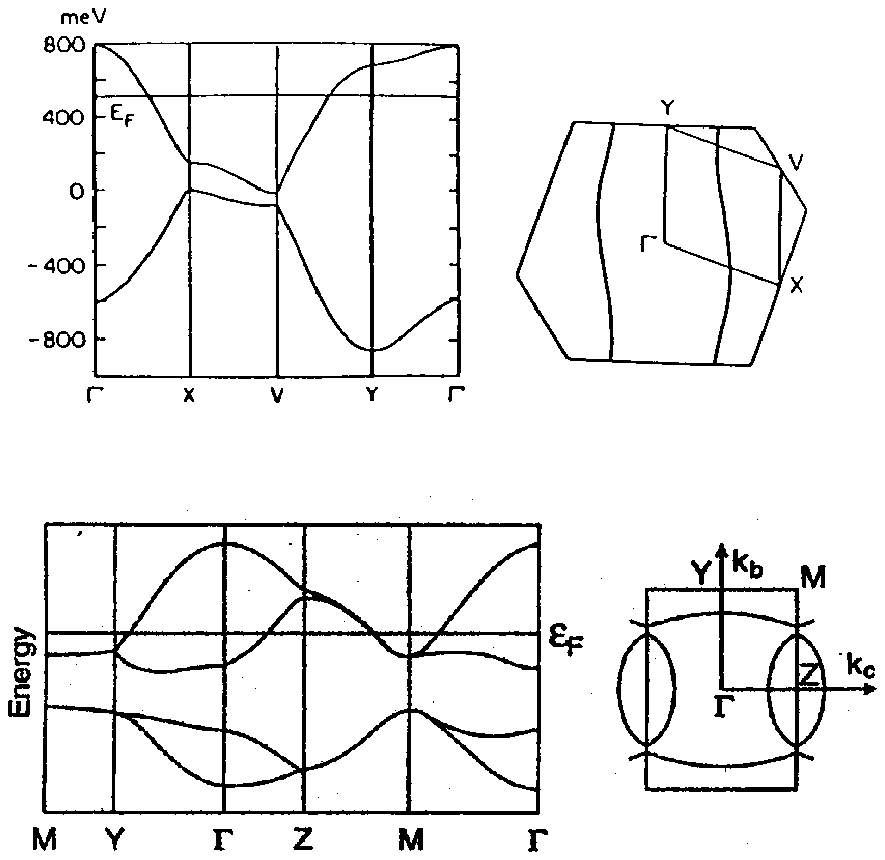}
\caption[]{Calculated energy dispersion and Fermi surface of (TMTSF)$_2$X \cite{Grant 83}
(upper panel) and \cuncs\ \cite{Oshima 88a,Mori 90} (lower panel).\\}\label{FS}
\end{figure}

Based on the above approximations, Grant et al.\ have calculated a model band structure
\index{band structure} for the quasi-1D materials (TM)$_2$X \cite{Grant 83}, see upper
panel of Fig.~\ref{FS}. The FS consists of two open sheets which are slightly corrugated
due to weak interactions perpendicular to the stacking axis. While the standard
magnetic-quantum-oscillation studies cannot be used for these quasi-1D metals, some
important information on the FS can still be derived from angular-dependent
magnetoresistance measurements, see e.g.\ \cite{Danner 94}. Of crucial importance are the
topological aspects of the FS, i.e.\ the nesting \index{nesting} properties, and the
\index{band filling} band filling. The conduction band can accommodate four electrons per
(TM)$_2$ unit. Due to the weak structural dimerization \index{dimerization} which is more
pronounced in the TMTTF compared to the TMTSF salts, a dimerization \index{dimerization}
gap splits the conduction band into two parts. Therefore, removing one electron per unit
cell in the
charge-transfer \index{charge transfer} process results in a half-filled \index{band filling} conduction band.\\

The FS topology of the quasi-2D materials has been studied in great detail employing
measurements of the de Haas-van Alphen (dHvA) and Shubnikov-de Haas (SdH) effect, the
angular-dependent magnetorestistance (AMRO) and the cyclotron resonance, see
\cite{Wosnitza 96,Singleton 00}. These results clearly demonstrate the presence of a
well-defined Fermi surface \index{Fermi surface} and quasiparticle excitations in
accordance with the Fermi-liquid \index{Fermi liquid} theory.\\ The lower panel of
Fig.~\ref{FS} shows the results of EHA band-structure calculations for the superconductor
\cuncs.
Since any {\em inter}layer electron transfer has been neglected in these calculations,
the resulting FS \index{Fermi surface} is strictly two dimensional. Despite the various
simplifications employed, the main features of the so-derived FS are generally found to
be in remarkable agreement with the experimental results \cite{Wosnitza 96}, although a
more elaborated analysis reveals certain details which
are not adequately described \cite{Harrison 99,Singleton 02}.\\
The four bands correspond to the four ET molecules in the unit cell, each represented by
its \index{highest occupied molecular orbital (HOMO)} HOMO. Due to the lack of a
center-of-inversion symmetry, an energy gap opens at the Z-M zone boundary. As a
consequence, the FS consists of closed hole-like quasi-2D orbits ($\alpha$-pockets) and a
pair of open quasi-1D corrugated sheets. According to a charge transfer \index{charge
transfer} of one electron per pair of ET molecules, the conduction band is three quarters
filled. Due to the strong dimerization \index{dimerization} of the ET molecules in the
$\kappa$-type structure, the conduction bands split up so that the upper band becomes
\index{band filling} half filled. Band-structure calculations based on high-temperature
crystallographic data reveal FS topologies which are very similar among the various
$\kappa$-(ET)$_2$X systems \cite{Whangbo 90,Geiser 91}, except for the degeneracy of the
upper two bands along the Z-M zone boundary for the linear anion X = I$_3$. While the
$\alpha$- and $\kappa$-phase (ET)$_2$X salts still combine quasi-1D and -2D bands, the FS
of $\beta$-type salts is even more simple. It is of almost cylindrical shape and closed
within the first Brillouin zone, reflecting the isotropic in-plane interactions between
adjacent ET molecules.

\subsubsection{Fermi-surface studies}
\begin{figure}[b]
\sidecaption
\includegraphics[width=.45\textwidth]{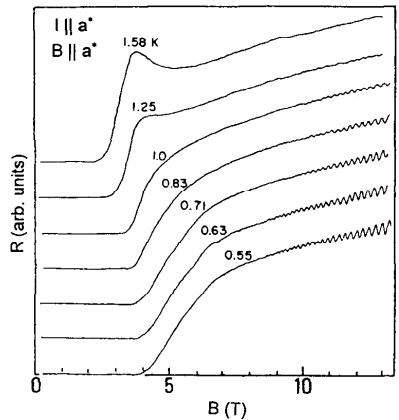}
\caption[]{Magnetorestistance of \cuncs\ at different temperatures. Electrical current
and magnetic field were applied along the $a^\ast$-axis, i.e.\ perpendicular to the
conducting planes. Shubnikov-de Haas oscillations starts to become visible below 1\,K.
Taken from \cite{Oshima 88a}.\\}\label{SdH}
\end{figure}
In a magnetic field $B$, the transverse motion of electrons becomes quantized and their
allowed states in \boldmath $k$\unboldmath-space are confined to so-called Landau levels.
Periodic oscillations in magnetization and resistivity as a function of $1/B$ arise from
the oscillatory behavior of the density of states \index{density of states} at the Fermi
level $E_F$ as the Landau levels pass through the Fermi surface. From the oscillation
period $\Delta(1/B)$ the extremal area of the FS cross section, $S_F$, perpendicular to
the magnetic-field direction can be derived \cite{Onsager 52}:
\begin{equation}
S_F = \frac{2 \pi e}{\hbar} \frac{1}{\Delta (1/B)},
\end{equation}
where $-e$ is the electron charge and $\hbar$ the Planck constant. A quantitative
description of the oscillatory magnetization was given by Lifshitz and Kosevich
\cite{Lifshitz 55}. According to their work, the amplitude of the oscillations is given
by:
\begin{equation}
A \propto \frac{T}{\sqrt{B}} \frac{\exp(-\frac{\lambda m_c^\ast T_D}{m_e
B})}{\sinh(\frac{\lambda m_c^\ast T}{m_e B})},
\end{equation}
where the effect on the electron spin has been neglected. $m_e$ denotes the free electron
mass and $m_c^\ast$ the cyclotron effective mass, $\lambda = 2 \pi^2 m_e k_B / (e \hbar)$
with $k_B$ being the Boltzmann constant. $T_D = \hbar / (2 \pi k_B \tau)$ is the Dingle
temperature \index{Dingle temperature} which accounts for the broadening of the Landau
levels due to scattering of the electrons where $\tau$ is the relaxation time averaged
over a cyclotron orbit. For state-of-the-art crystals of the \etzx\ salts, the Dingle
temperatures are usually far below about 1\,K, as e.g.\ $T_D \sim 0.5$\,K as
reported for \cuncs \cite{Caulfield 94}, which reflect the high quality of these materials.\\
Figure~\ref{SdH} shows early magnetoresistance data on \cuncs\ \cite{Oshima 88a}; see
also \cite{Singleton 00} for more recent data. At the low-field side of the data sets,
the transition from the superconducting to the normal state is visible. With increasing
the field and at temperatures below 1\,K, Subnikov-de Haas (SdH) oscillations caused by
the closed $\alpha$-orbits of the FS \index{Fermi surface} are superimposed. As expected
from the simple FS topology (lower panel of Fig.~\ref{FS}), a single frequency according
to only one extremal orbit ($\alpha$-orbit) dominates the oscillatory behavior at lower
fields. At higher magnetic fields, however, a second high-frequency component becomes
superimposed \cite{Sasaki 90,Sasaki 91}. The latter corresponds to the so-called magnetic
breakdown effect which is due to tunnelling of charge carriers across the energy gaps at
the FS. It is common to refer to the magnetic-breakdown orbit which encompasses the whole
FS as the $\beta$-orbit. For a detailed description of the FS studies on quasi-1D and
quasi-2D charge-transfer salts, see \cite{Wosnitza 96,Singleton 00} and references
therein.

\subsubsection{Effective masses and renormalization effects}
The experimentally derived effective cyclotron masses $m_c^\ast$ for the various
(ET)$_2$X salts are significantly larger than the band masses $m_b$ predicted by the
above band-structure calculations, which are of the order of the free-electron mass $m_e$
(see below). For the \cuncs\ salt, for example, experiments reveal $m_c^\ast = (3.5 \pm
0.1)\,m_e$ for the $\alpha$-orbit and $(6.9 \pm 0.8)$ for the magnetic-breakdown
$\beta$-orbit \cite{Sasaki 90}. A mass enhancement of comparable size is observed also
for the thermodynamic effective mass $m_{th}^\ast$ as determined by measurements of the
specific heat $C(T)$. For a quasi-2D material consisting of stacks of metallic planes with
interlayer spacing $s$, the Sommerfeld coefficient $\gamma^{\rm 2D} = C/T$ is given by:
\begin{equation}
\gamma^{\rm 2D} = \frac{\pi k^2_B}{3} \frac{m_{th}^\ast}{\hbar^2} \frac{1}{s}.
\end{equation}
For \cuncs\ one finds $\gamma = (23 \pm 1)$\,mJ/mol\,K$^2$ \cite{Mueller 02,Wosnitza 02}
which corresponds to $m_{th}^\ast = (4.7 \pm 0.2)\,m_e$. Both $m_c^\ast$ and
$m_{th}^\ast$ are renormalized compared to the band mass $m_b$. The latter takes into
account the fact that the electrons are moving in a periodic potential associated with
the crystal lattice. The band masses estimated from tight-binding calculations and
interband optical measurements are of the order of the free electron mass: Caulfield et
al.\ applied the effective dimer \index{dimerization} model to \cuncs\ and found
$m_b^\alpha = 0.64\,m_e$ and $m_b^\beta = 1.27\,m_e$ \cite{Caulfield 94} corresponding to
a width of \index{bandwidth} the conduction band of $W = 0.5 \sim 0.7$\,eV. These values
have to be compared with $m_b^\alpha = (1.72 \pm 0.05)\,m_e$ and $m_b^\beta = (3.05 \pm
0.1)\,m_e$ as derived from first-principles self-consistent local-density calculations
\cite{Xu 95}. For a discussion on the band masses derived from band-structure
calculations and their relation to the cyclotron effective \index{effective masses}
masses, see e.g.\ \cite{Merino 00}. The substantial enhancement of the cyclotron masses
compared to the band masses suggest an appreciable quasiparticle renormalization due to
many-body \index{many-body effects} effects, i.e.\ electron-electron
\index{electron-electron correlations} and electron-phonon \index{electron-phonon
interaction} interactions. It has been proposed that a direct tool to determine the
relative role of electron-electron correlations \index{electron-electron correlations} in
the mass enhancement is provided by cyclotron resonance measurements \cite{Singleton 92}.
According to the Kohn theorem the effective mass determined by cyclotron resonance
experiments, $m_{cr}^\ast$, is independent of the electron-electron
\index{electron-electron correlations} interactions. As a consequence, the experimental
finding of $m_{cr}^\ast \approx m_b$ has been attributed to a dominant role of the
Coulomb interaction for the mass renormalization \cite{Caulfield 94}. However, recent
studies on different \etzx\ systems along with theoretical calculations showed that the
general applicability of the Kohn theorem for the quasi-2D organic superconductors is
questionable,
see \cite{Singleton 00,Singleton 02}.\\
On the other hand, various experiments such as optical studies \cite{Eldridge 96,Eldridge
97,Pedron 97,Pedron 99,Faulques 00,Girlando 00}, thermal conductivity \cite{Belin
98,Izawa 02,Wosnitza 02} as well as inelastic neutron scattering experiments
\cite{Pintschovius 97} indicate a substantial coupling of the charge carriers to the
lattice vibrations. Taken together, it is likely that for the present molecular
conductors both electron-electron \index{electron-electron correlations} as well as
electron-phonon interactions \index{electron-phonon interaction} are responsible for the
mass-renormalization.

By means of pressure-dependent Shubnikov-de Haas experiments a striking interrelation
between the suppression of superconductivity and changes in the effective masses
\index{effective masses} have been found \cite{Caulfield 94}.
\begin{figure}[t]
\sidecaption
\includegraphics[width=.6\textwidth]{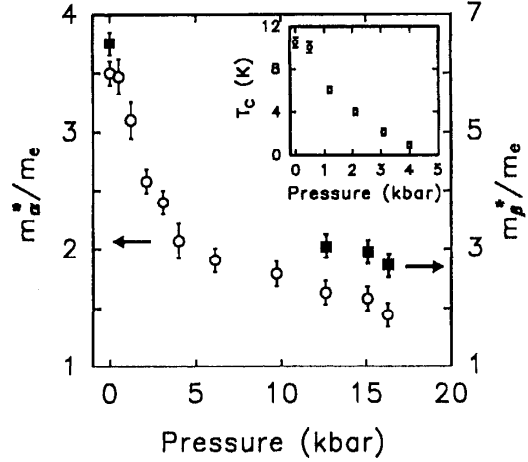}
\caption[]{Cyclotron effective masses of the $\alpha$- (open circles, left scale) and
$\beta$- (full squares and right scale) orbits as a function of hydrostatic pressure for
\cuncs. The inset shows $T_c$ against pressure. Taken from \cite{Caulfield
94}.\\}\label{meffp}
\end{figure}
Figure~\ref{meffp} shows the hydrostatic-pressure dependence of the effective masses
\index{effective masses} for \cuncs\ \cite{Caulfield 94}. As the pressure increases,
$m_c^\ast$ rapidly decreases, an effect which has been observed also for other \etzx\
compounds \cite{Weiss 99}. Above some critical pressure of about $4 \sim 5$\,kbar the
rate of suppression of $m_c^\ast$ becomes much weaker. As this is about the same pressure
value above which superconductivity becomes completely suppressed (see inset of
Fig.~\ref{meffp}), an intimate interrelation between mass enhancement and
superconductivity has been suggested \cite{Caulfield 94}. This is consistent with recent
results of reflectivity measurements which showed that the pressure dependence of the
'optical masses', which are closely related to the bare band masses, do not show such a
crossover behaviour \cite{Klehe 00}. Consequently, the pressure-induced reduction of the
effective cyclotron masses has to be associated with a decrease in the strength of the
electron-electron \index{electron-electron correlations} and/or electron-phonon
\index{electron-phonon interaction} interactions.

\subsection{Transport and optical properties}\label{Transport and optical properties}
\subsubsection{Electrical resistivity}
The organic superconductors discussed in this article are fairly good metals at room
temperature with resistivities that vary over wide ranges depending on the particular
compound and the current direction in respect to the crystal axes. The pronounced
anisotropies found in the electrical properties are direct manifestations of the strongly
directional-dependent overlap \index{overlap integral} integrals.\\ For the (TM)$_2$X
series (cf.\ Fig.~\ref{PF6}) one typically finds $\rho_a$ : $\rho_b$ : $\rho_c$ of the
order of $1$ : $200$ : $30.000$, where $a$ is along the stacking axis. These numbers
correspond to a ratio of the overlap \index{overlap integral} integrals $t_a$ : $t_b$ :
$t_c$ of about $10$ : $1$ : $0.1$ with $t_a \approx 0.1 \sim 0.24$\,eV and $0.36$\,eV for
the TMTTF and TMTSF compounds, respectively \cite{Jerome 02,Jerome 94}.
\begin{figure}[]
\sidecaption
\includegraphics[width=.625\textwidth]{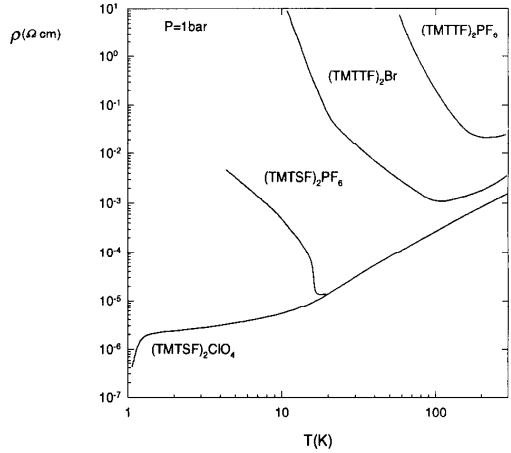}
\caption[]{Resistivity vs temperature for various (TMTTF)$_2$X and (TMTSF)$_2$X salts at
ambient-pressure conditions in a double-logarithmic plot \cite{Wzietek
93}.\\}\label{restm2}
\end{figure}\\
Figure~\ref{restm2} compiles temperature profiles of the resistivity for various (TM)$_2$X
compounds. Below a temperature $T_\rho$, depending on the anion, the resistivity of the
sulfur-containing (TMTTF)$_2$X compounds changes from a metallic-like high-temperature
into a thermally-activated low-$T$ behavior. Upon further cooling through $T_{SP} \simeq
20 K < T_\rho$ (not shown), the (TMTTF)$_2$PF$_6$ salt undergoes a phase transition into a
spin-Peierls (SP)-distorted nonmagnetic ground state, see e.g.\ \cite{Wzietek 93} and
references cited therein. With the application of moderate pressure both $T_\rho$ and
$T_{SP}$ were found to decrease. By increasing the pressure to $p \geq 10$\,kbar, the
spin-Peierls ground state becomes replaced by an antiferromagnetic N\'{e}el state similar
to the one found at normal conditions in (TMTTF)$_2$Br.\\ In the selenium-containing
(TMTSF)$_2$PF$_6$ salt the metallic range extends down to lower temperatures until the
sudden increase in the resistivity indicates the transition into an insulating SDW ground
state. Above a critical pressure of about 6\,kbar, the SDW state of (TMTSF)$_2$PF$_6$
becomes unstable giving way to superconductivity at lower temperatures, cf.\ the phase
diagrams Figs.~\ref{bechpd} and \ref{bechpd2} in section~\ref{phase diagrams}.\\
Interestingly enough, when the spin-Peierls salt (TMTTF)$_2$PF$_6$ is exposed to
sufficiently high pressure in excess of 43.5\,kbar, a superconducting state can be
stabilized \cite{Adachi 00,Wilhelm 01} which completes - for a single compound - the
sequence of ground states indicated in the generic phase diagram in Fig.~\ref{bechpd}.\\
In their metallic regime the resistivity of the (TM)$_2$X salts along the most conducting
direction decreases monotonically with a power-law temperature dependence $\rho \propto
T^\alpha$, where the exponent $\alpha$, depending on the temperature interval, varies
between $1$ and $2$, see e.g.\ \cite{Cooper 94,Jerome 94}. For (TMTSF)$_2$PF$_6$ for
example, $\alpha \approx 1.8$ between $300$ and $100$\,K and approaches approximately $2$
at lower temperatures down to the metal-SDW transition \cite{Jerome 94}. A $T^2$
dependence in the resistivity has been frequently observed not only in the quasi-1D
\cite{Cooper 94} but also for the various \etzx\ salts, see below.\\ A question of high
current interest for the present quasi-1D conductors concerns the nature of their
low-energy excitations. Is a Fermi-liquid \index{Fermi liquid} approach still adequate or
do we have to treat these materials within the framework of a Tomonaga-Luttinger
\index{Luttinger liquid} liquid - a concept which has been proposed for dimensionality
${\rm D} = 1$ \cite{Tomonaga 50,Luttinger 63,Schulz 91}. Arguments in favor of a
Luttinger-liquid behavior have been derived from various observations, not all of which
have been generally accepted. Undisputed are, however, the non-Fermi-liquid features in
the sulfur compounds (TMTTF)$_2$X: these materials undergo a charge localization at
elevated temperatures $T_\rho = 250$\,K for X = PF$_6$ and $100$\,K for X = Br (cf.\
Fig.~\ref{restm2}) which leaves the static magnetic susceptibility unaffected \cite{Jerome
02,Jerome 94}. This apparent separation of spin and charge degrees of freedom is one of
the hallmarks of a Luttinger \index{Luttinger liquid} liquid, see e.g.\ \cite{Bourbonnais
98}. Indications for a spin and charge separation have been reported also for other
(TM)$_2$X salts from optical- and thermal-conductivity experiments \cite{Vescoli
98,Lorenz 02}. The other signatures of a Luttinger liquid are (ii) a power-law decay at
long distances of the spin or charge correlation functions which suppresses long-range
order in $1$D systems and (iii) the absence of any discontinuity in the distribution
function for electron states at the Fermi energy. Indications for (ii) and (iii) have
been reported from NMR \cite{Jerome 94}, photoemission \cite{Dardel 93,Vescoli 00} as
well as transverse ($c$-axis) dc-resistivity \cite{Moser 98} measurements.\\

\begin{figure}[t]
\sidecaption
\includegraphics[width=.6\textwidth]{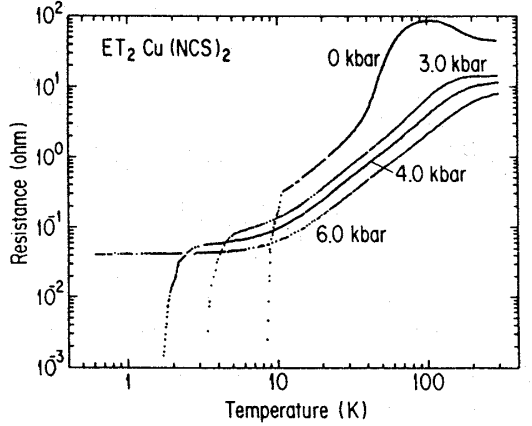}
\caption[]{Temperature dependence of the resistivity measured along the in-plane $b$-axis
of \cuncs\ at various pressures. Taken from \cite{Murata 88}.\\}\label{rkcuncs}
\end{figure}

The resistivity for the various \etzx\ and related compounds can be roughly classified
into two distinct types of temperature dependences. While some of the materials show a
more or less normal metallic-like $T$ behavior, i.e.\ a monotonic decrease of $\rho(T)$
upon cooling, a pronounced $\rho(T)$ maximum \index{resistance maximum} above about
$80$\,K has been found for a number of \etzx\ compounds. Among them are the
$\kappa$-\etzx\ salts with polymere-like anions such as X = Cu(NCS)$_2$ or
Cu[N(CN)$_2$]Br \cite{Kini 90,Murata 88,Sato 91,Schirber 91}, the
$\alpha$-(ET)$_2$NH$_4$Hg(SCN)$_4$ \cite{Oshima 90} as well as the $\kappa$- and
$\lambda$-type BETS salts \cite{Kobayashi 93a}. The occurrence of the same kind of
$\rho(T)$ anomaly in (DMET)$_2$AuBr$_2$ \cite{Kikuchi 88} demonstrates that (i) this
feature is not a property specific to \index{ET} ET- or BETS-based \index{BETS} salts and
(ii) does not rely on the presence of Cu ions. Fig.~\ref{rkcuncs} shows the in-plane
resistivity of \cuncs\ as a function of temperature at various pressures \cite{Murata
88}. With decreasing temperatures, $\rho(T)$ first increases to a maximum
\index{resistance maximum} at around $100$\,K before a metallic behavior sets in at lower
temperatures. Under hydrostatic pressure, the maximum shifts to higher temperatures and
becomes progressively suppressed. This is accompanied by a significant reduction of $T_c$
(see also inset of Fig.~\ref{meffp}). The origin of the anomalous $\rho(T)$ hump has been
discussed by many authors and various explanations have been suggested including the
formation of small polarons \cite{Yamaji 88}, a metal-metal phase transition
\cite{Gaertner 88}, a valence instability of Cu \cite{Toyota 91}, an order-disorder
\index{disorder} transition of the terminal \index{ethylene endgroups} ethylene groups of
the ET molecules \cite{Parker 89,Kund 93,Tanatar 99} as well as a crossover from
localized small-polaron to coherent large-polaron behavior \cite{Wang 00}. In this
context it is interesting to note that for the \brom\ system this maximum
\index{resistance maximum} has been found to be sample dependent: using a different
synthesis route, Thoma et al.\ \cite{Thoma 96} and Montgomery et al.\ \cite{Montgomery
99} succeeded in preparing superconducting crystals which lack the anomalous resistance
hump.\\ A closer look on the resistivity of \cuncs\ below the maximum \index{resistance
maximum} discloses an abrupt change in the slope at temperatures around $45 \sim 50$\,K
\cite{Murata 90}. A similar behavior is found for \brom\ \cite{Sushko 91} and has been
interpreted as a crossover from a regime of antiferromagnetic fluctuations
\index{antiferromagnetic spin fluctuations} of localized spins at high temperatures to a
low-$T$ Fermi-liquid \index{Fermi liquid} regime \cite{Kanoda 97a,Kanoda 97b}. More
recent results, however, suggest that this temperature marks a density-wave-type
\index{density wave} {\em phase transition} \cite{Mueller
02a,Lang 02,Sasaki 02}, see section~\ref{thermal and magnetic} for a detailed discussion.\\
At temperatures below the inflection point, the resistivity turns into an approximate
$\rho(T) = \rho_0 + A T^2$ behavior until superconductivity sets in around 10\,K. As
mentioned above, a resistivity roughly following a $T^2$ law - even at elevated
temperatures - is not an exception in the present molecular conductors, see e.g.\
\cite{Ishiguro 98}. In some high-quality crystals of $\beta$- and $\kappa$-type
(ET)$_2$I$_3$, it has been observed over an extraordinarily wide temperature range up to
temperatures as high as 100\,K \cite{Weger 93}. It has been argued that the $T^2$
dependence of the resistivity indicates a dominant role of electron-electron scattering
in these materials \cite{Bulaevskii 88}. On the other hand, for such a mechanism to
predominate the resisitivity at temperatures as high as $45$\,K for the X = Cu(NCS)$_2$
and even $100$\,K for the X = I$_3$ salt implies that there is only a minor contribution
from electron-phonon scattering. In light of the considerable electron-phonon coupling in
these materials as proved by various experiments, such a scenario appears questionable.
Alternatively, the $T^2$ law has been attributed to the scattering of electrons by
phonons via electron-libron \cite{Weger 81} or a novel electron-phonon scattering
mechanism proposed for the high-$T_c$ cuprates \cite{Weger 93} which invokes
electron-electron \index{electron-electron correlations} interactions \cite{Weger
95,Weger 95a}. For the discussion of the temperature dependence of the resisitivity, it
is important to bear in mind, however, that due to the large pressure coefficients of the
resisitivity of about $\partial \ln \rho /
\partial p \simeq - 20$\,\%/kbar at room temperature together with the extraordinary
strong thermal contraction, it is difficult to make a comparison with theoretical
predictions. Since the theory usually describes the temperature dependence at constant
volume, a detailed comparison is meaningful only after transforming the constant-pressure
into constant-volume profiles by taking into account the thermal expansion of the
material.

Similar to the quasi-1D (TM)$_2$X salts, the room-temperature resistivities of the
quasi-2D (BEDT-TTF)$_2$X materials are generally rather high. For the \cuncs\ salt for
example, one finds $\rho_{b} \approx 6 \cdot 10^4\,{\mu \Omega {\rm cm}}$ and $\rho_{c}
\approx 3 \cdot 10^4\,{\mu \Omega {\rm cm}}$ \cite{Buravov 89}, which exceed the values
for Cu by several orders of magnitude. This is partly due to the relatively low
charge-carrier concentration \index{carrier concentration} of only about $10^{21}\,{\rm
cm}^{-3}$.
\\ In accordance with their quasi-2D electronic structure, a
pronounced in-plane vs out-of-plane anisotropy has been observed which amounts to
$10^{-3} \sim 10^{-5}$ \cite{Singleton 00}. In this respect it is interesting to ask
whether under these conditions the interlayer transport is coherent or not, i.e.\ whether
there is a coherent motion of band states associated with well-defined wave vectors or if
the motion from layer to layer is diffusive and a Fermi velocity perpendicular to the
layers cannot be defined \cite{McKenzie 98}. This question has been addressed in recent
magnetoresistance studies on the \cuncs\ salt \cite{Singleton 02a}. Here the
interlayer-transfer integral \index{overlap integral} has been estimated to be $t_\perp
\approx 0.04$\,meV \cite{Singleton 02a} as compared with $t_\parallel \sim 150$\,meV for
the intralayer transfer \cite{Caulfield 94}. According to this work, the Fermi surface
\index{Fermi surface} is extended along the interlayer direction corresponding to a
coherent transport.

\subsubsection{Optical conductivity}
Optical investigations by means of infrared and Raman measurements provide important
information on the electronic parameters such as the plasma frequency, the optical masses
and also the bandwidths and collision times for the carriers. In addition, they permit an
investigation of vibrational properties and their coupling to the charge carriers. Using
polarized light it is also possible to look for anisotropies in these quantities, as
e.g.\ in the effective \index{effective masses} masses. The optical properties of
quasi-1D and -2D organic conductors have been reviewed by several authors \cite{Jerome
02,Jerome 94,Jacobsen 87,Graja 94,Dressel 97,Degiorgi 00}, see also \cite{Ishiguro
98,Singleton 02}. For a detailed discussion on the normal- and superconducting-state
optical properties of the \etzx\ salts see \cite{Dressel 00}. A summary of Raman results
on \etzx\ salts is
given in \cite{Eldridge 98,Lin 01}.\\
First extensive optical studies of the electronic properties of (TM)$_2$X by Jacobsen et
al.\ \cite{Jacobsen 83} provided information on the energy of charge-transfer processes
and on the electron-phonon \index{electron-phonon coupling} coupling: the large absorption
features observed in the optical conductivity of the TMTTF salts have been assigned to
intra- and intermolecular vibrations \cite{Pedron 94}. These studies have been
supplemented by a series of more detailed investigations covering also the low frequency
range, see \cite{Degiorgi 00} and references therein. In accordance with the expectations
for a strongly anisotropic material with open Fermi \index{Fermi surface} surface, the
optical response of the Bechgaards salts (TMTSF)$_2$X was found to deviate strongly from
that of a simple metal. The main features are a gap-like structure around 25\,meV for X =
PF$_6$ and a zero-frequency mode which grows upon decreasing the temperature. The latter
contribution, having only a small spectral weight corresponding to 1\,\% of the carriers,
is responsible for the metallic conductivity of the compound \cite{Degiorgi 00}. At high
frequencies, i.e.\ at energies in excess of the interchain transfer integral
\index{overlap integral} $t_b$ but below the intraband width $4 t_a$, the data for the
optical conductivity follow a $\sigma_1(\omega) \propto \omega^{-\gamma}$ dependence with
$\gamma = 1.3$ \cite{Degiorgi 00}. Here $\gamma = 4 n^2 K_{\rho} - 5$, where $K_{\rho}$ is
the Luttinger-liquid-correlation \index{Luttinger liquid} parameter and $n$ the degree of
commensurability. From the experimentally derived $\gamma$ value and assuming $n = 2$,
i.e.\ a dominant quarter-filled band Umklapp scattering, $K_{\rho} \simeq 0.23$ has been
determined which agrees reasonably well with photoemission \cite{Dardel 93,Zwick 97,Zwick
98} and transport data \cite{Moser 98}. These observations are consistent with a
dimensional crossover in (TMTSF)$_2$PF$_6$ from a high-temperature Luttinger-liquid
\index{Luttinger liquid} phase to a
low-temperature (anisotropic) 3D Fermi liquid \index{Fermi liquid} induced by interchain coupling.\\

For the various ET salts, the reflectance spectra are generally characterized by
intensive sharp features due to molecular vibrations superimposed on a broad electronic
\begin{figure}[b]
\center
\includegraphics[width=\textwidth]{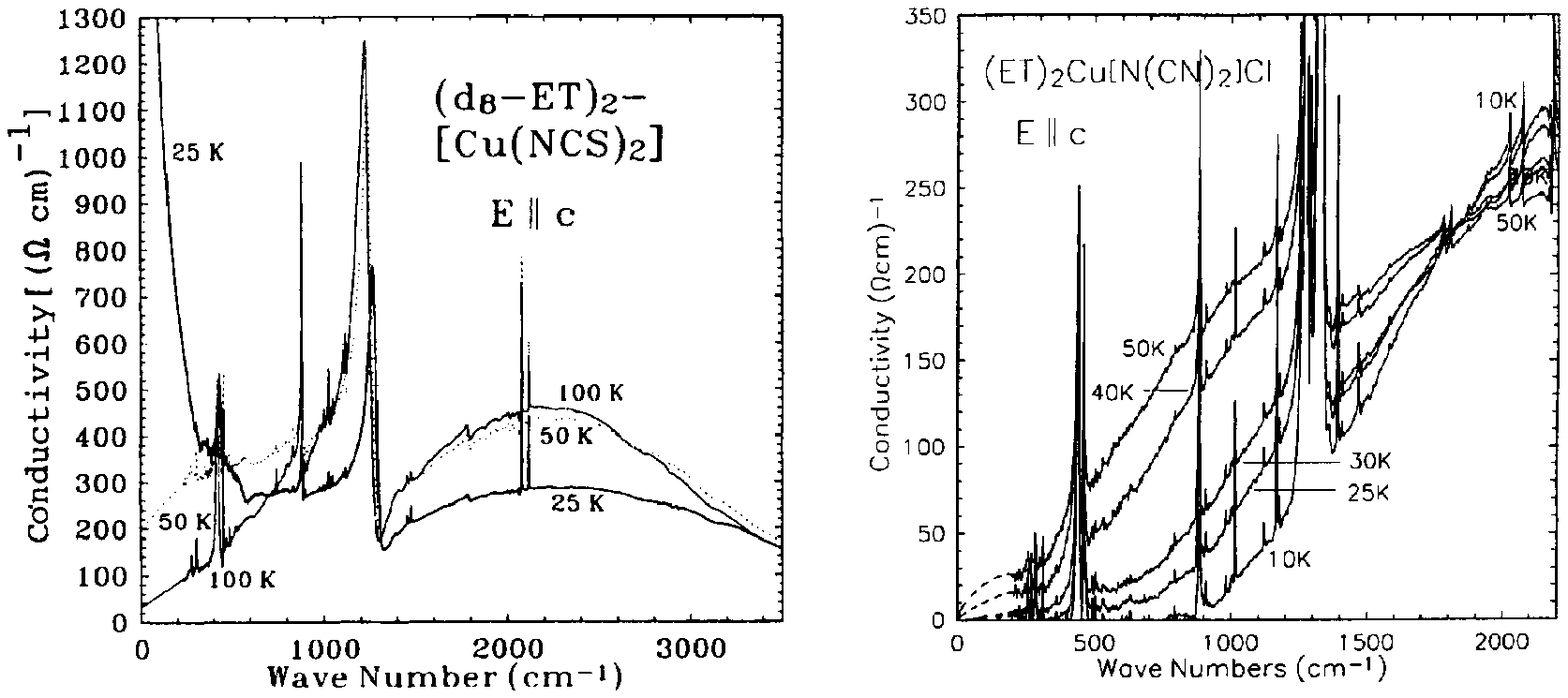}
\caption[]{Optical conductivity of \cuncsd\ (left panel) and \cl\ (right panel) at various
temperatures for $E \parallel c$. Taken from \cite{Kornelsen 91} and \cite{Kornelsen 92},
respectively.}\label{optics}
\end{figure}
background with the plasma edge at a frequency of about $4500 \sim 5000$\,cm$^{-1}$.
Fig.~\ref{optics} shows optical conductivity data of superconducting deuterated \cuncsd\
(left panel) and insulating \cl\ (right panel) at various temperatures obtained from
reflectivity measurements after a Kramers-Kronig analysis \cite{Kornelsen 91,Kornelsen
92}. The far-infrared conductivities have been found to agree reasonably well with the
dc-conductivities in showing a rapid increase below 50\,K for \cuncs. The low-frequency
feature has been interpreted as a Drude peak which increases with decreasing temperature
or increasing pressure, i.e.\ when the metallic character of the material increases. The
data are in good agreement with results of other studies, see \cite{Dressel 00,Wang
00,Singleton 02} and references cited therein. However, the interpretation of the spectra
may vary from author to author: Kornelsen et al.\ attributed the mid-infrared peak ($1000
\sim  4000\,{\rm cm}^{-1}$) to interband transitions superimposed on the free-carrier
tail \cite{Kornelsen 91}. Wang et al.\ \cite{Wang 00} argued that the sharp Drude peak in
the conductivity that develops at low temperature, together with the large mid-infrared
spectral weight indicate polaronic effects. According to this interpretation, the change
from non-metallic to metallic behavior around $90 \sim 100$\,K as observed in the
resistivity is due to a crossover from localized small-polaron to coherent large-polaron
behavior \cite{Wang 00}. In contrast to the metallic/superconducting \cuncs\ salt, the
optical conductivity in the low-energy region for insulating \cl\ rapidly decreases below
50\,K, indicating a large temperature-dependent semiconducting energy gap with a value of
about $900$\,cm$^{-1}$ at 10\,K \cite{Kornelsen 92}.

\subsubsection{Electron-phonon coupling}
Infrared reflectivity measurements can also be used to obtain information on the
electron-phonon \index{electron-phonon interaction} interaction. In molecular crystals,
the coupling between the conduction electrons and the phonons is twofold. One kind of
interaction, the so-called electron-molecular-vibration (EMV)
\index{electron-molecular-vibration (EMV) coupling} coupling, involves the {\em
intra}molecular vibrations which are characteristic of the molecular structure. This has
to be distinguished from the electron-{\em inter}molecular-vibration coupling
\index{intermolecular electron-phonon interaction} which refers to the interaction of the
charge carriers with motions of almost rigid molecules around their equilibrium positions
and orientations (translational and librational modes).

\paragraph{Electron-molecular-vibration coupling}
It is well known \cite{Lipari 77} that electrons in the HOMO's of \index{highest occupied
molecular orbital (HOMO)} the TTF molecule and its derivatives couple strongly to the
totally symmetric ($A_g$) molecular vibrations via the modulation of the HOMO
\index{highest occupied molecular orbital (HOMO)} energy, $E_{\rm HOMO}$, by the atomic
displacements. The linear EMV coupling constant $g_i$ for mode $i$ is defined as:
\begin{equation}
g_i = \frac{1}{h \nu_i} \frac{\partial E_{\rm HOMO}}{\partial Q_i},
\end{equation}
where $Q_i$ is an intramolecular normal coordinate and $\nu_i$ the mode frequency. The
effective electron-intramolecular-phonon coupling constant $\lambda_i$ can then be
calculated using $\lambda_i = 2 g_i^{2} h \nu_i N(E_F)$ where $N(E_F)$ is the density of
states \index{density of states} at the Fermi level.\\ The sharp features superimposed on
the electronic background in Fig.~\ref{optics} can be attributed to molecular vibrations.
An assignment of these features to the various vibrational modes \cite{Link 3d} is
feasible by comparing spectra of different isotopically labelled salts with calculations
based on a valence-force-field model by Kozlov et al.\ \cite{Kozlov 87,Kozlov 89}. For
the ET molecule, the modes with the strongest coupling constants are those involving the
central carbon and sulfur atoms \cite{Kozlov 87,Kozlov 89,Meneghetti 86} at which the
HOMO's \index{highest occupied molecular orbital (HOMO)} have the largest amplitudes
\cite{Mori 84}. The frequencies of these C=C stretching and ring-breathing modes are
$\nu_2 = 1465\,{\rm cm}^{-1} (g_2=0.165), \nu_3 = 1427\,{\rm cm}^{-1} (g_3=0.746)$ and
$\nu_9 = 508\,{\rm cm}^{-1} (g_9=0.476)$, where the calculated coupling constants are
given in the brackets. Despite these sizable coupling constants, several studies -
especially those of the mass isotope shifts on $T_c$ for the ET salts \cite{Kini 96} -
indicate that the EMV \index{electron-molecular-vibration (EMV) coupling} coupling seems
to play only a minor role in mediating the attractive electron-electron interaction, cf.\
section~\ref{Gretchenfrage}.

\paragraph{Electron-intermolecular-vibration coupling}
While much experimental data are available on the EMV \index{electron-molecular-vibration
(EMV) coupling} coupling, relatively little is known about the coupling of the charge
carriers to the low-lying intermolecular phonons. This interaction is provided by the
modulation of the charge-transfer integrals \index{overlap integral} $t_{\rm eff}$
between neighbouring molecules \index{intermolecular electron-phonon interaction} during
their translational or librational motions. Within the Eliashberg theory, the
dimensionless electron-inter\-mo\-le\-cu\-lar-phonon coupling constant $\lambda$ is given
by
\begin{equation}
\lambda = 2 \int \frac{\alpha^2(\omega)}{\omega}F(\omega)d\omega,
\end{equation}
where $\alpha(\omega)$ is the electron-phonon-coupling \index{electron-phonon coupling
constant} constant, $\omega$ the phonon frequency and $F(\omega)$ the phonon density of
states. The Eliashberg function $\alpha^2(\omega)F(\omega)$ can, in principle, be derived
from tunneling characteristics of strong-coupling superconductors or via point-contact
measurements. The latter experiments have been carried out by Nowack et al.\ on the
$\beta$-type (ET)$_2$X salts with X = I$_3$ and AuI$_2$ \cite{Nowack 86,Nowack 87}
yielding $\lambda \simeq 1$. Some of the frequencies of the intermolecular modes have
been determined by employing Raman and far-infrared measurements \cite{Sugai 93,Pokhodnia
93,Dressel 92,Girlando 00}. More recent studies including inelastic-neutron
\cite{Pintschovius 97} and Raman-scattering \cite{Pedron 97,Pedron 99,Faulques
00,Girlando 00} have focused on investigation of the
role of intermolecular phonons for superconductivity. 
These experiments yielded quite sizable superconductivity-induced
phonon-re\-nor\-ma\-li\-za\-tion effects which clearly indicate a significant coupling of
the superconducting charge carriers to the inter\-mo\-le\-cu\-lar phonons and suggest an
important role of these modes in the pairing interaction, cf.\
section~\ref{Gretchenfrage}.

\subsection{Thermal and magnetic properties}\label{thermal and magnetic}
Common to both the quasi-1D and -2D charge-transfer salts is the variety of ground states
the systems can adopt depending on parameters such as the chemical composition or
external pressure. Most remarkable is the fact that for both families the superconducting
phase shares a common phase boundary with a long-range magnetically ordered state.
\newline
For the (TMTSF)$_2$PF$_6$ salt, cooling at ambient pressure leads to a metal-insulator
transition at $T_{MI} \sim 12$\,K. The insulating ground state in the (TMTSF)$_2$X series
has been identified via NMR \cite{Andrieux 82,Barthel 93} and susceptibility
\cite{Mortensen 82} measurements as a spin-density-wave \index{density wave} ordering
whose wave vector \boldmath $q$\unboldmath\ = 2 \boldmath $k_F$\unboldmath\ is determined
by the optimum nesting \index{nesting} condition of the quasi-1D Fermi \index{Fermi
surface} surface \cite{Jerome 02}. The application of hydrostatic pressure to the X =
PF$_6$ salt suppresses the magnetic state by destroying the nesting \index{nesting}
properties and instead stabilizes superconductivity below about $T_c = 1.1$\,K at
6.5\,kbar \cite{Greene 80}. As has been shown again by NMR experiments, short-range
spin-fluctuations \index{antiferromagnetic spin fluctuations} with the same wave vector
\boldmath $q$\unboldmath\ remain active up to $\sim 100$ K, i.e.\ far above the SDW
ordering in this salt. Moreover, these fluctuations are present also in the normal state
of the ClO$_4$ compound despite its
superconducting ground state \cite{Jerome 91}.\\

For the quasi-$2$D $\kappa$-phase \etzx\ compounds, the nesting \index{nesting}
properties are expected to be less strongly pronounced, cf.\ the Fermi \index{Fermi
surface} surface in
\begin{figure}[b] \sidecaption
\includegraphics[width=.55\textwidth]{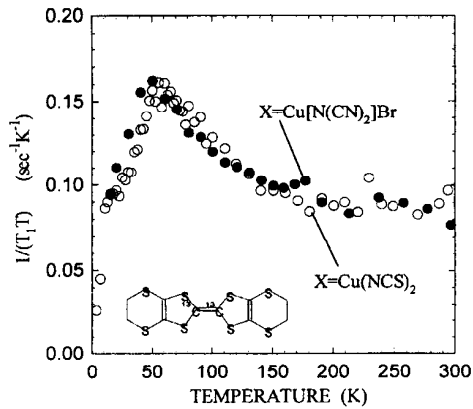}
\caption[]{Spin-lattice relaxation rate divided by temperature, $(T_1 T)^{-1}$, for
\cuncs\ and \brom\ as a function of temperature, taken from \cite{Kawamoto 95a}. For the
$^{13}$C-NMR measurements, the $^{12}$C atoms of the central ${\rm C} = {\rm C}$ double
bond in the ET molecule have been replaced by $^{13}$C.\\}\label{magnetic1}
\end{figure}
Fig.~\ref{FS}. Nevertheless, as in the case of the quasi-1D salts, superconductivity lies
next to an antiferromagnetic insulating state in the pressure-temperature phase diagram,
see section~\ref{phase diagrams}.\ While the compounds with the complex anions
X=Cu(NCS)$_{2}$ and Cu[N(CN)$_2$]Br are superconductors with $T_{c}$ values of 10.4\,K
and 11.2\,K, respectively, $\kappa$-(ET)$_{2}$Cu[N(CN)$_{2}$]Cl is an antiferromagnetic
insulator with $T_{N}=27\,{\rm K}$ which can be transformed into a superconductor with
$T_{c}=12.8\,{\rm K}$ by the application of a small hydrostatic pressure of only
$300$\,bar \cite{Williams 90}. Likewise seen in the quasi-$1$D salts, the metallic state
above $T_c$ in these quasi-2D systems reveals indications for magnetic fluctuations. This
has been demonstrated by NMR measurements on the various $\kappa$-\etzx\ salts
\cite{Mayaffre 94,Mayaffre 95a,Kawamoto 95a,Kawamoto 95b,de Soto 95}. As shown in
Fig.~\ref{magnetic1} the spin-lattice relaxation rate divided by temperature, $(T_1
T)^{-1}$, for the superconductors \cuncs\ and \brom\ behaves quite differently from what
would be expected for a simple metal and realized to a good approximation in some other
organic superconductors as, e.g. $\alpha$-(ET)$_2$NH$_4$Hg(SCN)$_4$ \cite{Kawamoto
95a,Kanoda 97b}. For both $\kappa$-phase compounds, the $(T_1 T)^{-1}$ values at higher
temperatures are enhanced by a factor $5 \sim 10$ compared to a conventional
Korringa-type behavior. Upon cooling, $(T_1 T)^{-1}$ gradually increases down to a
temperature $T^\ast \simeq 50$\,K, below which a steep decrease sets in. Both the overall
enhancement of $(T_1 T)^{-1}$ along with its anomalous peak around $50$ K have been
assigned to the effect of strong antiferromagnetic spin fluctuations
\index{antiferromagnetic spin fluctuations} with a finite wave vector \cite{Mayaffre
94,Wzietek 96,Kawamoto 95a,Kanoda 97b}.
\begin{figure}[t]
\sidecaption
\includegraphics[width=.55\textwidth]{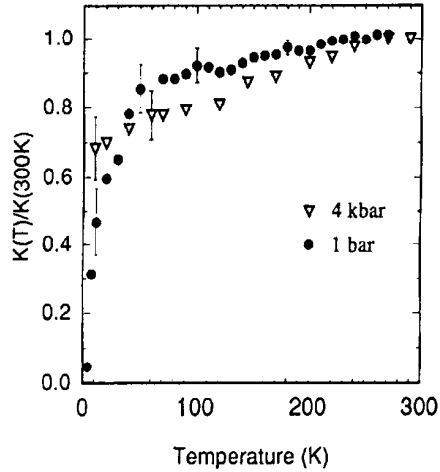}
\caption[]{Knight-shift, $K_S$, for \brom\ as a function of temperature at ambient
pressure and $p = 4$\,kbar, taken from \cite{Mayaffre 94}.\\}\label{magnetic2}
\end{figure}
The latter might be related to the ordering wave vector that characterizes the AF phase
of \cl\ \cite{Mayaffre 94,Wzietek 96,Kanoda 97b}. It has been emphasized by these authors
that despite the rapid drop below $T^\ast \simeq 50$\,K the overall enhancement of $(T_1
T)^{-1}$ persists until the onset of superconductivity indicative of the
highly-correlated nature of the metallic state in these compounds \cite{Kawamoto
95a,Kanoda 97b}. NMR investigations performed at various pressures revealed that with
increasing the pressure, the maximum at $T^\ast$ shifts to higher temperatures while its
size becomes progressively reduced. At pressures above $4$\,kbar the peak is replaced by a
normal Korringa-type behavior, i.e. $(T_1 T)^{-1} = \rm{const.}$ \cite{Mayaffre
94,Mayaffre 95a}. In \cite{Kanoda 97a,Kanoda 97b} the abrupt reduction of $(T_1 T)^{-1}$
below $T^\ast \simeq 50$ has been linked phenomenologically to a spin-gap behavior as
discussed also in connection with the high-$T_c$ cuprates. Indeed, the formation of a
pseudogap was first proposed by Kataev et al.\ \cite{Kataev 92} on the basis of their ESR
measurements. These authors observed a reduction of the spin susceptibility around
$50$\,K, indicating a decrease of the electronic density of states \index{density of
states} at the Fermi level $N(E_F)$ near $T^\ast$. The same conclusion has been drawn
from results of the static magnetic susceptibility \cite{Kawamoto 95a} and Knight-shift
($K_S$) measurements \cite{Mayaffre 94}. Figure~\ref{magnetic2} shows $K_S$ as a function
of temperature for the \brom\ salt. While the data at ambient pressure reveal a clear
drop below about $50$K, a rather smooth behavior with a gradual reduction upon cooling
was found at $4$\,kbar \cite{Mayaffre 94}. The variation of the spin susceptibility in the
high-temperature region has been attributed to the lattice contraction \cite{Wzietek 96}.
Although these experiments show clear evidence for a reduction of the density of states
\index{density of states} at $E_F$ below $T^\ast \simeq 50$, the nature of this
phenomenon and its interrelation to superconductivity are still unclear.

Cooling through $T^\ast \simeq 50$ does not only cause anomalies in the above magnetic
properties but also leads to clear signatures in transport, acoustic, optical and
thermodynamic quantities. As mentioned above, for both superconducting compounds a
distinct peak shows up in the temperature derivative of the electrical resistivity ${\rm
d} \rho / {\rm d} T$ \cite{Murata 90,Sushko 91} indicating a change in the density of
states \index{density of states} at $E_F$. A pronounced softening of ultrasound modes for
\brom\ and \cuncs\ with distinct minima at $T^\ast \simeq 38$\,K and $46$\,K,
respectively, have been attributed to a coupling between acoustic phonons and
antiferromagnetic fluctuations \index{antiferromagnetic spin fluctuations} \cite{Frikach
00,Shimizu 00}. An interaction between the phonon system and magnetism has also been
suggested by Lin et al.\ based on their Raman scattering experiments \cite{Lin 98,Lin
01}. Recent theoretical studies have attempted to explain both the acoustic and Raman
experiments by a correlation-induced crossover from a coherent Fermi liquid \index{Fermi
liquid} at low temperatures to an incoherent bad metal at high temperatures \cite{Merino
00a,Merino 00b}. According to this work, pronounced phonon anomalies as well as anomalous
transport and thermodynamic properties are expected to occur at the crossover temperature
$T^\ast$. Based on their NMR results, Kawamoto et al.\ \cite{Kawamoto 95b} and Kanoda
\cite{Kanoda 97a,Kanoda 97b} argued that $T^\ast$ marks the crossover temperature from a
region of antiferromagnetic fluctuations \index{antiferromagnetic spin fluctuations} of
localized spins at high $T$ to a Fermi-liquid \index{Fermi liquid} regime at low
temperatures. This differs from the interpretation given by the Orsay group
\cite{Mayaffre 94,Mayaffre 95a,Wzietek 96} who analyzed their NMR results in terms of
strong antiferromagnetic fluctuations \index{antiferromagnetic spin fluctuations}
enhanced by Coulomb repulsion and the nesting \index{nesting} properties of the Fermi
\index{Fermi surface} surface.

More insight into the nature of the anomaly at $T^\ast$ and its interrelation with
superconductivity can be obtained by
\begin{figure}[t]
\center
\includegraphics[width=.875\textwidth]{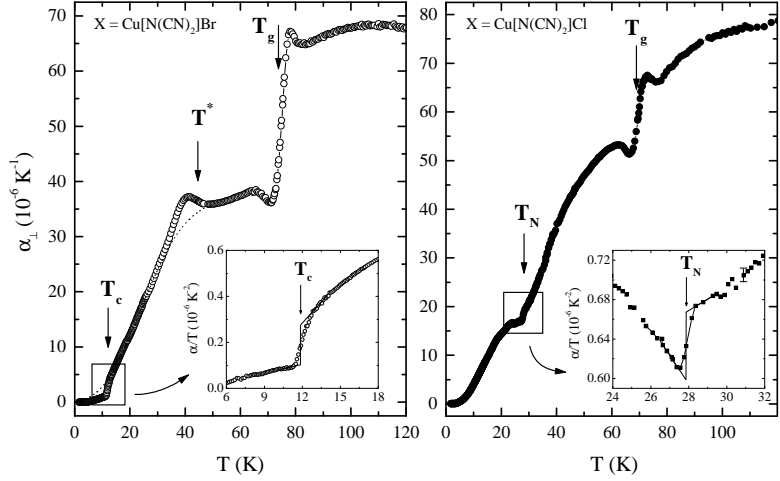}
\caption[]{Linear thermal expansion coefficient perpendicular to the planes of highest
conductivity, $\protect\alpha_{\perp}$, as a function of temperature for superconducting
$\kappa$-(ET)$_2$Cu[N(CN)$_2$]Br (left panel) and insulating
$\kappa$-(ET)$_2$Cu[N(CN)$_2$]Cl (right panel). The inset shows details of
$\protect\alpha_{\perp}$ for both salts as $\protect\alpha_{\perp} / T$ vs $T$ at the
superconducting (X=Cu[N(CN)$_2$]Br) and antiferromagnetic (X=Cu[N(CN)$_2$]Cl) phase
transition. Arrows indicate different kinds of anomalies as explained in the text. Taken
from \cite{Mueller 02a}.}\label{alpha1}
\end{figure}
studying the coupling to the lattice degrees of freedom. This has been done by employing
high-resolution thermal expansion measurements which also allow for studying
directional-dependent effects \cite{Kund 93,Kund 94,Mueller 02a}. Figure~\ref{alpha1}
compares the linear coefficient of thermal expansion, $\alpha (T)=
\partial \ln l(T) /
\partial T$, where $l(T)$ is the sample length, perpendicular to the planes for the
superconducting \brom\ (left panel) with that of the non-metallic \cl\ salt (right
panel). For both compounds, various anomalies have been observed as indicated by the
arrows \cite{Mueller 02a,Mueller 00}. These are (i) large step-like anomalies at $T_{g} =
70 \sim 80$\,K which are due to a kinetic, glass-like transition \index{glass-like
transition} associated with the \index{ethylene endgroups} ethylene endgroups, cf.\
section~\ref{glassy phenomena} and (ii) a distinct peak in $\alpha (T)$ at $T^{\ast}$.
The latter feature, also observed for the superconductor \cuncs\ (not shown) is absent in
the non-metallic \cl\ salt, cf.\ right panel of Fig.~\ref{alpha1} \cite{Mueller 02a}. As
demonstrated in the insets of Fig.~\ref{alpha1}, the transitions into the superconducting
($T_c = 11.8$\,K) and antiferromagnetic ($T_N = 27.8$\,K) ground states for the
X=Cu[N(CN)$_2$]Br and X=Cu[N(CN)$_2$]Cl salts, respectively, are accompanied by distinct
second-order phase
transition anomalies in the coefficient of thermal expansion.\\
Figure~\ref{alpha2} shows the anomalous contribution, $\delta \alpha_i (T) = \alpha_i (T)
- \alpha_{ib} (T)$, to the uniaxial thermal expansion coefficients along the principal
\begin{figure}[t] \center
\includegraphics[width=.9\textwidth]{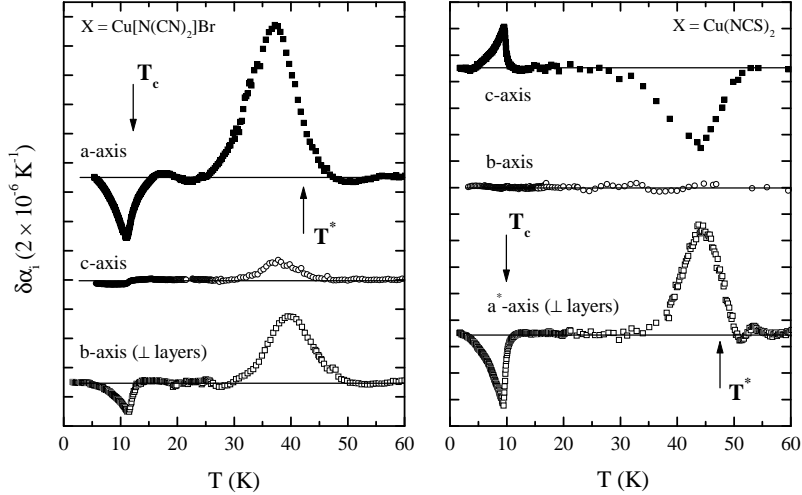}
\caption[]{Anomalous contributions, $\delta \alpha_i (T) = \alpha_i (T) - \alpha_{ib}
(T)$, to the uniaxial thermal expansion coefficients, $\alpha_i (T)$, along the three
principal axes for the superconducting salts \brom\ (left panel) and \cuncs\ (right
panel). Taken from \cite{Lang 02}.}\label{alpha2}
\end{figure}
axes, $\alpha_i (T)$, at $T_c$ and $T^{\ast }$ for the superconducting salts
X=Cu[N(CN)$_2$]Br (left panel) and X=Cu(NCS)$_2$ (right panel) obtained after subtracting
a smooth background $\alpha_{ib}$ (dotted line in the left panel of Fig.~\ref{alpha1})
\cite{Lang 02}. Judging from the shape of the anomalies at \tst, i.e.\ their sharpness
and magnitude, it has been suggested that this feature be assigned to a second-order
phase transition \cite{Lang 02}. Figure ~\ref{alpha2} uncovers an intimate interrelation
between the phase-transition anomalies at \tc\ and \tst: while both features are
correlated in size, i.e.\ a large (small) anomaly at \tc\ complies with a large (small)
one at \tst, they are anticorrelated in sign. A positive peak at \tc\ goes along with a
negative anomaly at \tst\ and vice versa. According to the Ehrenfest relation
\begin{equation}
\left( \frac{\partial T^{\star}}{\partial p_{i}}\right) _{p_{i}\rightarrow 0}=V_{{\rm
mol}}\cdot T^{\star}\cdot \frac{\Delta \alpha _{i}}{\Delta C}, \label{eq:ehrenfest}
\end{equation}
which relates the uniaxial-pressure dependence of a second-order phase-tran\-si\-ti\-on
temperature $T^{\star}$ to the discontinuities in $\alpha_i$, $\Delta \alpha_i$, and that
of the specific heat, $\Delta C$, the above findings imply that the uniaxial-pressure
coefficients of \tc\ and \tst\ are strictly anticorrelated. In \cite{Mueller 02a,Lang 02}
it has been argued that the transition at \tst\ is not of structural but of electronic
origin and related to the Fermi-surface \index{Fermi surface} topology. Based on the
above uniaxial-pressure results, it has been proposed that \tc\ and \tst\ mark competing
instabilities on disjunct parts of the Fermi surface \cite{Mueller 02a,Lang 02}: while
the instability at \tst\ most likely involves only the minor quasi-$1$D fractions (see
Fig.~\ref{FS} and \cite{Mayaffre 94}), the major quasi-$2$D parts are subject to the
superconducting instability at lower temperatures.
\begin{figure}[t]
\sidecaption
\includegraphics[width=.6\textwidth]{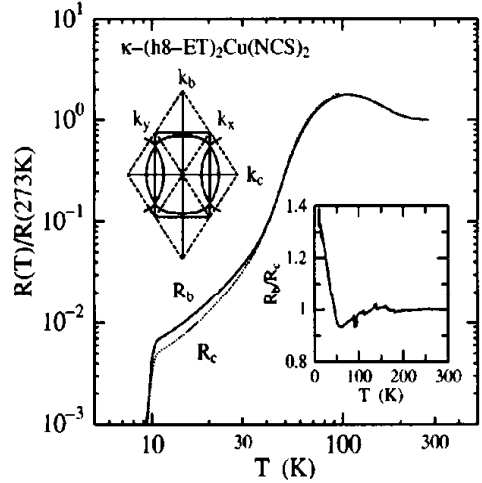}
\caption[]{Temperature dependence of the normalized resistance along the in-plane $b$-
and $c$-axis of \cuncs. The top left inset shows the FS and the Brillouin zone. Taken
from \cite{Sasaki 02}.\\}\label{resaniso}
\end{figure}
As a result, these studies hint at the opening of a real gap associated with \tst\ on the
small $1$D-parts of the FS as opposed to a pseudogap on the major quasi-$2$D fractions.
The condensation of parts of the FS into a density-wave \index{density wave} below
$T^\ast$ would imply the onset of anisotropies in magnetic and transport properties. In
fact, this has been found in recent orientational-dependent studies on both compounds
\cite{Sasaki 02}: cooling through $T^{\ast }$ is accompanied by the onset of a small but
distinct anisotropy in the magnetic susceptibility. As can be seen in
Fig.~\ref{resaniso}, $T^\ast$ affects also the charge degrees of freedom where below
$T^\ast$ the $b$-axis transport becomes more resistive compared to that along the
$c$-axis \cite{Sasaki 02}. These authors proposed that below \tst\ a static or
fluctuating charge-density-wave \index{density wave} (CDW) on minor parts of the FS
coexists with the metallic phase on the remaining quasi-2D fractions.

The above discussion on the nature of the anomalies at \tst\ for the superconducting
salts \brom\ and \cuncs\ poses the question whether this phenomenon is related to the
magnetic signatures found in the non-metallic \cl. Earlier magnetization measurements on
\cl\ revealed a shallow decrease below 45\,K which was interpreted as the onset of an
\index{antiferromagnetic order} antiferromagnetic order \cite{Welp 92}. In addition,
indications were found for a weak ferromagnetic state at 22\,K with a small saturation
moment of $8 \cdot 10^{-4}\,\mu_{B}/{\rm dimer}$. However, according to more recent NMR
experiments, the spins order in a commensurate antiferromagnetic structure below $T_N
\approx 27$\,K with a sizable magnetic moment of $(0.4-1.0)\,\mu_{B}/{\rm dimer}$
\cite{Miyagawa 95}. From these measurements, along with magnetization studies, it has
been inferred that the easy magnetic axis is aligned perpendicular to the planes and that
a small canting of the spins causes a weak ferromagnetic moment parallel to the planes
below about $22 \sim 23$\,K, see also \cite{Pinteric 99}. Recent $^{13}$C-NMR experiments
confirmed the commensurate character
of the magnetic structure yielding a moment of $0.45\,\mu_{B}/({\rm ET})_2$ \cite{Miyagawa 00}.\\
Three different proposals have been put forward on the origin of the magnetic moments and
the nature of the antiferromagnetic insulating state in \cl: (i) electron localization due
to lattice disorder \index{disorder} accom\-pa\-nied by an incomplete compensation of
their spins, i.e.\ an inhomogeneous frozen-in magnetic state \cite{Posselt 94}, (ii) an
itinerant SDW-type magnetism associated with the good nesting \index{nesting} properties
of the quasi-1D parts of the Fermi \index{Fermi surface} surface \cite{Kornelsen
92,Wzietek 96,Tanatar 97} and (iii) a correlation-induced Mott-Hubbard type
metal-insulator tran\-si\-ti\-on \index{Mott-Hubbard insulating state} leading to a
magnetic
state characterized by localized spins \cite{Miyagawa 95}.\\
Although proposal (i) has been ruled out by a recent thermal expansion study providing
clear thermodynamic evidence for a phase transition at $T_{N}$ \cite{Mueller 02a} (see
inset of the right panel of Fig.~\ref{alpha1}), the nature of the ordered state is still
unclear. So far the results of optical, thermal and magnetic properties seem to indicate
that certain elements of the models (ii) and (iii) would be applicable to \cl.

\subsection{Anion ordering and glassy phenomena}\label{glassy phenomena}
In discussing molecular conductors and superconductors, an important issue which should
not be overlooked is disorder \index{disorder} and its possible implications on the
electronic properties. In this respect we have to distinguish between different kinds of
imperfections. The {\em extrinsic} disorder, i.e.\ impurity \index{impurities}
concentrations, contaminations or crystal defects can be vastly controlled in the
preparation process although some aspects remain puzzling, see e.g.\ the discussion on
the resistivity maximum \index{resistance maximum} in section~\ref{Transport and optical
properties}. In a study of the alloy series $\beta$-(ET)$_2$X with X =  
(I$_3$)$_{1-x}$(IBr$_2$)$_x$ ($0 \leq x \leq 1$) - where the salts with the two limiting
compositions with $x=0$ and $x=1$ are superconductors - a clear correlation between the
residual-resistivity ratio (RRR) and \tc\ was found \cite{Tokumoto 87}. These experiments
show that superconductivity is very sensitive to the induced random potentials which lead
to electron localization. The effect of random potentials created by radiation damage
effects - resulting in a suppression of superconducivity - has been studied for the
Bechgaard salts \index{Bechgaard salts} as well as for $\beta$-(ET)$_2$I$_3$. For more
details, the reader is referred to \cite{Ishiguro 98} and references therein.

\subsubsection{(TM)$_2$X salts}
However, certain kinds of {\em intrinsic} disorder are unavoidable and can be of
particular importance for experiments attempting to explore superconducting-state
properties. The latter type of imperfections concerns materials where, by symmetry,
certain structural elements can adopt one of two possible orientations
which are almost degenerate in energy \cite{Pouget 96,Ravy 88}.\\
This can be seen in the (TM)$_2$X salts with non-centrosymmetric anions
\index{non-centrosymmetric anion} such as tetrahedral ClO$_4$. As a result, these anions
are disordered \index{disorder} at room temperature with an equal occupation for both
orientations. Upon cooling, entropy is gained by a more or less perfect ordering of the
anions, depending on how fast the system is cooled through the ordering temperature
$T_{AO}$. A perfect long-range \index{anion ordering} anion ordering, realized to a good
approximation when cooled sufficiently slowly, then introduces a new periodicity of the
lattice. Depending on the anion this can have quite different implications on the
electronic properties: for the (TMTSF)$_2$ClO$_4$ salt, for example, the anion ordering
\index{anion ordering} below 24\,K is accompanied by a doubling of the periodicity along
the $b$-axis, i.e.\ perpendicular to the stacking axis which leaves the conducting
properties almost unaffected. In contrast, the anion ordering \index{anion ordering} for
(TMTSF)$_2$ReO$_4$ opens up a large gap at the Fermi level leading to an insulating
ground state.
\begin{figure}[t]
\center
\includegraphics[width=.875\textwidth]{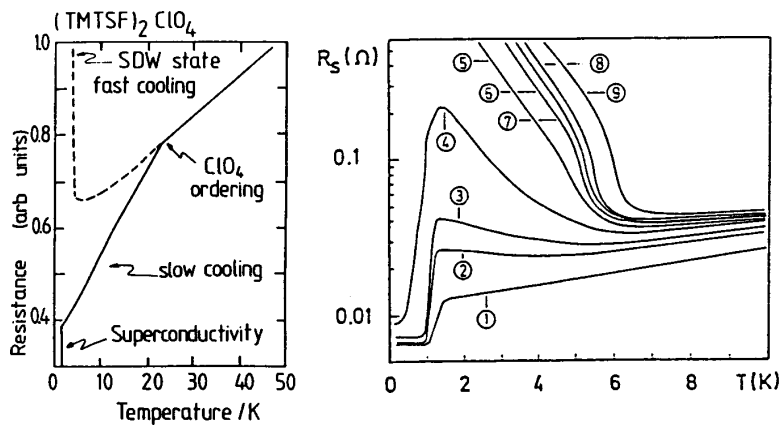}
\caption[]{Left panel: Effects of anion ordering \index{anion ordering} in
(TMTSF)$_2$ClO$_4$ at $T_{AO}=24$\,K, on the resistivity, from \cite{Tomic 83}. Right
panel: Semilogarithmic plot of the resistance vs temperature in states with various
degrees of disorder characterized by $T_Q$ numbered from 1 to 9 (see text), from
\cite{Schwenk 84}.\\}\label{anord}
\end{figure}
If the compound had been cooled quickly through $T_{AO}$, a disordered \index{disorder}
"quenched" state is adopted at low temperatures whose properties are quite different from
the "relaxed" state obtained after slow cooling.\\ The left panel of Fig.~\ref{anord}
shows the effect of anion ordering \index{anion ordering} on (TMTSF)$_2$ClO$_4$. For a
slowly cooled system, the anion-ordering phase transition occurs at $T_{AO} = 24$\,K.
This is accompanied by a decrease of the resistivity due to the reduction of scattering by
randomly distributed anion potentials. Upon further cooling, superconductivity sets in
with $T_c = (1.2 \pm 0.2)$\,K. In contrast, for a crystal cooled rapidly, the
high-temperature disordered \index{disorder} state becomes quenched and the system
undergoes a metal-to-insulator transition at $T_{MI} = 6.05$\,K. The insulating quenched
state has been identified via NMR and ESR measurements \cite{Takahashi 82,Mortensen 83}
as a SDW state with an energy gap $2\,\Delta_0 / k_B T_{SDW} = 3.64$ close to the
mean-field value of 3.52 \cite{Schwenk 84}. The phase transition at $T_{SDW}$ has been
explored recently by specific heat measurements \cite{Yang 00}. The right panel of
Fig.~\ref{anord} shows that the low-temperature resistivity may change over several
orders of magnitude depending on the thermal history of the sample. Intermediate states
with various degrees of frozen anion disorder have been produced by rapidly cooling the
crystal from different temperatures $T_Q \geq T_{AO}$. Recent specific heat measurements
showed that there is no difference between the specific heat anomalies at $T_{AO}$ of
slowly cooled and those of quenched-cooled samples, i.e.\ the structural transition is
independent of the kinetic conditions. From these results, the authors concluded that the
reordering transition of the anions is of second order and occurs within the experimental
timescales \cite{Yang 00}. This is contrary to what would be expected for a
\index{glass-like transition} glass-like transition. Such a glassy behavior, however, has
been observed at the ordering temperatures of the compounds with X = ReO$_4$ ($T_{AO} =
180$\,K) and FSO$_3$ ($T_{AO} = 86$\,K). For more details see \cite{Jerome 94,Ishiguro
98} and references therein.

\subsubsection{$\kappa$-(BEDT-TTF)$_2$X salts}
\begin{figure}[t]
\center
\includegraphics[width=.8\textwidth]{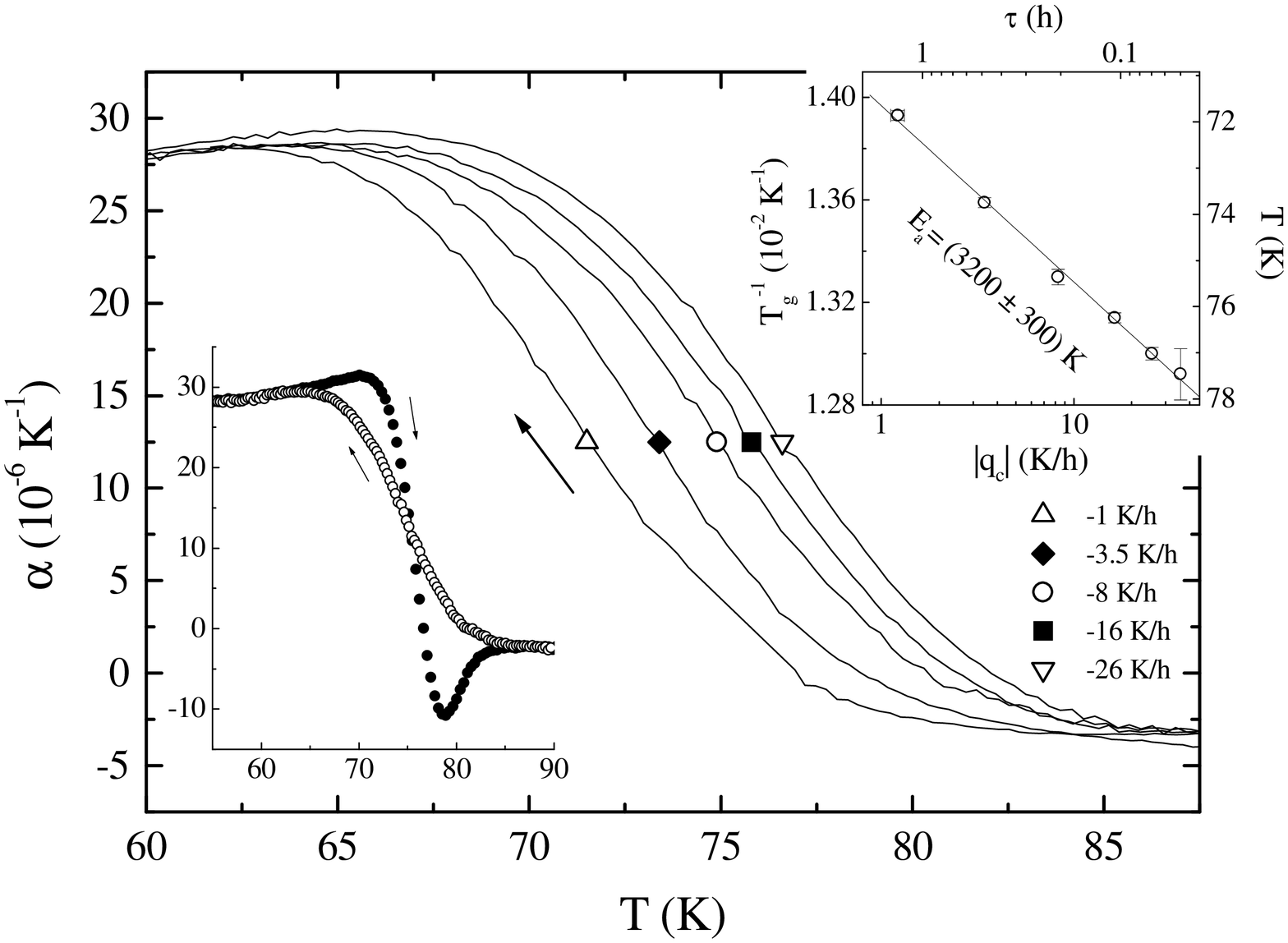}
\caption[]{Linear thermal expansion coefficient, $\protect\alpha$, vs $T$ measured
parallel to the conducting planes of $\protect\kappa$-(ET)$_2$Cu[N(CN)$_2$]Br in the
vicinity of the glass transition defined as the midpoint of the step-like change in
$\alpha$ for varying cooling rates $q_c$. Insets: hysteresis between heating and cooling
curves around $T_g$ (left side) and Arrhenius plot of $T_{{\rm g}}^{-1}$ vs $|q_c|$ and
$\protect\tau$ (right side), where $|q_c|$ is the cooling rate and $\protect\tau$ the
relaxation time. Taken from \cite{Mueller 02a}.}\label{alpha3}
\end{figure}

Indications for frozen-in disorder \index{disorder} have been also reported for the
quasi-$2$D salts of the $\kappa$-(ET)$_2$X family. In an ac-ca\-lo\-ri\-metry study, a
glass-like transition \index{glass-like transition} has been found for \brom\ and \cl\
\cite{Saito 99,Akutsu 00,Sato 01}. The authors observed step-like anomalies in the heat
capacity around 100\,K, which have been attributed to a freezing out of the
intramolecular motions of the ethylene endgroups \index{ethylene endgroups} at the ET
molecules. Clear evidence for a glass-like transition have also been derived from thermal
expansion measurements \cite{Mueller 02a} which are extremely sensitive to structural
rearrangements such as those involved in the glass-like \index{glass-like transition}
freezing process. The outcome of this study confirmed, on the one hand, the
above-mentioned specific heat results and clarified on the other, the nature of the
thermal expansion anomalies previously reported by Kund et al.\ \cite{Kund 93,Kund 94}.
In addition, this study showed that a glass-like transition also exists for the
$\kappa$-(ET)$_2$Cu(NCS)$_2$ salt. A glass transition \index{glass-like transition} is
due to a relaxation process where, below a characteristic temperature \tg\, the
relaxation time $\tau(T)$ of certain structural elements or molecules becomes so large
that they can no longer reach thermodynamic equilibrium. As a result, a short range
order, which is characteristic for this temperature \tg, becomes frozen in.
Figure~\ref{alpha3} shows exemplarily the linear coefficient of thermal expansion for
\brom\ measured parallel to the conducting planes at temperatures near \tg. As discussed
in \cite{Mueller 02a} the anomaly at $T_g \simeq 75$\,K shows all the characteristics
expected for a glassy transition \cite{Cahn 91}, i.e.\ a step in the thermal expansion
coefficient, a pronounced hysteresis between heating and cooling (left inset) and a
cooling-rate dependent \index{cooling-rate dependence} characteristic temperature
$T_{g}$. The inset on the right side of Fig.~\ref{alpha3} shows in an Arrhenius plot the
inverse of the \index{glass-like transition} glass-transition
temperatures, $T_{g}^{-1}$, vs the cooling rate $|q_{c}|$. The data nicely follow a
linear behavior as expected for a thermally activated relaxation time \cite{deBolt
76,Nagel 00}:
\begin{equation}
\tau (T)=\nu_{0}^{-1}\cdot \exp\left(\frac{E_{\rm a}}{k_{B}T}\right), \label{eq:tau}
\end{equation}
where $E_{{\rm a}}$ denotes the activation energy \index{activation energy} barrier. The
pre\-factor represents an attempt frequency $\nu_{0}$. A linear fit to the
data of Fig.~\ref{alpha3} yields $E_{{\rm a}}=(3200\pm 300)$\,K.\\
The characteristic activation energy \index{activation energy} of the [(CH$_{2}$)$_{2}$]
conformational \index{ethylene conformation} motion (cf.\ Fig.~\ref{Ethylen}) was
determined to $E_{{\rm a}}=2650$\,K by $^{1}$H-NMR measurements \cite{Wzietek 96}. The
similar size of the activation energy \index{activation energy} derived from
Fig.~\ref{alpha3} along with the observation of a mass-isotope shift \index{isotope
effect, - substitution} when replacing the hydrogen atoms in [(CH$_{2}$)$_{2}$] by
deuterium provide clear evidence that the ethylene endgroups \index{ethylene endgroups}
are the relevant entities involved in the relaxation process \cite{Mueller 02a}.

An influence of the thermal history of the samples around $70 \sim 80$\,K on the
\begin{figure}[t] \center
\includegraphics[width=.95\textwidth]{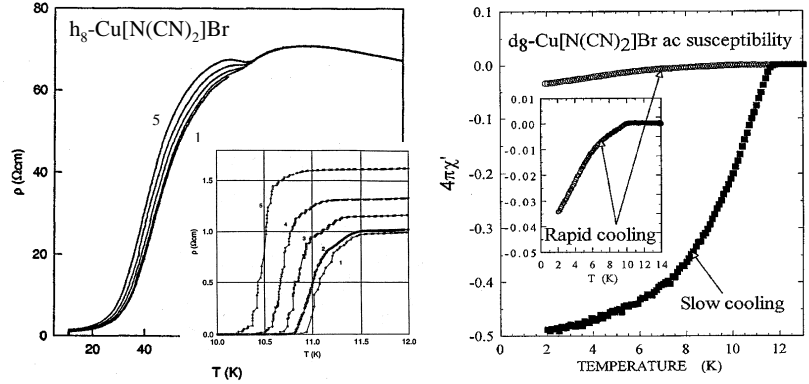}
\caption[]{Left panel: resistivity as a function of temperature for a sample of \bromh\
cooled at different rates ranging from about 0.5\,K/min, lower curve (1), to 60\,K/min
upper curve (5). The inset shows an expansion of the data near the superconducting
transition. Reproduced from \cite{Su 98a}. Right panel: ac-susceptibility of \bromd\
after slow and rapid cooling. Taken from \cite{Kawamoto 97}.}\label{rbr1}
\end{figure}
electronic properties had been realized by various authors and interpreted in different
ways. Based on resistance measurements of structural relaxation kinetics on \brom, Tanatar
et al.\ \cite{Tanatar 99} claimed that the ethylene-endgroup ordering is associated with
a sequence of first-order phase transitions around 75\,K. A pronounced kink in the
resistivity at 75\,K accompanied by hysteresis between heating and cooling had been
reported early on for \brom\ \cite{Watanabe 91,Sato 91}. Similar behavior in the vicinity
of \tg\ was observed also for \cuncs\ \cite{Parker 89}. Besides that, more recent
measurements on \brom\ yielded interesting time dependences affecting not only the
electronic properties around $70 \sim 80$\,K but also the properties at lower
temperatures: Su et al.\ reported relaxation effects in $R(T)$ and a separation of the
curves below about $80$\,K as a function of the \index{cooling-rate dependence} cooling
rate $q_{c}$. As shown in the inset of Fig.~\ref{rbr1} (left panel) the residual
resistivity increases with increasing $|q_{c}|$ \cite{Su 98a,Su 98b}. The inset also
shows that the way of cooling \index{cooling-rate dependence} through $70 \sim 80$\,K may
influence the superconducting properties such that $T_{c}$ decreases on increasing
$|q_{c}|$. In addition, magnetization measurements revealed that with increasing
$|q_{c}|$, a growing amount of disorder \index{disorder} is induced causing an enlarged
penetration \index{penetration depth} depth \cite{Aburto 98}. In reference \cite{Pinteric
02} an ac-susceptibility investigation of the magnetic penetration depths
\index{penetration depth} and their dependence on the cooling-rate-dependent intrinsic
disorder have been performed. The authors found that the superconducting-state
properties are critically determined by the time scale of the experiment around $T_g$.\\
For the deuterated \brom\ salt, it has been reported that rapid cooling
\index{cooling-rate dependence} through $80$\,K drives the superconducting ground state
into an insulating antiferromagnetic state \cite{Kawamoto 97,Ito 00}.\,\footnote{The
ground state of deuterated \bromd\ is strongly sample dependent; there are both
superconducting as well as non-superconducting samples. In \cite{Kawamoto 97} it is
claimed that the crystals always contain superconducting and non-superconducting
components, whereas the latter have a magnetic character, possibly similar to that of
\cl. It is thus believed that the system is situated in the critical region of the phase
diagram just between the superconducting and antiferromagnetic phases, see
section~\ref{phase diagrams}.} As shown in the right panel of Fig.~\ref{rbr1} the
ac-susceptibility data reveal a strong suppression of the superconducting volume fraction
with increasing cooling \index{cooling-rate dependence} rate \cite{Kawamoto 97}. It is
tempting to assign the apparent deterioration of superconductivity to the frozen disorder
\index{disorder} at $T_{g}$: via the C-H\thinspace $\cdots $\thinspace donor and
C-H\thinspace $\cdots $\thinspace anion contact interactions, disorder \index{disorder}
in the ethylene groups introduces a random potential that may alter the effective
transfer integrals \index{overlap integral} $t_{{\rm eff}}$ and, by this, may destroy
superconductivity, see also \cite{Geiser 91}.

\subsection{Phase diagrams}\label{phase diagrams}
\subsubsection{(TM)$_2$X salts}
\begin{figure}[]
\center
\includegraphics[width=.8\textwidth]{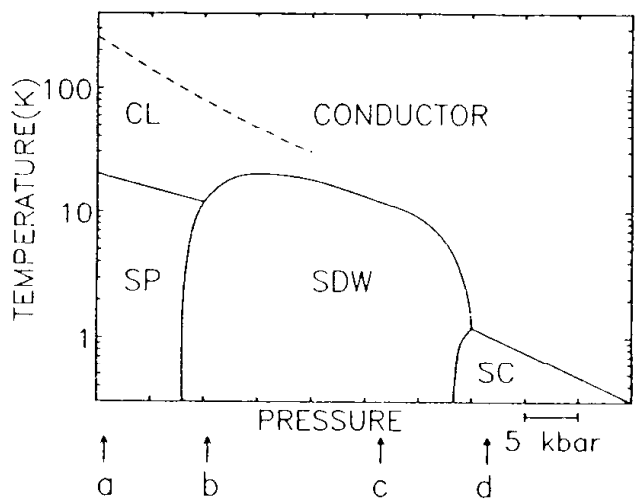}
\caption[]{Generalized phase diagram for the quasi-1D (TM)$_2$X salts as proposed first
by J$\acute{\rm e}$rome \cite{Jerome 91}. Arrows indicate the positions of the salts with
different anions X at ambient pressure: (a) (TMTTF)$_2$PF$_6$, (b) (TMTTF)$_2$Br, (c)
(TMTSF)$_2$PF$_6$, (d) (TMTSF)$_2$ClO$_4$. The following abbreviations are used:
charge-localised insulator (CL), spin Peierls (SP), incommensurate spin-density-wave
(SDW) and superconductivity (SC). Taken from \cite{Balicas 94}.}\label{bechpd}
\end{figure}

\begin{figure}[t]
\center \sidecaption
\includegraphics[width=.6\textwidth]{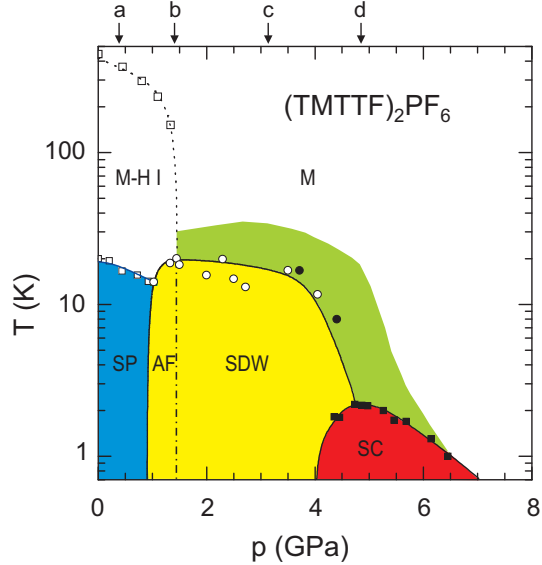}
\caption[]{Temperature-hydrostatic-pressure phase diagram for (TMTTF)$_2$PF$_6$. The
abbreviations are: Mott-Hubbard \index{Mott-Hubbard insulating state} insulating state
(M-H I), metallic (M) and superconducting (SC) state, spin-Peierls (SP), commensurate
(AF) and incommensurate (SDW) antiferromagnetic spin-density-wave \index{density wave}
state. Arrows: ambient-pressure
ground-state location of other salts; 
(a) (TMTTF)$_2$BF$_4$, (b) (TMTTF)$_2$Br, (c) (TMTSF)$_2$PF$_6$, (d) (TMTSF)$_2$ClO$_4$.
Taken from \cite{Wilhelm 01}.\\}\label{bechpd2}
\end{figure}

Figure~\ref{bechpd} comprises in a pressure-temperature plane results of various
experiments on the quasi-1D charge-transfer salts (TMTTF)$_2$X and their sulfur analogues
(TMTSF)$_2$X. The arrows indicate the position of the various salts at ambient pressure
(cf.\ also \ Fig.~\ref{restm2}). The generic character of the phase diagram - first
proposed by J$\acute{\rm e}$rome et al.\ \cite{Jerome 91} - has been demonstrated
recently by pressure studies on the (TMTTF)$_2$PF$_6$ salt \cite{Adachi 00,Jaccard
01,Wilhelm 01} for which the ambient-pressure ground state is a dimerized spin-Peierls
state. With increasing pressure, the system was found to pass through the whole sequence
of ground states as shown in Fig.~\ref{bechpd2} and eventually becoming superconducting
at high pressures above 43.5\,kbar \cite{Jaccard 01,Wilhelm 01}.
\\
On the low-pressure (left) side of the phase diagram  in Figs.~\ref{bechpd} and
\ref{bechpd2} the molecular stacks can be considered as only weakly-coupled chains, i.e.\
the system is close to be truly 1D. In fact upon cooling, the (TMTTF)$_2$PF$_6$ compound
behaves very much like canonical 1D conductors where spin and charge degrees of freedom
are decoupled: below $T_\rho = 250$\,K the resistivity increases by several orders of
magnitude due to charge localization while the spin susceptibility remains unaffected
\cite{Jerome 02,Jerome 94,Moser 98}. The phase below $T_\rho$ has been interpreted as a
Mott \index{Mott-Hubbard insulating state} insulating state. Upon further cooling to
$T_{SP} = 19$\,K a spin gap opens and the system enters a distorted spin-Peierls ground
state. In (TMTTF)$_2$Br, position (b) Figs.~\ref{bechpd} and \ref{bechpd2}, a long-range
magnetic order \index{antiferromagnetic order} is established below the spin-density-wave
\index{density wave} transition at $T_{SDW} \approx 13$\,K.
Toward the right side of the phase diagram, which can be carried out either by varying the
anion or by the application of hydrostatic pressure, inter-stack interactions become more
important. In this region of the phase diagram the electron-phonon interaction
\index{electron-phonon interaction} is less dominant and electron-electron
\index{electron-electron correlations} interactions along with the good nesting
\index{nesting} properties of the Fermi \index{Fermi surface} surface (cf.\
Fig.~\ref{FS}) lead to a spin-density-wave \index{density wave} ground state as observed,
e.g.\ in the Bechgaard salt (TMTSF)$_2$PF$_6$ at ambient pressure. After suppression of
the SP phase in (TMTTF)$_2$PF$_6$ with increasing pressure, a commensurate
antiferromagnetic state is adopted before an incommensurate SDW phase is stabilized. With
increasing pressure, $T_{SDW}$ becomes progressively reduced until, above some critical
pressure, the systems remain metallic and superconductivity replaces the SDW ground
state. The effect of pressure is to increase the $\pi$-orbital \index{$\pi$-electrons,
-orbital} overlap also in the transverse direction, i.e.\ perpendicular to the stacking
axis. As a result the almost perfect nesting \index{nesting} properties are destroyed and
the systems become more 3D in character.\,\footnote{Recent measurements under uniaxial
strain revealed that the SDW transition is most strongly suppressed and superconductivity
can be induced by the strain along the stacking $a$-axis \cite{Guo 00,Hirai 01}.
Considerably smaller effects were observed for strain along the $b'$- and $c'$-axes -
although the former direction is directly associated with the denesting \index{nesting}
of the quasi-1D Fermi surface \cite{Guo 00}. This has been interpreted in terms of a
pressure-induced decrease in the density of states \index{density of states} at the FS
\cite{Miyazaki 00}.} According to NMR experiments \cite{Jerome 94,Wzietek 93} and recent
transport measurements under hydrostatic pressure \cite{Jaccard 01,Wilhelm 01} strong SDW
correlations are still active in the metallic state even when the SDW instability is
replaced by superconductivity in (TMTSF)$_2$ClO$_4$. The range of strong SDW correlations
for (TMTTF)$_2$PF$_6$ derived from these experiments is indicated in
Fig.~\ref{bechpd2} by the shaded region above the SDW and SC phase boundaries. \\

\subsubsection{$\kappa$-(BEDT-TTF)$_2$X salts}
\begin{figure}[t]
\center \sidecaption
\includegraphics[width=.525\textwidth]{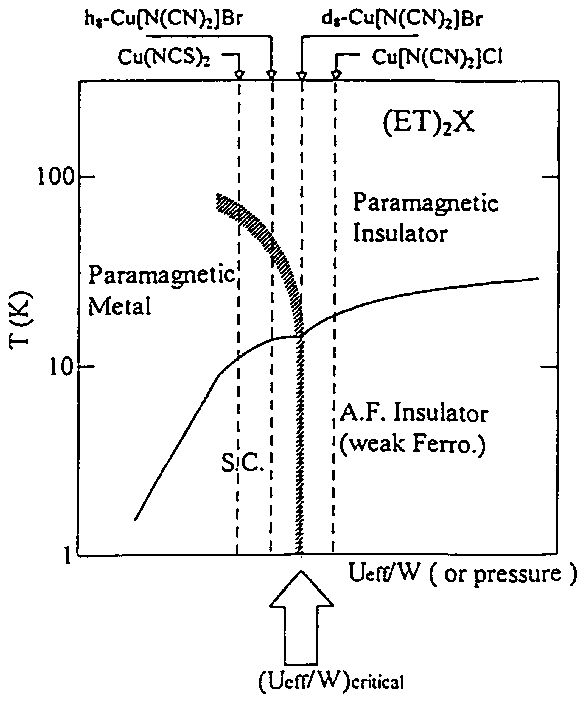}
\caption[]{Conceptual phase diagram for $\kappa$-phase \etzx\ as proposed by Kanoda
\cite{Kanoda 97b}. Note that hydrostatic pressure decreases the ratio $U_{\rm eff}/W$,
i.e.\ the low pressure side is on the right end of the phase diagram. The arrows indicate
the ambient-pressure position of the various systems and AF and SC denote an
antiferromagnetic insulator and superconductor, respectively.
\\}\label{kanopd}
\end{figure}
Figure~\ref{kanopd} shows a conceptual phase diagram proposed by Kanoda for the dimerized
\index{dimerization} $\kappa$-type BEDT-TTF salts. Here it has been assumed that it is the
effective on-site (dimer) Coulomb interaction $U_{{\rm eff}}$ normalized to the
\index{bandwidth} bandwidth $W$ which is the key factor associated with the various phases
and phase transitions \cite{Kanoda 97a,Kanoda 97b}. The positions of the various salts are
determined by their ambient-pressure ground-state properties. The deuterated
$\kappa$-(D$_{8}$-ET)$_{2}$Cu[N(CN)$_{2}$]Br salt is situated right at the AFI/SC border.
The system lies in between the antiferromagnetic insulating X = Cu[N(CN)$_{2}$]Cl and the
superconducting hydrogenated $\kappa$-(H$_{8}$-ET)$_2$Cu[N(CN)$_{2}$]Br salts. It has
been proposed that a {\em partial} substitution of the $2 \times 4$ H-atoms by D-atoms
allows for fine-tuning the \brom\ system across the AFI/SC border \cite{Kawamoto
98,Taniguchi 99,Nakazawa 00}. The close proximity of an antiferromagnetic insulating to a
superconducting phase has been considered - in analogy to the high-$T_c$ cuprates - as a
strong indication that both phenomena are closely connected to each other \cite{McKenzie
97}.

\begin{figure}[t] \center
\includegraphics[width=.75\textwidth]{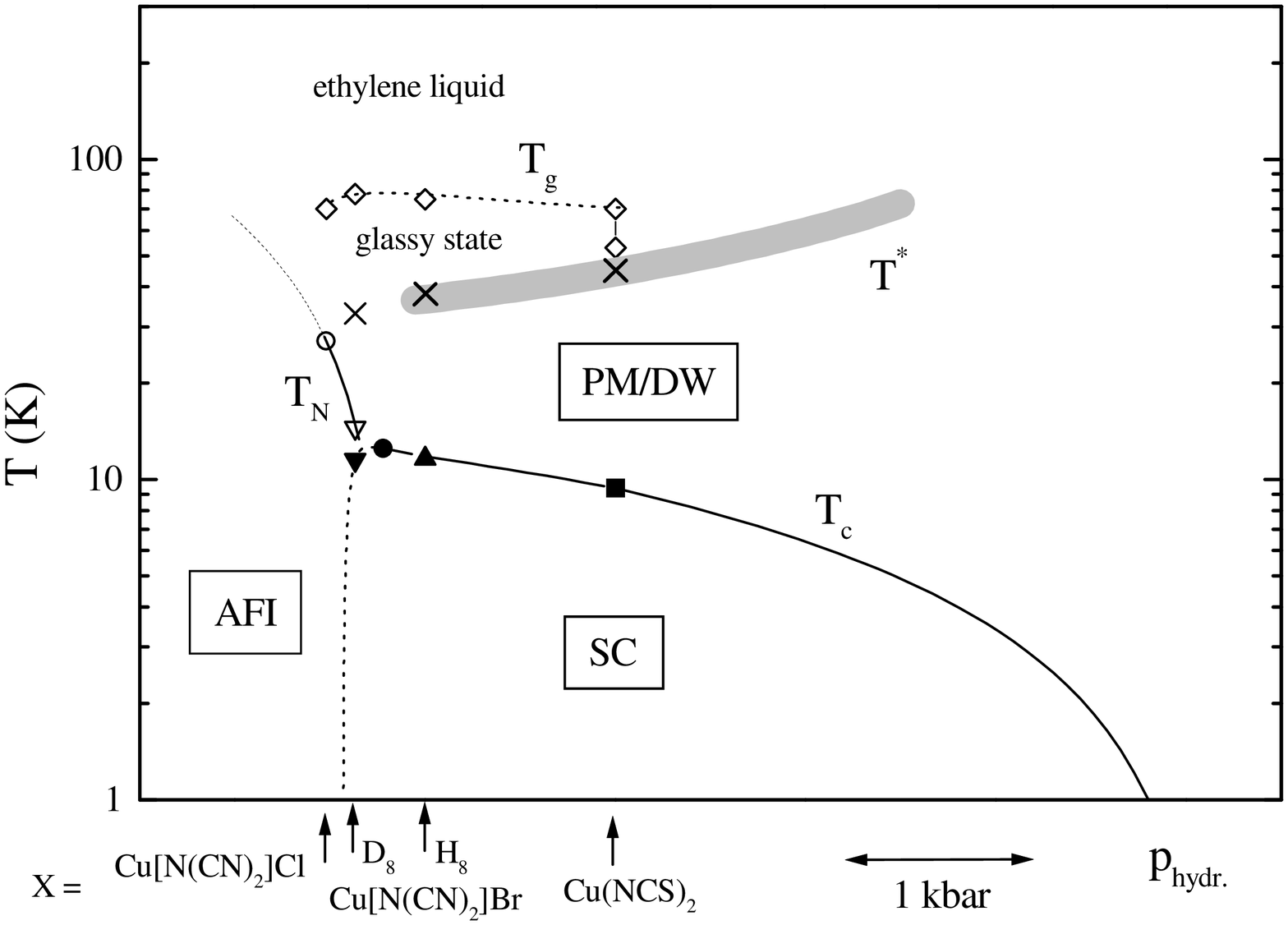}
\caption[]{Temperature-hydrostatic-pressure phase diagram for the $\kappa$-\etzx\
compounds. Arrows indicate the positions of the various compounds at ambient pressure.
Circles correspond to results on $\protect\kappa$-(ET)$_2$Cu[N(CN)$_2$]Cl while down and
up triangles indicate phase-transition temperatures on deuterated and hydrogenated
$\protect\kappa$-(ET)$_2$Cu[N(CN)$_2$]Br, respectively, and squares stand for results on
$\protect\kappa$-(ET)$_2$Cu(NCS)$_2$. The transitions into the superconducting and
antiferromagnetic states are represented by closed and open symbols, respectively.
Diamonds denote the glass-like \index{glass-like transition} freezing of ethylene
disorder and crosses the density-wave (DW) transition on minor parts of the Fermi
\index{Fermi surface} surface. These anomalies coincide with various features observed in
magnetic, transport and acoustic properties (shaded area, see section \ref{thermal and
magnetic}). Taken from \cite{Mueller 02a,Mueller 01}.}\label{jenspd}
\end{figure}
Figure~\ref{jenspd} summarizes experimental data of a detailed thermodynamic study on the
various $\kappa $-(ET)$_{2}$X compounds in a pressure-temperature phase diagram
\cite{Mueller 02a,Mueller 01}. The positions of the various salts at ambient pressure are
indicated by the arrows.\,\footnote{Note that hydrostatic pressure has been used as an
abscissa for the purpose of compatibility with the conceptual phase diagram in
Fig.~\ref{kanopd}. It has been found, however, that the uniaxial-pressure dependences for
the various phase boundaries are strongly anisotropic \cite{Mueller 00,Mueller 02a,Lang
02} with a non-uniform behavior for the uniaxial-pressure coefficients of both the
density-wave instability at \tst\ and those of \tc, cf.\ Figs.~\ref{alpha2} and
\ref{alpha5}.} The solid lines representing the phase boundaries between the paramagnetic
(PM) and the superconducting (SC) or antiferromagnetic insulating (AFI) states refer to
the results of hydrostatic-pressure studies of
$T_{c}$ and $T_{N}$ \cite{Schirber 88,Schirber 90,Schirber 91}.\\
At elevated temperatures, a glass-like transition \index{glass-like transition} at a
temperature $T_g$ (dotted line) has been identified. It marks the boundary between an
ethylene-liquid at $T>T_g$ and a glassy state at $T < T_g$. While at temperatures above
$T_g$, the motional degrees of freedom of the ethylene endgroups \index{ethylene
endgroups} are excited with an equal occupancy for the two possible ethylene
\index{ethylene conformation} conformations, a certain disorder \index{disorder} becomes
frozen in at temperatures below $T_g$. The glass-like transition which is structural in
nature has been shown to cause time dependences in electronic properties and may have
severe implications on the ground-state properties of the
$\kappa$-(ET)$_2$Cu[N(CN)$_2$]Br salt
depending on the cooling \index{cooling-rate dependence} rate employed at $T_{g}$ (see section \ref{glassy phenomena}).\\
At intermediate temperatures $T^{\ast}$, anomalies in the coefficient of thermal
expansion have been found and assigned to a density-wave \index{density wave} transition
involving only the quasi-1D parts of the Fermi \index{Fermi surface} surface. These
anomalies coincide with various features observed in magnetic, transport and acoustic
properties (thick shaded line in Fig.~\ref{jenspd}, see also section \ref{thermal and
magnetic}). In \cite{Mueller 02a,Lang 02} it has been proposed that instead of a
pseudogap on the quasi-2D parts of the Fermi surface, a real gap associated with a
density wave \index{density wave} opens on the minor quasi-1D parts below $T^{\ast}$, see
also \cite{Sasaki 02}. This scenario implies that the density wave \index{density wave}
and superconductivity involve disjunct parts on the Fermi surface \index{Fermi surface}
and compete for stability.

\begin{figure}[h]
\center 
\includegraphics[width=.65\textwidth]{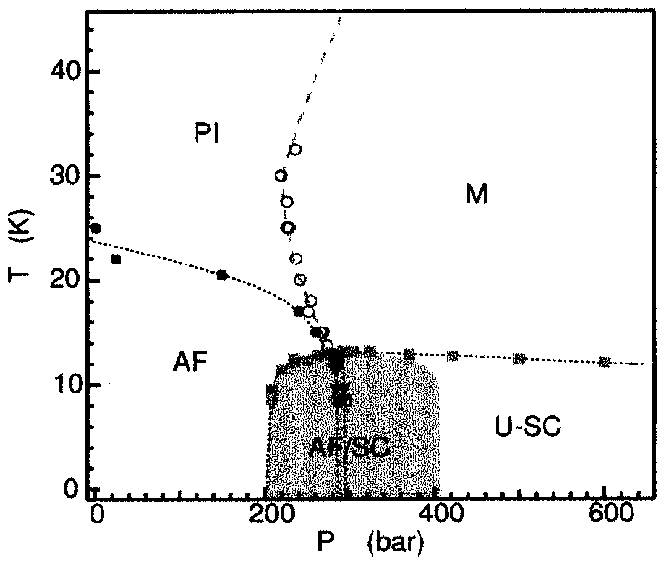}
\caption[]{Pressure-temperature phase diagram for \cl. The antiferromagnetic (AF)
transition temperature $T_N(p)$ (closed circles) were determined from the NMR relaxation
rate while $T_c(p)$ (closed squares) and $T_{MI}(p)$ (open circles) were obtained from
ac-susceptibility measurements. U-SC denotes unconventional superconductivity; the AF-SC
boundary line separates two regions of inhomogeneous phase coexistence (shaded area).
Taken from \cite{Lefebvre 00}.} \label{lefebpd}
\end{figure}
Details of the pressure-temperature phase diagram of the antiferromagnetic insulating salt
\cl\ have been reported by Ito et al.\ \cite{Ito 96} and Lefebvre et al.\ \cite{Lefebvre
00}. In the latter work, it has been shown that the superconducting and antiferromagnetic
phases overlap through a first-order boundary that separates two regions of an
inhomogeneous phase coexistence \cite{Lefebvre 00}. It has been argued that this boundary
curve merges with a first-order line of the metal-insulator transition and that this line
ends at a critical point at higher temperature, see Fig.~\ref{lefebpd}. The figure also
suggests the existence of a point-like region where the metallic, insulating,
antiferromagnetic as well as superconducting phases all meet. This would imply the
absence of a boundary between metallic and complete antiferromagnetic phases which would
be incompatible with an itinerant type of magnetism \cite{Lefebvre 00}.\\

\subsubsection{(BEDT-TSF)$_2$X salts}
\begin{figure}[t]
\center \sidecaption
\includegraphics[width=.55\textwidth]{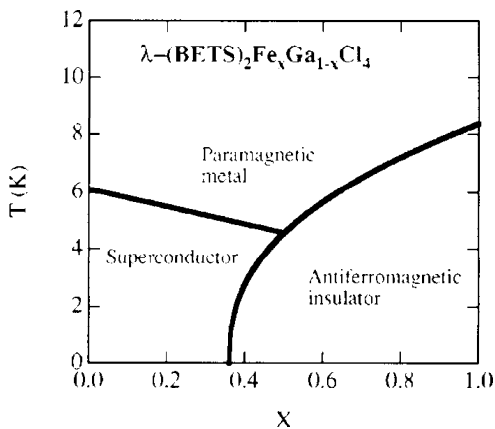}
\caption[]{Phase diagram for $\lambda$-(BETS)$_2$Fe$_x$Ga$_{1-x}$Cl$_4$ taken from
\cite{Uji 02}.\\} \label{ujipd1}
\end{figure}

Figure~\ref{ujipd1} shows the phase diagram for the quasi-2D alloy system
$\lambda$-(BETS)$_2$Fe$_x$Ga$_{1-x}$Cl$_4$ which has recently gained strong interest due
to its interesting magnetic and superconducting properties \cite{Sato 98}. The system is
based on the donor molecule BEDT-TSF \index{BEDT-TSF} or simply BETS \index{BETS} which
represents the Se analogue to \index{BEDT-TTF} BEDT-TTF, see section \ref{Kapitel2}.
$\lambda$-(BETS)$_2$GaCl$_4$ on the left side is a nonmagnetic salt which becomes
superconducting at $T_c = 6$\,K \cite{Kobayashi 97}. A magnetic field of 13\,T aligned
parallel to the highly conducting planes destroys superconductivity and stabilizes a
paramagnetic metallic state. Conversely, $\lambda$-(BETS)$_2$FeCl$_4$ shows a
metal-insulator (M-I) transition around 8\,K which is accompanied by an antiferromagnetic
order \index{antiferromagnetic order} of the Fe$^{3+}$ moments in the anion layers
\cite{Kobayashi 93}. Applying a magnetic field in excess of about 10\,T destabilizes the
insulating phase and the paramagnetic phase recovers \cite{Goze 95,Brossard 98}. In the
mixed series $\lambda$-(BETS)$_2$Fe$_x$Ga$_{1-x}$Cl$_4$, the M-I transition becomes
suppressed as the concentration $x$ of magnetic Fe ions decreases and a superconducting
ground state is formed for $x \leq 0.35$. A striking feature is the
metal-superconductor-insulator transition for $0.35 \leq x \leq 0.5$, see
Fig.~\ref{ujipd1}. Apparently, the various phases contained in the above phase diagram
originate from an intimate coupling between the magnetic moments of the Fe$^{3+}$ $3 d$
electrons and the $\pi$-conduction-electron spins of the BETS \index{BETS} molecule, see
e.g.\ \cite{Brossard 98,Hotta 00,Uji 02}.

\section{Superconducting-state properties}\label{Kapitel4}
Since the discovery of superconductivity in pressurized (TMTSF)$_2$PF$_6$ in 1979
\cite{Jerome 80}, continuing efforts to design new potential donor and acceptor molecules
have led to more than about 80 organic superconductors.\,\footnote{The materials
discussed in this article have to be distinguished from another class of molecular
superconductors - the alkali-metal-doped fullerenes discovered in 1991 \cite{Hebard 91} -
which are usually not referred to as organic materials as they contain only carbon
atoms.} In the vast majority of cases, the donor molecules are derivatives of the
prototype TMTSF molecule \index{TMTSF} including \index{BEDT-TTF} BEDT-TTF as well as the
selenium and oxygen substituted variants BEDT-TSF \index{BEDT-TSF} and BEDO-TTF,
respectively. In addition, superconductors have been derived using asymmetric hybrides
such as DMET and MDT-TTF. For a comprehensive list of organic superconductors the reader
is referred to \cite{Ishiguro 98}.\\ In the following discussion of superconducting
properties we will confine ourselves to a few selected examples. These are the
(TMTSF)$_2$X and the quasi-2D (BEDT-TTF)$_2$X and (BEDT-TSF)$_2$X salts which are
representative for a wide class of materials.

\subsection{The superconducting phase transition}
Organic superconductors are characterized by a highly anisotropic electronic structure, a
low charge carrier concentration \index{carrier concentration} and unusual lattice
properties. As will be discussed below, the combination of these unique material
parameters lead to a variety of remarkable phenomena of the superconducting state such as
pronounced thermal fluctuations, an extraordinarily high sensitivity to external pressure
and anomalous mixed-state properties.

\subsubsection{Superconducting anisotropy}
The abrupt disappearance of the electrical resistance is one of the hallmarks that
manifests the transition from the normal into the superconducting state for usual 3D
superconductors. For the present low-dimensional organic superconductors - as in the
layered high-$T_c$ cuprates - however, strong fluctuations of both the amplitude and
phase of the superconducting order parameter \index{order parameter} may cause a
substantial broadening of the superconducting transition. This becomes particularly clear
when a strong magnetic field is applied. As a consequence, for these materials
zero-resistance is no longer a good measure of the superconducting transition temperature
in a finite magnetic field. Therefore, for a precise determination of the upper critical
fields, thermodynamic investigations such as magnetization, specific heat or thermal
expansion measurements are
necessary.\\
For materials with strongly directional-dependent electronic properties, a high\-ly
anisotropic superconducting state is expected as well. In an attempt to account for these
anisotropies, the phenomenological Ginzburg-Landau and London models have been extended
by employing an effective-mass tensor \cite{Kogan 81}. In the extreme case of a quasi-2D
superconductor characterized by a superconducting coherence \index{coherence length}
length perpendicular to the planes, $\xi_{\bot}$, being even smaller than the spacing
between the conducting layers, $s$, these anisotropic 3D models are no longer valid.
Instead, the superconductor has to be described by a model that takes the discreteness of
the structure into account. Such a description is provided by the phenomenological
Lawrence-Doniach model \cite{Lawrence 71} which encloses the above anisotropic
Ginzburg-Landau and London theories as limiting cases for $\xi_{\bot} > s$. This model
considers a set of superconducting layers separated by thin insulating sheets implying
that the 3D phase coherence is maintained by Josephson currents running across the
insulating layers. In fact, the presence of an intrinsic Josephson effect
\index{Josephson effect} has been demonstrated for several layered superconductors
including some of the high-$T_c$ cuprates and the present \cuncs\ salt \cite{Kleiner
92,Mueller 94}.

To quantify the degree of anisotropy, it is convenient to compare the results of
orientational-dependent measurements with the above anisotropic models. For layered
systems such as the present (BEDT-TTF)$_2X$ compounds, it is customary to use the
effective-mass ratio $\Gamma = m^\ast_\perp / m^\ast_\parallel$, where $m^\ast_\perp$ and
$m^\ast_\parallel$ denote the effective masses \index{effective masses} for the
superconducting carriers moving perpendicular and parallel to the conducting planes,
respectively. In the London and Ginzburg-Landau theories, $\Gamma$ is directly related to
the anisotropies in the magnetic penetration depths $\lambda$ \index{penetration depth}
and coherence \index{coherence length} lengths $\xi$ by
\begin{equation}
\sqrt{\Gamma} = \frac{\lambda_\perp}{\lambda_\parallel}=\frac{\xi_\parallel}{\xi_\perp}.
\end{equation}

\begin{figure}[t]
\center
\includegraphics[width=\textwidth]{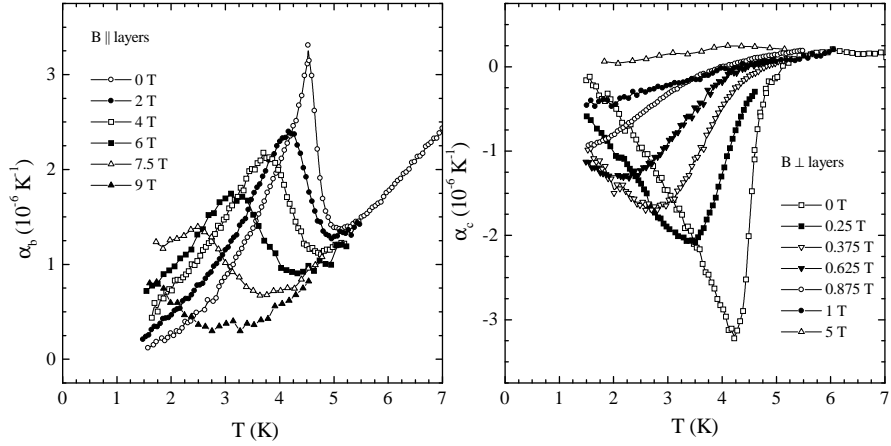}
\caption[]{Coefficient of thermal expansion of $\beta''$-(ET)$_2$SF$_5$CH$_2$CF$_2$SO$_3$
measured parallel (left panel) and perpendicular (right panel) to the conducting planes
in varying fields applied along the measuring directions. Taken from \cite{Mueller
00}.}\label{alpha4a}
\end{figure}

As an example for the highly anisotropic response of the superconducting transition to a
magnetic field, we show in Fig.~\ref{alpha4a} results of the coefficient of thermal
expansion $\alpha(T)$
near $T_c$ for \sfcho.\,\footnote{This salt of the $\beta''$-type structure contains
large discrete anions and is unique in being the first superconductor free of any metal
atoms \cite{Geiser 96,Schlueter 97}.} While for fields parallel to the planes (left
panel) the phase transition in $\alpha(T)$ is still visible even in $B$ = 9\,T, a field
of 1\,T applied perpendicular to the conducting planes is sufficient to suppress almost
completely superconductivity (right panel of Fig.~\ref{alpha4a}). In addition, for this
field orientation, a pronounced rounding of the phase-transition anomaly even for very
small fields can be observed. From these measurements the upper critical fields,
$B_{c_2}$, \index{$B_{c_2}$} can be determined permitting an estimate of the anisotropy
parameter \index{anisotropy parameter} $\Gamma$:
\begin{eqnarray}
B_{c_2}^{\perp '}=\left| \frac{{\rm d} B_{c_2}^{\perp}}{{\rm d} T} \right|=
\frac{\phi_0}{2 \pi \xi_{\parallel}^2 T_c} \qquad \mbox{and} \qquad \frac{B_{c_2}^{\perp
'}}{B_{c_2}^{\parallel '}}= \frac{\xi_{\perp}}{\xi_{\parallel}}=\frac{1}{\sqrt{\Gamma}},
\label{Orlando}
\end{eqnarray}
where $B_{c_2}^{{\perp}'}$ and $B_{c_2}^{{\parallel}'}$ are the initial slopes of the
upper critical fields \index{$B_{c_2}$} for $B$ perpendicular and parallel to the
conducting planes, respectively \cite{Orlando 79,Tinkham 96} and $\phi_0$ is the flux
quantum. For \sfcho\ one finds $\xi_{\parallel}=(144 \pm 9)\,{\rm \AA}$,
$\xi_{\perp}=(7.9 \pm 1.5)\,{\rm \AA}$ and $\Gamma \approx 330$ \cite{Mueller 00} (cf.\
Table~\ref{Strukturtabelle} in section~\ref{Structural aspects} and Table~\ref{Tabelle4c}
in section~\ref{Superconducting parameters}), which underlines the quasi-2D character of
the superconducting state in this material. The large anisotropy parameter
\index{anisotropy parameter} $\Gamma \approx 330$ exceeds the value of $\Gamma \approx
100$ found for
the $\kappa$-(ET)$_2$Cu(NCS)$_2$ salt in dc-magnetization experiments \cite{Lang 94}.\\
The so-derived $\Gamma$ values, however, may serve only as a rough estimate of the actual
\index{anisotropy parameter} anisotropy parameters. The latter can be probed most
sensitively by employing torque-mag\-ne\-to\-metry. For \cuncs\ for example, $\Gamma$
values ranging from $200$ to $350$ have been reported \cite{Farrell 90,Kawamata 94} which
place this material in the same class of quasi-2D superconductors as the most anisotropic
high-$T_c$ cuprates with $\Gamma = 150 \sim 420$ for Bi$_2$Sr$_2$CaCu$_2$O$_{8+x}$
\cite{Martinez 92,Matsuda 97}.

\subsubsection{Fluctuation effects}\label{fluctuations}
\begin{figure}[b]
\sidecaption
\includegraphics[width=.65\textwidth]{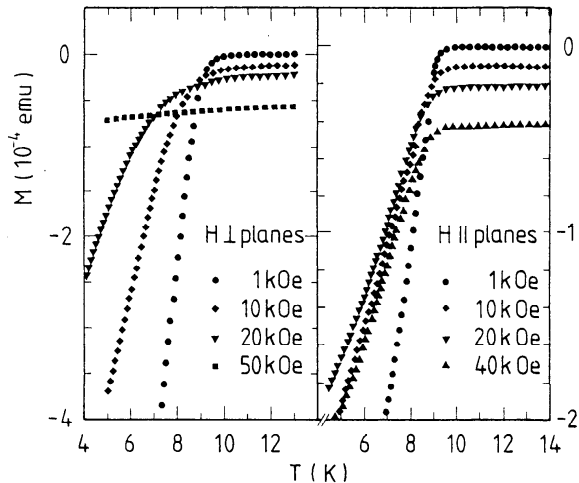}
\caption[]{Raw data of the dc-magneti\-zation of single crystal\-line \cuncs\ in various
fields perpendicular (left panel) and parallel (right panel) to the conducting planes.
The offset of each curve is due to contributions from the core electrons, the spin
susceptibility as well as a small background signal. Taken from \cite{Lang
94}.\\}\label{Michael1}
\end{figure}

The highly anisotropic response of the present quasi-2D superconductors to a magnetic
field is also demonstrated in Fig.~\ref{Michael1} where the temperature dependence of the
magnetization around the superconducting transition is shown for \cuncs. While for fields
aligned perpendicular to the planes (left panel) the transition considerably broadens
with increasing field strength, there is only a little effect on the transition for
fields parallel to the layers (right panel) \cite{Lang 94}, cf.\ also Fig.~\ref{alpha4a}.
This behavior is quite different from that which is found in a usual 3D superconductor
and indicates the presence of strong superconducting fluctuations \index{superconducting
fluctuations} which are strongly enhanced in systems with reduced dimensionality
\cite{Skocpol 75}.\\ A measure of the strength of thermal order-parameter fluctuations is
provided by the so-called Ginzburg number \index{Ginzburg number}
\begin{eqnarray}
G = \frac{\mid T-T_c \mid}{T_c} = \frac{1}{2} \left( \frac{k_B T}{B_{c_{th}}^2(0)
\xi_{\parallel}^2 \xi_{\perp}} \right)^2, \label{eq:GL-Zahl}
\end{eqnarray}
where $B_{c_{th}} (0)$ is the thermodynamic critical \index{$B_{c_{th}}$} field. $G$
measures the ratio of thermal energy to the condensation energy per coherence volume. For
classical 3D superconductors
like niobium $G$ amounts to about $G \sim 10^{-11}$. In contrast, for the present
compounds and some of the high-T$_c$ cuprates one finds $G \sim 10^{-2} - 10^{-3}$
\cite{Lang 94,Welp 91,Pasler 00}. Here, the relatively high transition temperatures
together with a low \index{carrier concentration} charge-carrier concentration - the
latter results in a small Fermi velocity, $v_F$, and thus a short coherence length
\index{coherence length} $\xi \propto \hbar v_F / (k_B T_c)$ - enhance the effect of
superconducting \index{superconducting fluctuations} fluctuations. The strong rounding of
the phase-transition anomaly with increasing magnetic fields aligned perpendicular to the
conducting planes is then understood to be a result of a field-induced dimensional
crossover: while the electronic state in small fields is quasi-2D, the confinement of the
quasiparticles to their lower Landau levels in high fields leads to a
quasi-zerodimensional situation \cite{Bergmann 69,Lee 72}. As a result, the relatively
sharp phase-transition anomaly in zero field becomes progressively rounded and smeared
out with increasing field. Instead of a well defined phase boundary between normal and
superconducting states, the high-field range of the $B$-$T$ phase diagram is
characterized by a crossover behavior with extended critical fluctuations. Here, the
assertion of a mean-field transition \index{mean-field transition temperature}
temperature, $T_c^{\rm mf}(B)$, from the raw data is difficult and a fluctuation analysis
has to be invoked.\\ The effect of fluctuations on transport and thermodynamic properties
has been studied by several authors \cite{Ikeda 90,Ullah 91}. Assuming the
lowest-Landau-level approximation and taking into account only non-interacting Gaussian
fluctuations, Ullah and Dorsey obtained an expression for a scaling function
\index{scaling behavior} of various thermodynamic quantities as the magnetization $M$ or
the specific heat $C$:
\begin{equation}\label{Ullah}
\Xi_i = F_i \left( A\frac{T-T_c^{\rm mf}(B)}{(TB)^n} \right),
\end{equation}
with $\Xi_i=M/(TB)^n$ or $C/T$ \cite{Welp 91}. $F_i$ is an unknown scaling \index{scaling
behavior} function, $A$ a temperature- and field-independent coefficient characterizing
the transition width and $n=2/3$ for anisotropic 3D materials and $n=1/2$ for a 2D
system. Thus from a scaling \index{scaling behavior} analysis both the actual
dimensionality as well as the mean-field-transition \index{mean-field transition
temperature} temperature $T_c^{\rm mf}(B)$ can be determined.
\begin{figure}[t]
\center
\includegraphics[width=\textwidth]{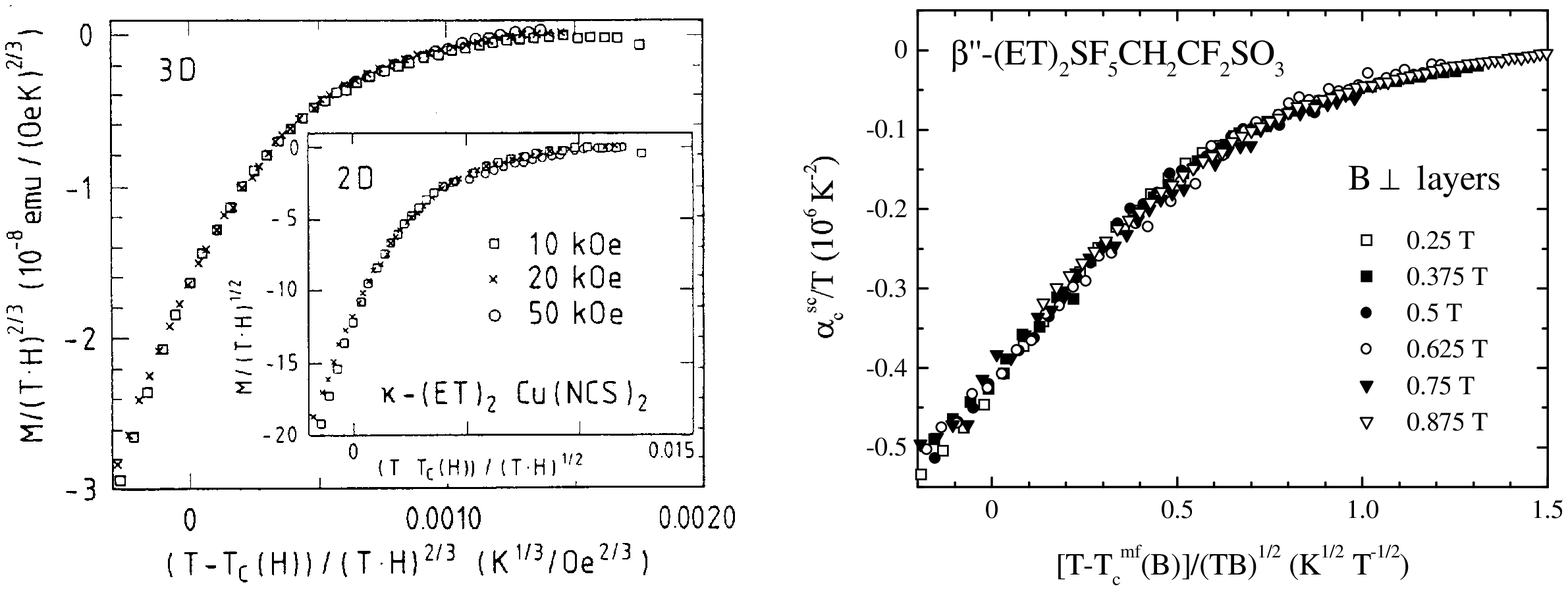}
\caption[]{Scaling behavior \index{scaling behavior} of the superconducting contribution
to the magnetization of \cuncs\ (left panel) \cite{Lang 94} and the linear coefficient of
thermal expansion of \sfcho\ (right panel) \cite{Mueller 00} in magnetic fields applied
perpendicular to the planes.}\label{scaling}
\end{figure}

Figure~\ref{scaling} shows the data of the dc-magnetization of \cuncs\
(Fig.~\ref{Michael1}) and thermal expansion of \sfcho\ (Fig.~\ref{alpha4a}) taken in
varying fields in the proper scaling \index{scaling behavior} forms $M/(TH)^{(n)}$ vs
$(T-T_c^{\rm mf}(B))/(T B)^{n}$ and $\alpha_{c}^{sc}/T$ vs $(T-T_c^{\rm mf}(B))/(T
B)^{1/2}$, respectively, where $\alpha_{c}^{sc}$ denotes the superconducting contribution
to the coefficient of thermal expansion.\,\footnote{Since the volume coefficient of
thermal expansion, $\beta (T)$, is related to the specific heat via the Gr\"uneisen
relation $\beta (T) = \gamma \cdot \frac{\kappa_T}{V_{mol}}\cdot C_V(T)$, where
$\kappa_T$ denotes the isothermal compressibility, $V_{mol}$ the molar volume and
$\gamma$ a field- and temperature-independent Gr\"uneisen parameter, the scaling form
holds also for $\alpha/T$.} As shown in Fig.~\ref{scaling} the various field curves
$\alpha_{c}^{sc} (T,B)$ show the 2D scaling \index{scaling behavior} over a rather wide
temperature and field range, see also \cite{Welp 91,Li 96}. According to the scaling
\index{scaling behavior} analysis of the high-field magnetization in Fig.~\ref{scaling}
as well as the high-field conductivity in \cite{Friemel 96}, \cuncs\ is at the threshold
from being a strongly anisotropic 3D to a 2D superconductor. On the other hand, a
distinct 2D behavior has been claimed from a scaling \index{scaling behavior} analysis of
low-field magnetization data by Ito et al.\ \cite{Ito 92}.

\subsubsection{Pressure dependence of \tc}
By applying pressure to a superconductor, one can study the volume dependence of the
pairing interaction through changes of $T_c$. For the (TMTSF)$_2$X and \etzx\
superconductors, one generally finds an extraordinarily high sensitivity to external
pressure and, in the vast majority of cases, a rapid decrease of \tc\ with pressure.
\begin{figure}[h]
\sidecaption
\includegraphics[width=.6\textwidth]{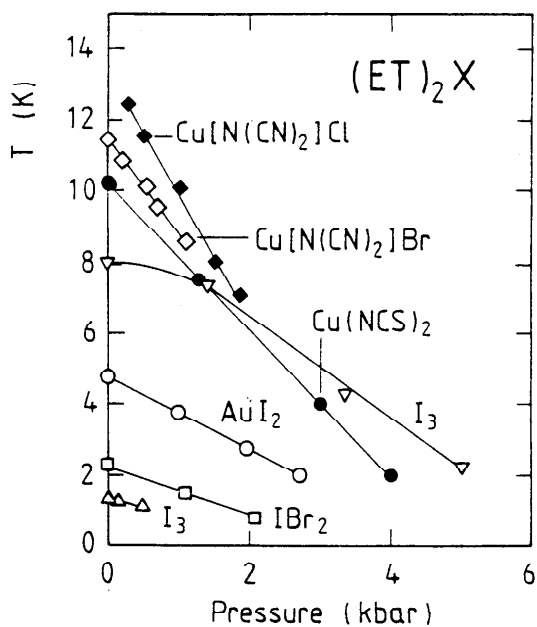}
\caption[]{Hydrostatic-pressure dependence of \tc\ for various $\beta$- and $\kappa$-type
\etzx\ superconductors. Reproduced from \cite{Ishiguro 98}.\\}\label{tcvonp}
\end{figure}\\
Figure~\ref{tcvonp} shows the variation of $T_c$ for a selection of $\beta$- and
$\kappa$-type \etzx\ salts under hydrostatic-pressure conditions. The initial slope of
the pressure dependence of \tc, $\left(\partial T_c /
\partial p \right)_{p \rightarrow 0}$, determined via resistivity measurements ranges
from $-0.25$\,K/kbar for $\alpha$-(ET)$_2$NH$_4$Hg(SCN)$_4$ ($T_c = 1$\,K) \cite{Brooks
95} to $-3.2$\,K/kbar for \cl\ ($T_c = 12.8$\,K at 0.3\,kbar) \cite{Schirber 91}. For
(TMTSF)$_2$PF$_6$, one finds $-(8 \pm 1) \cdot 10^{-2}$\,K/kbar ($T_c = 1.1$\,K at
6.5\,kbar) \cite{Greene 80}. At first glance a strong pressure dependence of \tc\ appears
not surprising in view of the weak van der Waals bonds between the organic molecules,
giving rise to a highly compressible crystal lattice. In fact, the isothermal
compressibility $\kappa_T = -\partial \ln V /
\partial p$ for \cuncs\ of $\kappa_T = (122\,{\rm kbar})^{-1}$
\cite{Chasseau 91,Rahal 97} exceeds the values found for ordinary metals by about a
factor of five. To account for this "lattice effect" one should, therefore, consider the
physically more meaningful volume dependence of $T_c$:
\begin{equation}
\frac{\partial \ln T_c}{\partial \ln V}=\frac{V}{T_c} \cdot \frac{\partial T_c}{\partial
V}= - \frac{1}{\kappa_T \cdot T_c} \cdot \frac{\partial T_c}{\partial p}.
\end{equation}
Using the above isothermal compressibility, one finds $\partial \ln T_c / \partial \ln V
\approx 40$ for $\kappa$-(ET)$_2$Cu(NCS)$_2$ \cite{Mueller 00} which exceeds the values
found for ordinary metallic superconductors, as e.g.\ for Pb with $\partial \ln T_c /
\partial \ln V = 2.4$ \cite{Gladstone 69}, or the layered copper-oxides with $-(0.36 \sim
0.6)$ reported for YBa$_2$Cu$_3$O$_7$ \cite{Meingast 90}
by $1 \sim 2$ orders of magnitude.\\

\begin{figure}[t]
\center
\includegraphics[width=.625\textwidth,angle=-90]{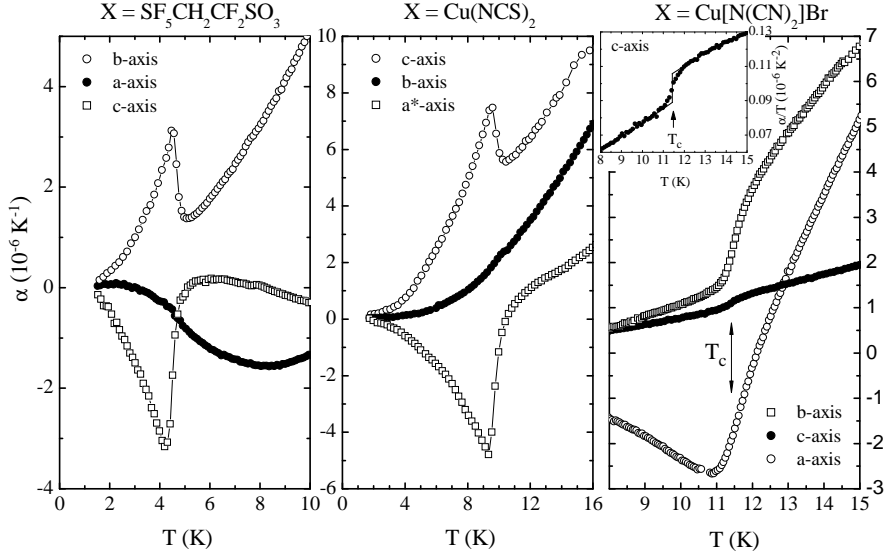}
\caption[]{Uniaxial coefficients of thermal expansion, $\alpha_i$, vs temperature around
the superconducting transition for \etzx\ with X = SF$_5$CH$_2$CF$_2$SO$_3$ (left),
Cu(NCS)$_2$ (middle) and Cu[N(CN)$_2$]Br (right panel), taken from \cite{Mueller
00,Mueller 02a,Mueller 01}. Open squares indicate $\alpha$ data perpendicular to the
planes; open and closed circles correspond to the in-plane expansion coefficients.}
\label{alpha5}
\end{figure}

For strongly anisotropic superconductors like those discussed here, even more information
on the relevant microscopic couplings can be obtained by studying the effect of uniaxial
pressure on \tc. Different techniques have been employed to determine the
uniaxial-pressure coefficients of $T_c$ for the (TMTSF)$_2$X and (ET)$_2$X salts
including measurements under uniaxial strain or stress \cite{Kusuhara 90,Campos
95,Kagoshima 01,Choi 01} or by using a thermodynamic analysis of ambient-pressure thermal
expansion and specific heat data \cite{Kund 94b,Kund 95,Mueller 00,Mueller 02a}. The
latter approach is based on the Ehrenfest relation which connects the pressure
coefficients of $T_c$ for uniaxial pressure along the $i$-axis (in the limit of vanishing
pressure) to the phase-transition anomalies at $T_c$ in the coefficient of thermal
expansion, $\Delta \alpha_i$, and specific heat, $\Delta C$:
\begin{equation}
\left (\frac{\partial T_c}{\partial p_i}\right )_{p_i \rightarrow 0}=V_{mol} \cdot T_c
\cdot \frac{\Delta \alpha_i}{\Delta C}, \label{eq:Ehrenfest}
\end{equation}
with $V_{mol}$ being the molar volume. Figure~\ref{alpha5} shows results of the linear
thermal expansion coefficients along the three principal crystal axes of the
superconductors \sfcho, \cuncsd\ and \brom.
In all three cases, the uniaxial expansion coefficients are strongly anisotropic with
striking similarities in the $\alpha_i$'s for the $\beta''$ and \cuncs\ salts.\\
Using equation (\ref{eq:Ehrenfest}), the uniaxial-pressure coefficients can be derived
\cite{Mueller 00,Mueller 02a}. For \brom\, one finds $\partial T_{{c}}/\partial
p_{b}=-(1.26\pm 0.25)\,{\rm K/kbar}$ for the out-of-plane coefficient and $\partial
T_{c}/\partial p_{a} = - (1.16 \pm 0.2)\,{\rm K/kbar}$ and $\partial T_{{\rm c}}/\partial
p_{c} = - (0.12 \pm 0.05)\,{\rm K/kbar}$ for the in-plane coefficients, employing a jump
height in the specific heat $\Delta C$ as reported by \cite{Elsinger 00}. In the same
way, one obtains for \sfcho\ $ - (5.9 \pm 0.25)\,{\rm K/kbar}$ along the out-of-plane
$c$-axis and $+(3.9 \pm 0.15)\,{\rm K/kbar}$ and $+(0.39 \pm 0.1)\,{\rm K/kbar}$ for the
in-plane coefficients along the $b$- and $a$-axis, respectively, using the $\Delta C$
value given in \cite{Sadewasser 97}. To check for consistency, the hydrostatic-pressure
dependences can be calculated by summing over the uniaxial-pressure coefficients yielding
$\partial T_{c}/\partial p_{hydr}=\sum_{i}(\partial T_{c}/\partial p_{i})=-(2.54 \pm
0.5)\,{\rm K/kbar}$ and $-(1.6 \pm 0.5)\,{\rm K/kbar}$ for the $\kappa$ and the $\beta''$
salt, respectively. These values are found to be in good agreement with the results
obtained by hydrostatic-pressure experiments, i.e.\ $-(2.4 \sim 2.8)\,{\rm K/kbar}$ for
\brom\ \cite{Schirber 90,Sushko 91} and $-1.43$\,K/ kbar for \sfcho\ \cite{Sadewasser
97}.\,\footnote{Figure~\ref{alpha5} shows that there is a strikingly similar behavior for
the uniaxial thermal expansion coefficients and thus the uniaxial-pressure coefficients
of \tc\ of \sfcho\ and \cuncsd. The same observation was made also for the hydrogenated
\cuncsh\ compound \cite{Kund 95,Mueller 01}. Due to the lack of specific heat data for the
deuterated salt, the uniaxial-pressure coefficients of \cuncs\ can only be discussed
qualitatively. According to a recent comparative study on the pressure dependences of the
normal- and superconducting-state properties of hydrogenated and deuterated
\cuncs, the latter compound reveals an even stronger pressure dependence of \tc\ \cite{Biggs 02}.}\\
An obvious step towards a microscopic understanding of this class of superconductors is
to trace out those uniaxial-pressure effects which are common to all systems and thus
reflect a generic property. Such information would provide a most useful check of
theoretical models attempting to explain superconductivity and its interrelation with the
various other instabilities in the pressure-temperature plane. As Fig.~\ref{alpha5}
demonstrates, an extraordinarily large negative uniaxial-pressure coefficient of $T_c$ for
uniaxial pressure perpendicular to the conducting planes is common to all three
superconductors shown there. Apparently, it is this huge component which predominates the
large response of $T_c$ under hydrostatic pressure. On the other hand, and in contrast to
what has been frequently assumed, the systems behave quite non-uniformly concerning the
in-plane pressure effects. While for \cuncs\ the in-plane pressure coefficients of $T_c$
are either vanishingly small or positive, they are both negative for the related
\brom\ system \cite{Mueller 02a}. \\
The above finding of a large negative uniaxial-pressure coefficient of $T_c$ for pressure
perpendicular to the planes as the only universal feature common to the $\kappa$-\etzx\
family is supported by results on the related $\kappa$-(ET)$_2$I$_3$ salt, see
\cite{Mueller 01a}.\,\footnote{A different situation is encountered for the
$\alpha$-(ET)$_2$MHg(SCN)$_4$ salt, where uniaxial pressure perpendicular to the planes
is found to either induce superconductivity by suppressing an ambient-pressure
density-wave \index{density wave} ground state for $M$ = K, or enhance \tc\ for $M$ =
NH$_4$ \cite{Campos 95,Mueller 01}. This behavior is most likely related to the
exceptionally thick anion layers specific to this compound resulting in a strong
decoupling of the conducting layers.}\\ 
As has been discussed in \cite{Lang 96}, a large cross-plane pressure effect on $T_c$ may
arise from several factors: (i) Pressure-induced changes in the interlayer interaction.
This effect includes changes of both the interlayer coupling, i.e.\ the degree of
two-dimensionality, as well as changes in the electron-electron and electron-phonon
coupling \index{electron-phonon coupling constant} constants and (ii) changes in the
phonon frequencies. Likewise, changes in the vibrational properties could be of relevance
for the intraplane-pressure effects on $T_c$. In addition, in-plane stress effectively
modifies the electronic degrees of freedom by changing the transfer integrals
\index{overlap integral} between the HOMO's of \index{highest occupied molecular orbital
(HOMO)} the nearest-neighbour ET molecules. Most remarkably, for some compounds like
\cuncs\, the in-plane-stress effect is either positive or zero. This makes a purely
density-of-states \index{density of states} effect account for the pressure-induced $T_c$
shifts very unlikely: pressure-induced changes in the density-of-states should be
strongest for in-plane stress owing to the quasi-2D electronic \index{band structure}
band structure. According to the simple \index{relation} BCS relation \cite{Allen 75}
\begin{equation}\label{BCS}
T_c = 1.13\,\Theta_D \exp(-\frac{1}{\lambda}) \makebox[0.5cm]{} {\rm with}
\makebox[0.25cm]{} \lambda = \frac{N(E_F) \langle I^2 \rangle}{M {\bar{\omega}}^2},
\end{equation}
where $\Theta_D$ denotes the Debye temperature, $\langle I^2 \rangle$ the electron-phonon
matrix element averaged over the Fermi surface, $M$ the ionic mass and $\bar{\omega}$ an
average phonon energy, an in-plane-stress-induced increase in the $\pi$-orbital
\index{$\pi$-electrons, -orbital} overlap, i.e.\ a reduced density-of-states
\index{density of states} at the Fermi level $N(E_F)$ is expected to cause a
reduction of $T_c$. This is in contrast to the experimental observations.\\
An important piece of information contained in the above uniaxial-pressure results is
that there is {\em no} uniform behavior in the intralayer-pressure effects on $T_c$ for
the various (ET)$_2$X superconductors. It is especially the results on \cuncs\ which show
that in-plane pressure can even cause an increase of $T_c$ \cite{Mueller 00,Choi 01}.
This is in contrast to what has been assumed in the 2D electronic models discussed in
\cite{Kino 96,Kondo 00}. In addition, the studies revealed a predominant effect of
uniaxial pressure perpendicular to the planes clearly demonstrating that attempts to
model the pressure-temperature phase diagrams by solely considering {\em in-plane}
electronic degrees of freedom are inappropriate, see also \cite{Painelli 01}.

\subsubsection{Isotope substitution}
Studying the effect of isotope substitutions \index{isotope effect, - substitution} on
the superconducting transition temperature is one of the key experiments to illuminate
the role of phonons in the pairing mechanism. For elementary superconductors, the
observation of a $M^{-1/2}$ dependence of \tc\, where $M$ is the isotopic mass, provided
convincing evidence that the attractive interaction between the electrons of a Cooper
pair is mediated by the exchange of lattice
deformations, i.e.\ by phonons.\\
For the $\kappa$-phase \etzx\ compounds, the mass-isotope effect \index{isotope effect, -
substitution} on \tc\ has been intensively studied, see \cite{Ishiguro 98}, including
isotope substitutions in both the ET donor molecule as well as the charge compensating
anions. A most comprehensive study has been performed by the Argonne group on \cuncs\
where overall seven isotopically labelled \index{BEDT-TTF} BEDT-TTF derivatives - with
partial substitutions of $^{13}$S, $^{34}$C and $^{2}$D - as well as isotopically labelled
anions [Cu($^{15}$N$^{13}$CS)$_{2}$]$^-$ have been used \cite{Kini 96}. As will be
discussed below in section \ref{Gretchenfrage}, these
studies revealed a genuine mass-isotope effect on \tc.\\
An "inverse" isotope effect \index{isotope effect, - substitution} on \tc\ has been
observed for \cuncs\ where \tc\ of deuterated samples \cuncsd\ was found to be higher
than that of hydrogenated salts, see \cite{Ishiguro 98}. This effect has been confirmed
and quantified by the above mentioned study where particular care has been taken to
guarantee otherwise comparable quality of both the labelled and unlabelled crystals
\cite{Kini 96}. The physical reason for the inverse isotope effect is still unclear. A
geometric H-D isotope effect has also been found for two other \etzx\ compounds
$\kappa_{\rm L}$-(ET)$_2$Ag(CF$_3$)$_4$(solvent) and \sfcho\ having different packing
motifs \index{packing} and anion structures. Although the \tc\ values vary considerably
among these salts ranging from 2.9\,K to 9.2\,K the investigations reveal an almost
identical "universal" shift of \tc\ of $\Delta T_c = +(0.26 \pm 0.06)$\,K \cite{Kini
01,Schlueter 01}. Taking into account the results of thermal expansion and X-ray studies
of the lattice parameters \cite{Mueller 00,Watanabe 97}, it has been proposed that the
inverse isotope effect is not directly related to the pairing mechanism. Instead it has
been attributed to a geometrical isotope \index{isotope effect, - substitution} effect,
i.e.\ changes in the internal chemical pressure: provided that the interlayer lattice
parameter is identical for both compounds, the effectively shorter C$-$D bond
\index{H-bonding} of the deuterated salt \cite{Williams 91} corresponds to a higher
chemical pressure perpendicular to the planes for the hydrogenated salt. The negative
values of $\partial T_c /
\partial p_{\perp}$ then result in a higher \tc\ for the deuterated compound \cite{Kini
01,Schlueter 01}, see also \cite{Mueller 01}.\\ An alternative explanation has been
proposed recently by Biggs et al.\ based on their measurements of the Shubnikov-de Haas
effect focusing on pressure-induced changes of the Fermi-surface \index{Fermi surface}
topology of deuterated and protonated \cuncs\ \cite{Biggs 02}. It has been suggested that
the superconducting mechanism is most sensitively influenced by the exact topology of the
Fermi surface. Since the latter has been found to change more drastically with pressure
in the deuterated salt, this effect might also be responsible for the inverse isotope
effect \cite{Biggs 02}. In addition from recent millimeter-wave magnetoconductivity
experiments it has been inferred that the quasi-one-dimensional FS sheets (see
Fig.~\ref{FS}) are more corrugated in the deuterated salt (higher \tc) suggesting that
the "nestability" \index{nesting} 
of the FS may be important for \tc\ \cite{Edwards 02}.

\subsection{Superconducting parameters}\label{Superconducting parameters}
\subsubsection{(TMTSF)$_2$X salts}
\begin{figure}[t]
\sidecaption
\includegraphics[width=.55\textwidth]{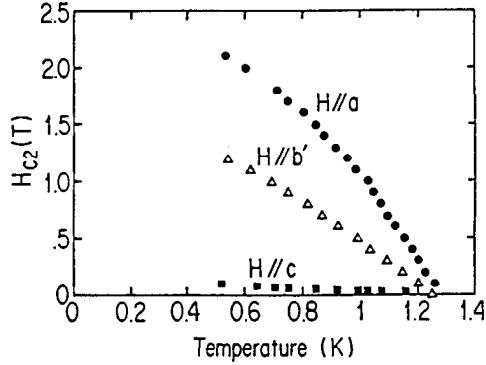}
\caption[]{Upper critical fields $B_{c_2}$ determined from resistivity measurements as a
function of temperature for (TMTSF)$_2$ClO$_4$ for the three principal crystal axes
\cite{Murata 87}.\\}\label{hc2tmtsf}
\end{figure}

As a consequence of the highly anisotropic electronic structure, strong directional
dependences are also expected for the superconducting-state properties such as the lower
and upper critical fields. Among the (TMTSF)$_2$X salts, the latter have been extensively
studied for the ambient-pressure superconductor X = ClO$_4$, see \cite{Jerome 94,Ishiguro
98} and, more recently, also for pressurised (TMTSF)$_2$PF$_6$, see also section
\ref{Gretchenfrage} below. For (TMTSF)$_2$ClO$_4$ the Meissner and diamagnetic shielding
effects have been examined for magnetic fields aligned along the three principal axes
\cite{Mailly 83}. From these experiments the lower critical \index{$B_{c_1}$} field values
$B_{c_1}$ (at 50\,mK) have been determined to 0.2, 1 and 10 (in units of $10^{-4}$\,T)
along the $a$-, $b$- and $c$-axis, respectively. The thermodynamic critical
\index{$B_{c_{th}}$} field, as estimated from the condensation energy, amounts to
$B_{c_{th}} = (44 \pm 2) \cdot 10^{-4}\,{\rm T}$ \cite{Garoche 82}. Figure~\ref{hc2tmtsf}
shows the temperature dependence of the upper critical \index{$B_{c_2}$} fields,
$B_{c_2}(T)$, for (TMTSF)$_2$ClO$_4$ as determined by early resistivity measurements
\cite{Murata 87}.
The $B_{c_2}(0)$ values of 2.8\,T, 2.1\,T and 0.16\,T for the $a$-, $b$- and $c$-axis,
respectively, have been obtained from the data of Fig.~\ref{hc2tmtsf} by an extrapolation
to zero temperature. The so-derived value for fields aligned parallel to the $a$-axis,
where $B_{c_2}$ \index{$B_{c_2}$} is the largest, is close to the Pauli-limiting field
\index{Pauli-limiting field} $B_{P}$ which is the critical value required to break
spin-singlet (S = 0) \index{spin-singlet state} Cooper pairs. For a non-interacting
electron gas this is the case when the Zeeman energy just equals the condensation energy,
i.e.\
\begin{equation}
B_{P} = \Delta_0 \frac{1}{\sqrt{2}\mu_B}, \label{Pauli}
\end{equation}
where $\mu_B$ denotes the Bohr magneton \cite{Clogston 62,Chandraskhar 62}. Assuming a
BCS ratio \index{ratio} for the energy gap $\Delta_0 = 1.76\,k_B T_c$ yields
\index{Pauli-limiting field} $B_{P}\,({\rm in \ Tesla}) = 1.84 \times T_c\,({\rm in \ K})
= 2.3\,{\rm T}$ for (TMTSF)$_2$ClO$_4$. The fairly good coincidence with the
experimentally derived critical field has been taken as an indication for a spin-singlet
\index{spin-singlet state} pairing state \cite{Murata 87,Greene 82,Chaikin 83}. As will
be discussed in more detail in section \ref{Gretchenfrage}, more recent resistivity
measurements on the pressurized X = PF$_6$ salt down to lower temperatures revealed upper
critical field \index{$B_{c_2}$} curves which show an upward curvature for $T \rightarrow
0$ with no sign of saturation down to 0.1\,K \cite{Lee 97a}. It has been argued in
\cite{Lee 97a} that this unusual enhancement of $B_{c_2}$ which exceeds the Pauli
paramagnetic limit \index{Pauli-limiting field} by a factor
of 4 is strongly suggestive of a spin-triplet (S=1) \index{spin-triplet state} pairing state.\\
Yet from the initial slopes of the upper critical field \index{$B_{c_2}$} curves in
Fig.~\ref{hc2tmtsf}, $B_{c_2}^{i '}$, the Ginzburg-Landau coherence lengths
\index{coherence length} $\xi_i(0)$ can be derived using the following relation:
$B_{c_2}^{i '} = \phi_0/(2 \pi \xi_{j} \xi_{k} T_c)$, where $i$, $j$ and $k$ can be $a$,
$b$ and $c$ \cite{Orlando 79,Tinkham 96}. The so-derived $\xi_c$ value of about
$20$\,\AA\ being much smaller than the numbers for the $a$- and $b$-axis coherence
lengths \index{coherence length} of 700\,\AA\ and 335\,\AA\, respectively, but comparable
to the lattice parameter $c = 13.5$\,\AA\ indicates that superconductivity has, in fact,
a more quasi-2D character. For the London penetration depth \index{penetration depth} for
$B$ parallel to the $a$-axis, the axis of highest conductivity, a value of $\bar{\lambda}
= 40\,\mu$m has been reported \cite{Schwenk 83}. This number exceeds the GL coherence
\index{coherence length} lengths by orders of magnitude indicating that the
present system is an extreme type-II superconductor.\\

\subsubsection{(BEDT-TTF)$_2$X and (BEDT-TSF)$_2$X salts}
Due to the strong effects of fluctuations in these superconductors of 
\begin{figure}[t]
\includegraphics[width=\textwidth]{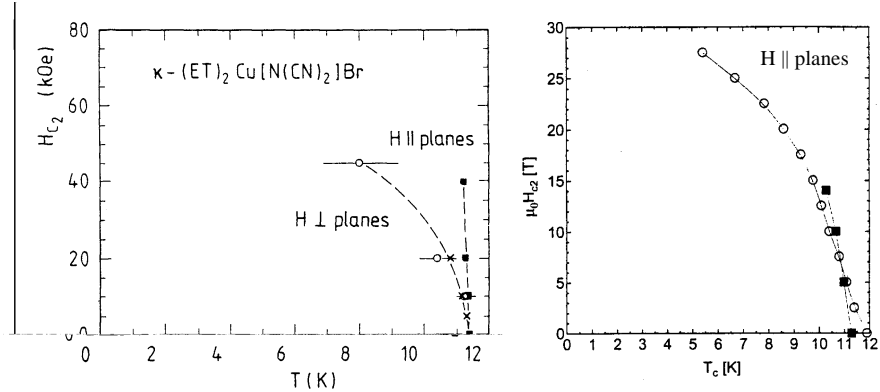}
\caption[]{Upper critical fields of \brom. Left panel: anisotropy of $B_{c_2}$ as a
function of temperature as determined from dc-magnetization measurements, taken from
\cite{Lang 94}. Right panel: $B_{c_2}(T)$ for fields aligned parallel to the conducting
planes as determined from resistivity measurements using different criteria (closed and
open symbols), reproduced from \cite{Shimojo 01}.}\label{hc2cuncs}
\end{figure}
reduced dimensionality, an accurate determination of the upper critical fields
\index{$B_{c_2}$} is difficult and in many cases not free of ambiguities. This holds true
in particular for resistivity measurements in finite fields aligned perpendicular to the
highly conducting planes as phase fluctuations \index{superconducting fluctuations} of
the order parameter \index{order parameter} give rise to a resistive state which tends to
descend far below the mean-field \index{mean-field transition temperature} transition
temperature. A more reliable way to determine the upper critical fields \index{$B_{c_2}$}
is provided by measuring thermodynamic properties and employing a fluctuation analysis as
described in section~\ref{fluctuations}. The left panel of Fig.~\ref{hc2cuncs} shows
$B_{c_2}$ curves for \brom\ as determined from dc-magnetization measurements \cite{Lang
94,Lang 96} using equation (\ref{Ullah}), cf.\ Figs.~\ref{Michael1} and \ref{scaling}.
The $B_{c_2} (T)$ curve \index{$B_{c_2}$} for $B$ aligned parallel to the highly
conducting planes as determined from resistivity measurements are shown in the right
panel of Fig.~\ref{hc2cuncs} over an extended field range. For the layered
superconductors with negligible in-plane anisotropy, the expression (\ref{Orlando}) can
be used to determine the GL coherence \index{coherence length} lengths perpendicular and
parallel to the conducting planes.

Table~\ref{Tabelle4c} compiles the $B_{c_2}$ values together with other superconducting
parameters for the above BEDT-TTF compounds. For comparison, the table also contains data
for the \index{BETS} BETS-based system $\lambda$-(BETS)GaCl$_4$. As indicated in the
table, the transition temperatures reported in the literature show a considerably large
variation depending on the method and criterion used to determine \tc. This may partly be
related to the fact that the superconducting transition is usually found to be relatively
broad. Even for high quality single crystals with an in-plane mean free path of typically
$\sim 2000$\,\AA\ \cite{Singleton 02a}, the transition can be broadened due to internal
strain fields as a consequence of the extraordinarily large pressure dependence of \tc. In
addition, the quasi-2D nature of the electronic structure gives rise to pronounced
fluctuations which cause a rounding of the transition \cite{Singleton 02}.\\ According to
recent magnetoresistivity studies, the in-plane Fermi surface \index{Fermi surface} of
$\lambda$-(BETS)GaCl$_4$ strongly resembles that of \cuncs\ with the effective masses
\index{effective masses} being almost identical for both compounds \cite{Mielke 01}.
However, the interplane transfer integral \index{overlap integral} of $t_\perp \approx
0.21$\,meV for the BETS salt is about a factor of $5$ larger
\begin{figure}[t]
\center
\includegraphics[width=0.95\textwidth]{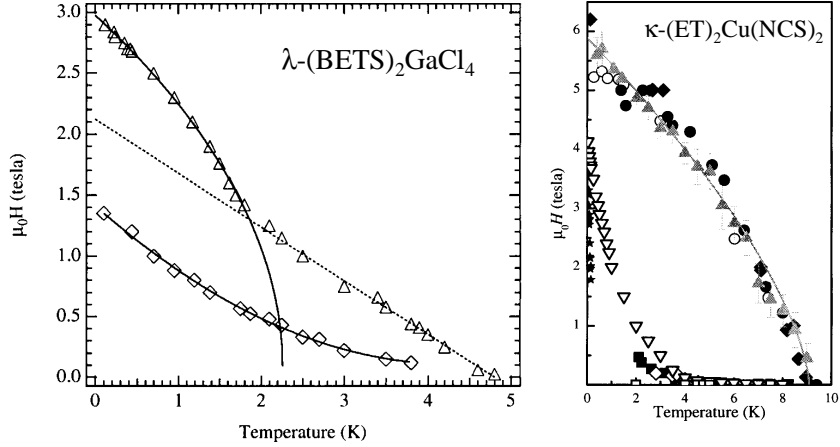}
\caption[]{Left panel: Upper \index{$B_{c_2}$} critical field for $B \perp$ to the planes
(triangles) of $\lambda$-(BETS)GaCl$_4$. Upper solid curve is $B_{c_2} \propto (T^\ast -
T)^{1/2}$; the dashed curve $B_{c_2} \propto (T_{c} - T)$ (see text). Diamonds and the
lower solid curve indicate the flux-lattice melting line (see also section \ref{Mixed
state}). Right panel: $B_{c_2} (T)$ for B $\perp$ to the planes of \cuncs\ as determined
from different experimental techniques, see \cite{Mielke 01}. The solid curve is $B_{c_2}
\propto (T_{c} - T)^{2/3}$ (see text). Also shown are transition and crossover lines in
the mixed state, cf.\ Figs.~\ref{vortexpd1} and \ref{vortexpd2} in section \ref{Mixed
state} below. Reproduced from \cite{Mielke 01}.}\label{pdcomp}
\end{figure}
\cite{Mielke 01}, indicating that \ $\lambda$-(BETS)GaCl$_4$ is more three-dimensional.
Figure~\ref{pdcomp} compares the magnetic field-temperature phase diagrams of the above
compounds for fields aligned perpendicular to the planes. A dimensional crossover has
been suggested to account for the unusual temperature dependence of $B_{c_2}^{\perp}$(T)
observed \index{$B_{c_2}$} for the BETS-based compound (left panel): $B_{c_2}$(T) shows a
3D-like linear behavior close to \tc\ which turns into a power-law dependence
characteristic for a 2D superconductor with weakly coupled layers below some crossover
temperature labelled as $T^\ast$ \cite{Mielke 01,Singleton 02}. In contrast, $B_{c_2}$(T)
for the \cuncs\ salt follows a $B_{c_2} (T) \propto (T - T_c)^{2/3}$ over the whole
temperature range. The additional crossover and transition lines in the mixed state below
the $B_{c_2} (T)$ curves indicated in Fig.~\ref{pdcomp} will be discussed in section
\ref{Mixed state}.\\
Important information on the spin state of the Cooper pairs can be gained by comparing
the experimentally determined $B_{c_2}(T)$ \index{$B_{c_2}$} for $T \rightarrow 0$ with
the Pauli-limiting field, $B_P$, \index{Pauli-limiting field} as defined in equation
(\ref{Pauli}). Using this formula, which neglects any orbital effects, and assuming a
weak-coupling BCS ratio \index{ratio} for the gap, i.e.\ $\Delta_0 = 1.76\,k_B T_c$,
$B_{P}$ amounts to $\sim 18$\,T and $\sim 21$\,T for \cuncs\ and \brom, respectively.
Apparently, these numbers are significantly smaller than the $B^\parallel_{c_2}(0)$
values found experimentally and listed in Table~\ref{Tabelle4c}. On the other hand, clear
evidence for a spin-singlet \index{spin-singlet state} pairing state has been inferred
from Knight shift measurements on \brom\ yielding a vanishingly small spin susceptibility
at low temperatures \cite{Mayaffre 95b}. These deviations might find a natural
\begin{table}[t]
\caption{Superconducting-state parameters of two representative $\kappa$-phase \etzx\
salts X = Cu(NCS)$_2$ and Cu[N(CN)$_2$]Br as well as $\lambda$-(BETS)GaCl$_4$.}
\renewcommand{\arraystretch}{1.4}
\setlength\tabcolsep{5pt}
\begin{tabular}{lccc}
\hline\noalign{\smallskip} & \cuncs & \brom & $\lambda$-(ET)$_2$GaCl$_4$
\\ \noalign{\smallskip} \hline \hline \noalign{\smallskip}
 $T_c$ & $8.7 \sim 10.4$ & $11.0 \sim 11.8$ & $5 \sim 6$  \\
 $B^\perp_{c_2}(0)$ (T) \footnotemark & 6 & $8 \sim 10$ & 3   \\
 $B^\parallel_{c_2}(0)$ (T) \footnotemark & $30 \sim 35$ & $> 30$ & 12   \\
 $B^\perp_{c_1}(0)$ (mT) \footnotemark & 6.5 & 3 &    \\
 $B^\parallel_{c_1}(0)$ (mT) \footnotemark & 0.2 &  &   \\
 $B_{c_{th}}(0)$ (mT) \footnotemark & 54 & 65 &     \\
 $\xi_\perp (0)$ (\AA) \footnotemark & $5 \sim 9$ & $5 \sim 12$ & $9 \sim 14$   \\
 $\xi_\parallel (0)$ (\AA) \footnotemark & $53 \sim 74$ & $28 \sim 64$ & 143   \\
 $\xi_\parallel (0)$ (\AA) \footnotemark & 74 & 60 & 105  \\
 $\lambda_\perp (0)$ ($\mu{\rm m}$) \footnotemark $^,$\footnotemark & $40 \sim 200$ & $38 \sim 133$ &  \\
 $\lambda_\parallel (0)$ (\AA) \footnotemark & $5100 \sim 20000$ & $6500 \sim 15000$ & 1500   \\
 $\kappa_\parallel$ \footnotemark & $100 \sim 200$ & $200 \sim 300$ & 107  \\  \hline
\end{tabular}
\label{Tabelle4c}
\end{table}
\index{$B_{c_1}$} \index{$B_{c_2}$} \index{$B_{c_{th}}$} \index{penetration depth}
\index{coherence length} \addtocounter{footnote}{-11}\footnotetext{\cite{Lang 96,Mayaffre
95b,Mielke 01,Kawasaki 01}} \stepcounter{footnote}\footnotetext{\cite{Shimojo 01,Kawasaki
01,Zuo 00,Singleton 00a,Tanatar 99a}} \stepcounter{footnote}\footnotetext{\cite{Wosnitza
96,Hagel 97}} \stepcounter{footnote}\footnotetext{\cite{Wosnitza 96}}
\stepcounter{footnote}\footnotetext{$B_{c_{th}}(0)= T_c \cdot \sqrt{\frac{\mu_0
\gamma}{2\,V_{\rm mol}}}$} \stepcounter{footnote}\footnotetext{\cite{Lang 94,Dressel
94,Lang 96,Wosnitza 96,Mueller 01,Mielke 01,Kawasaki 01}}
\stepcounter{footnote}\footnotetext{\cite{Lang 94,Dressel 94,Lang 96,Wosnitza 96,Kawasaki
01}} \stepcounter{footnote}\footnotetext{$\xi_\parallel (0) = \sqrt{\frac{\phi_0}{2\,\pi
B^\perp_{c_2}}}$} \stepcounter{footnote}\footnotetext{$\lambda_\perp$ and
$\lambda_\parallel$ denote the screening of supercurrents flowing perpendicular and
parallel to the conducting planes, respectively, and {\em not} the direction of the
magnetic field.} \stepcounter{footnote}\footnotetext{\cite{Dressel 94,Mansky 94,Pinteric
00}} \stepcounter{footnote}\footnotetext{\cite{Pinteric 00,Harshman 90,Lang 92,Lang
92b,Achkir 93,Dressel 94,Mielke 01}} \stepcounter{footnote}\footnotetext{\cite{Lang
96,Mielke 01}}
explanation by recalling that equation (\ref{Pauli}) is valid only in the
weak-coupling limit. For a strong-coupling superconductor, on the other hand, the density
of states \index{density of states} is renormalized leading to a Pauli field
\index{Pauli-limiting field} which is enhanced by a factor of $(1 + \lambda)^{1/2}$
\cite{Clogston 62,Orlando 79}, where $\lambda$ denotes the interaction parameter, see
equation (\ref{BCS}). Clear evidence for a strong-coupling type of superconductivity has
been found in specific heat experiments on the \brom\ and \cuncs\ salts \cite{Elsinger
00,Graebner 90,Mueller 02,Wosnitza 02}, see section \ref{Gretchenfrage}. As an
alternative mechanism to account for a $B_{c_2}^\parallel(0)$ \index{$B_{c_2}$} value in
excess of the Pauli field, a transition into a Fulde-Ferrell-Larkin-Ovchinnikov (FFLO)
state \index{Fulde-Ferrell-Larkin-Ovchinnikov (FFLO) state} \cite{Fulde 64,Larkin 65} has
been proposed \cite{Singleton 00a,Symington 01} and discussed controversially, see e.g.\
\cite{Ishiguro 00,Singleton 02}. In such a scenario, a superconductor with suitable
materials \index{$B_{c_1}$} parameter adopts a new \index{$B_{c_2}$} state at
sufficiently high fields where the order parameter \index{order parameter} is spatially
modulated. A more detailed discussion on this issue will be given in section~\ref{Mixed
state} below.\\ Besides \tc\ and the upper \index{$B_{c_{th}}$} critical fields,
Table~\ref{Tabelle4c} contains further super\-con\-duct\-ing-state parameters such as the
lower and the thermodynamic critical fields $B_{c_1}$ and $B_{c_{th}}$, respectively, as
well as the GL coherence \index{coherence length} lengths $\xi_{\parallel, \perp}$ and the
London penetration \index{penetration depth} depths $\lambda_{\parallel, \perp}$.
$B_{c_1}$ \index{$B_{c_1}$} is usually determined by measuring the magnetization as a
function of field under isothermal conditions where $B_{c_1}$ corresponds to the field
above which flux starts to penetrate the sample. Due to the smallness of $B_{c_1}$, the
plate-like shapes of the crystals and the peculiar pinning \index{pinning} properties of
these layered superconductors, an accurate determination of $B_{c_1}$ is difficult. A more
reliable way to determine $B^\perp_{c_1} (0)$ has been proposed by Hagel et al.\ based on
a model for thermally activated flux creep yielding $B^\perp_{c_1} (0) = (3 \pm 0.5)$\,mT
for \brom\ \cite{Hagel 97}. The values for the thermodynamic critical
\index{$B_{c_{th}}$} fields, $B_{c_{th}}$, in Table~\ref{Tabelle4c} are estimated from
specific heat results using $B_{c_{th}}(0) = T_c \cdot \sqrt{\mu_0\,\gamma/(2\,V_{\rm
mol})}$ where $\gamma$ is the Sommerfeld coefficient. These values roughly agree with
those calculated from $B_{c_{th}} = B_{c_2} / (\kappa \sqrt{2}) = B_{c_1} \sqrt{2} \kappa
/ \ln \kappa$, where $\kappa = \lambda / \xi$ is the GL parameter. The large numbers of
$\kappa$ reflect the extreme type-II character of these superconductors.\\
Also listed in Table~\ref{Tabelle4c} are values of the magnetic penetration
\index{penetration depth} depth $\lambda$, the characteristic length over which magnetic
fields are attenuated in the superconductor. While the absolute values for $\lambda$
derived from various experimental techniques are in fair agreement within a factor $4
\sim 5$, no consensus has yet been achieved concerning its temperature dependence, see
section \ref{Gretchenfrage} below.

\subsection{Mixed state}\label{Mixed state}
The peculiar material parameters of the present organic superconductors such as the
pronounced anisotropy of the electronic states, the small coherence \index{coherence
length} lengths and large magnetic penetration depths \index{penetration depth} give rise
to highly anomalous mixed-state properties and a rich $B$-$T$ phase diagram in the
superconducting state. Exploring the unusual features of extreme type-II layered
superconductors continues to be a subject of considerable interest owing to the variety
of exciting phenomena that has been found in these materials, see e.g. \cite{Blatter 94}.
Among them is a first-order melting transition of the Abrikosov vortex lattice into a
vortex-liquid phase \cite{Zeldov 95,Schilling 96} not known for usual 3D superconductors.
One of the striking early observations related to the anomalous mixed-state properties was
the appearance of a so-called irreversibility \index{irreversibility line} line, $B_{\rm
irr}$, which separates the $B$-$T$ plane into an extended range $B_{\rm irr} < B <
B_{c_2}$ where the magnetization
is entirely reversible from a magnetically irreversible state at $B < B_{\rm irr}$.\\
When a layered superconductor is exposed to a magnetic field $B > B_{c_1}$ aligned
perpendicular to the planes, the confinement of the screening currents to the
superconducting layers results in a segmentation of the flux lines into two-dimensional
objects, the so-called vortex pancakes \index{pancake vortices} \cite{Clem 91}. The
coupling between vortex segments of adjacent layers is provided by their magnetic
interaction and the \index{Josephson coupling} Josephson coupling. The latter interaction
drives tunnelling currents when two vortex segments are displaced relative to each other.
As a result of both effects, the vortex pancakes tend to align thereby forming extended
stacks.\\ A quite different situation arises for fields aligned parallel to the layers.
In the limiting case of a quasi-2D superconductor characterized by a cross-plane coherence
\index{coherence length} length $\xi_{\bot}$ being smaller than the interlayer distance
$s$, the vortex cores slip into the insulating layers where the superconducting order
parameter \index{order parameter} is small. Such a Josephson vortex \index{Josephson
vortices} has an elliptically deformed cross section and lacks a normal core. Since the
Josephson screening currents across the insulating layers are very weak, the material is
almost transparent for fields parallel to the planes corresponding to a large value for
the upper critical field \index{$B_{c_2}$} $B_{c_2}^{\parallel}$, cf.\
Table~\ref{Tabelle4c}. Below we will discuss some of the anomalous mixed-state properties
such as the vortex-lattice melting transition, the irreversibility \index{irreversibility
line} line, the lock-in transition \index{lock-in transition} as well as the possible
realization of an anomalous high-field state.

\begin{figure}[t]
\includegraphics[width=\textwidth]{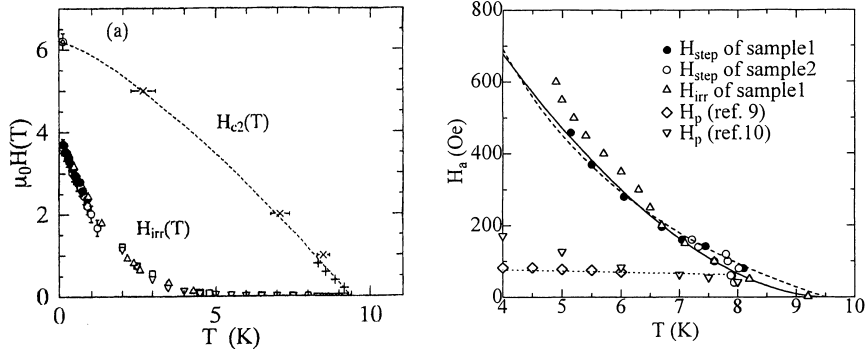}
\caption[]{Details of the $B$-$T$ phase diagram for fields perpendicular to the
conducting planes of \cuncs . Left panel: temperature dependence of the irreversibility
field \index{irreversibility line} and the upper \index{$B_{c_2}$} critical field taken
from \cite{Sasaki 98}. Right panel: first-order transition line (open and closed circles)
determined by a step in the local induction. Solid and dashed curves represent
theoretical predictions for the melting and decoupling transitions, respectively, see
text. Open diamonds and down triangles indicate an anomalous second peak in magnetization
associated with the dimensional-crossover field $B_{2D}$ of the vortex system, taken from
\cite{Inada 99}.}\label{vortexpd1}
\end{figure}
\begin{figure}[t]
\sidecaption
\includegraphics[width=.55\textwidth]{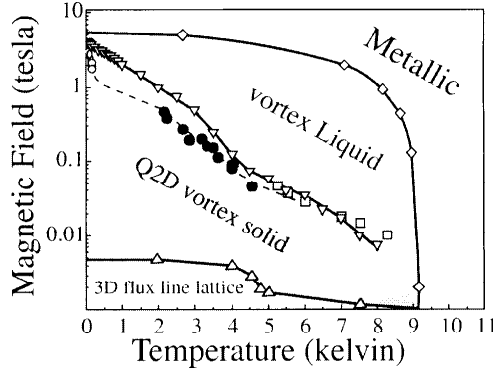}
\caption[]{Mixed-state $B$-$T$ phase diagram for \cuncs\ for B $\perp$ planes in a
semi-logarithmic representation. $B_{\rm irr}$ is indicated by down triangles, the
dimensional-crossover field by up triangles. Open squares denote a first order decoupling
and/or melting transition. Closed and open circles refer to thermal melting or depinning
\index{pinning} and quantum melting transition from a quasi-2D vortex lattice to a liquid
phase, respectively. Taken from \cite{Mola 01}.}\label{vortexpd2}
\end{figure}

Muon spin rotation measurements on the \cuncs\ salt have shown that a 3D flux-line
lattice exists only at very low fields $B < 7$\,mT \cite{Lee 97}.\,\footnote{Here the
sample was arranged so that the superconducting planes enclose an angle of $45^\circ$ in
respect to the magnetic field.} Using a decoration technique, Vinnikov et al.\ succeeded
in imaging the vortex lattice in the low-field range \cite{Vinnikov 00}. Upon increasing
the field to above some dimensional crossover field, $B_{2D}$, the vortex lattices of
adjacent layers become effectively decoupled. Theory predicts that the crossover field is
related to the anisotropy parameter \index{anisotropy parameter} $\Gamma$ and the
interlayer distance $s$ by $B_{2D} = \phi_0 / (\Gamma^2 s^2)$ \cite{Blatter 94}, which
for the \cuncs\ salt results in $B_{2D} = 7.3 \sim 30$\,mT \cite{Lee 97,Nishizaki 96}. An
anomalous second peak in magnetization curves indicating a redistribution of pancake
vortices \index{pancake vortices} at more suitable pinning \index{pinning} centers, has
been associated with $B_{2D}$ \cite{Nishizaki 96,Inada 99}. 
According to measurements of the interlayer Josephson-plasma resonance \cite{Mola 00}, a
long-range quasi-2D order among vortices within the individual layers characterizes the
state above $B_{2D}$ and persists up to the irreversibility \index{irreversibility line}
line. In this region of the $B$-$T$ plane the pancake vortices \index{pancake vortices} of
adjacent layers become effectively decoupled leading to a pinned quasi-2D vortex solid in
each layer with no correlations between the locations of vortices among the layers
\cite{Mola 01}. A somewhat different point of view is taken in \cite{Inada 99,Shibauchi
98}, where $B_{2D}$ marks the crossover from a 3D flux line lattice below $B_{2D}$ to a
state with less strong interlayer coupling on a
long range scale above, where a coupling between the layers is, to some extent, still present.\\
Another striking property common to the present quasi-2D superconductors is an extended
vortex-liquid phase above $B_{\rm irr}$. Here the magnetization behaves entirely
reversible upon increasing and decreasing the magnetic field, indicating that in this
range flux pinning \index{pinning} is ineffective. The abrupt onset of magnetic
hysteresis at $B \leq B_{\rm irr}$ indicates a drastic increase in the pinning
\index{pinning} capability. The temperature dependence of the irreversibility
\index{irreversibility line} field has been studied in detail for \cuncs\ and \brom\
using a variety of techniques including ac-susceptibility \cite{Hagel 97},
dc-magnetization \cite{Lang 94}, magnetic torque \cite{Nishizaki 96,Sasaki 98} or
Josephson-plasma-resonance experiments \cite{Mola 00}. Figure ~\ref{vortexpd1} shows on a
linear scale the irreversibility \index{irreversibility line} line in the $B$-$T$ phase
diagram of \cuncs\ deduced from torque-magnetometry measurements in fields perpendicular
to the planes \cite{Sasaki 98}. As demonstrated in the left panel of
Fig.~\ref{vortexpd1}, the irreversibility \index{irreversibility line} field at the
lowest temperatures lies well below the upper critical field. Thus quantum fluctuations
of the vortices as opposed to thermally driven motions are responsible for the vortex
liquid state in this region of the phase diagram \cite{Sasaki 98}. The crossover from
quantum to thermal fluctuations manifests itself in the temperature dependence of $B_{\rm
irr}(T)$. Below $\sim 1$\,K where quantum fluctuations are predominant, $B_{\rm irr}(T)$
varies linearly with temperature \cite{Sasaki 98}, whereas in the thermal fluctuation
regime an exponential behavior $B_{\rm irr} = B_0 \exp(-A T/T_c)$ has been observed above
the dimensional crossover field $B_{\rm 2D}$ in contrast to a power-law behavior in the
3D vortex line lattice region below $B_{\rm 2D}$ \cite{Lang 94,Nishizaki 96}. A similar
behavior has been observed also for \brom\ where the crossover in the temperature
dependence of the irreversibility line \index{irreversibility line} has been interpreted
as a crossover
from 2D to 3D pinning \index{pinning} \cite{Hagel 97}.\\
Indications for a first-order phase transition associated with a melting and/or a
decoupling of the quasi-2D vortex lattice near the irreversibility line
\index{irreversibility line} - similar to what has been found in some high-\tc\ cuprates
\cite{Zeldov 95,Schilling 96} - have been reported for \cuncs \cite{Inada 99,Yoneyama
02}. This refers to results from micro Hall-probe experiments where the local induction
as a function of temperature at constant fields shows step-like changes. As shown in the
right panel of Fig.~\ref{vortexpd1} the first-order transition line can be fitted equally
well by a melting or a decoupling transition. Steps in the equilibrium and local
magnetization have been observed also in SQUID and micro-hall-probe measurements,
respectively, for \brom\ \cite{Fruchter 97,Shibauchi 98}. In addition,
Josephson-plasma-resonance experiments revealed evidence that at the first-order
transition, the interlayer coherence becomes lost indicating that the melting and
the decoupling occur simultaneously \cite{Shibauchi 98}.\\
Figure~\ref{vortexpd2} shows in a semi-logarithmic plot the $B$-$T$ phase diagram for
\cuncs\ as proposed by \cite{Mola 01}. Note that here the crossover field $B_{\rm 2D}$
separating the 3D flux-line lattice from the quasi-2D vortex solid is temperature
dependent in contrast to the results shown in Fig.~\ref{vortexpd1}. A first-order melting
transition driven by quantum fluctuations has been inferred from
torque-magnetization measurements at very low temperatures \cite{Mola 01}.\\

An interesting situation arises when the magnetic field is aligned parallel to the planes
enabling the vortices to slip in between the superconducting layers where the order
parameter \index{order parameter} is small. For small tilt angles of the field in respect
to the exact alignment, the vortex lattice can gain energy by remaining in this parallel
"lock-in" configuration. This new state remains stable until the perpendicular field
component, $B_\perp$, exceeds a threshold field at which the energy required to expel
$B_\perp$ exceeds that associated with the creation of normal cores in the layers.
Evidence for coreless Josephson vortices \index{Josephson vortices} parallel to the
superconducting layers and a lock-in state has been reported from ac-susceptibility
measurements for \cuncs\ \cite{Mansky 94} and from torque magnetometry on \cuncs\ and
\brom\
\cite{Kawamata 94,Steinmeyer 94}, see also \cite{Lang 96}. 

Further interest in the behavior of the present quasi-2D organic superconductors in
fields precisely aligned parallel to the planes arose from the proposal that these
systems are possible candidates for a Fulde-Ferrell-Larkin-Ovchinnikov (FFLO) state
\index{Fulde-Ferrell-Larkin-Ovchinnikov (FFLO) state} \cite{Shimahara 97}. Under suitable
conditions, a spin-singlet \index{spin-singlet state} superconductor can reduce the
pair-breaking effect of a magnetic field by adopting a spatially modulated
order-parameter along the field direction \cite{Fulde 64,Larkin 65}. The wavelength of
the modulation is of the order of the coherence length which results in a periodic array
of nodal planes perpendicular to the vortices \cite{Fulde 64,Larkin 65}. In the case of
an anisotropic superconductor, calculations show that the FFLO state might lead to an
enhancement of the upper \index{$B_{c_2}$} critical field to between 1.5 and 2.5 times
the Pauli \index{Pauli-limiting field}
paramagnetic limit \cite{Shimahara 97,Shimahara 98}.\\
By studying the magnetic behavior and the resistivity of \cuncs\ in high magnetic fields
employing a tuned circuit differential susceptometer, changes in the rigidity of the
vortex arrangement at certain fields $B_L$ {\em within} the superconducting state have
been found for fields precisely aligned parallel to the planes \cite{Singleton
00a,Symington 01}.
\begin{figure}[t]
\sidecaption
\includegraphics[width=.5\textwidth]{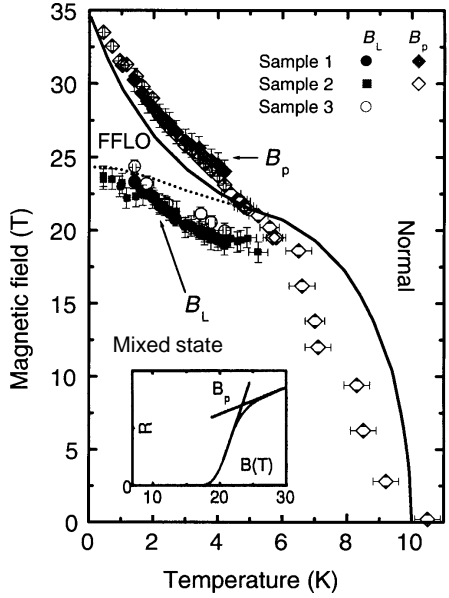}
\caption[]{Magnetic field-temperature phase diagram of \cuncs\ for $B$ aligned parallel
to the planes compared with the theoretical FFLO phase diagram discussed in
\cite{Shimahara 94}. $B_P$ as defined in the inset serves as a rough estimate for
$B_{c_2}$. $B_L$ indicates changes in the rigidity of the vortex arrangement within the
superconducting state. Reproduced from \cite{Singleton 00a,Singleton 02}.\\}\label{fflo}
\end{figure}
These effects have been interpreted as the manifestation of a phase transition from the
superconducting mixed state into an FFLO state.\,\footnote{According to \cite{Fulde
64,Larkin 65}, see also \cite{Tachiki 96,Gegenwart 96}, the stabilization of a FFLO state
requires: (i) a large electronic mean free path $l > \xi_0$, i.e. clean limit, (ii) a
Pauli limiting that predominates the orbital pair-breaking effect, (iii) a Zeeman energy
that overcompensates the loss of superconducting condensation energy and (iv) a short
coherence length (or a large GL parameter $\kappa = \lambda / \xi$). As already discussed
in section~\ref{Superconducting parameters}, the criteria (i) and (iv) are met for the
present $\kappa$-(ET)$_2$X salts. Condition (ii) is fulfilled since $H_P = \frac{1}{2}
H_c (\pi \chi_{\rm Spin})^{-1/2} \approx 17$\,T is considerably smaller than the orbital
critical field as derived from equation (\ref{Orlando}) taking the values from
Table~\ref{Tabelle4c}. 
To check for condition (iii) one has to compare the Zeemann energy with the condensation
energy. Using $\chi_{\rm Spin} = 4 \cdot 10^{-4}$\,emu/mol for \cuncs\ \cite{Toyota 91}
and the upper critical field of 35\,T yields a Zeeman-energy density of $E_Z =
\frac{1}{2} \chi_{\rm Spin} H_{c_2}^2 = 5$\,mJ/cm$^3$ ($ = 5 \cdot 10^4$\,erg/cm$^3$),
which exceeds the condensation-energy density
 of $E_c = H_c^2 / (8 \pi) = \frac{1}{4} \cdot \gamma/V_{\rm mol} \cdot T_c^2 = 1$\,mJ/cm$^3$
($= 10^4$\,erg/cm$^3$) calculated by employing the experimentally determined Sommerfeld
coefficient $\gamma = (23 \pm 1)$\,mJ/mol\,K$^2$ for \cuncs\ \cite{Mueller 02,Wosnitza
02}.} Figure~\ref{fflo} shows the temperature dependence of $B_L$ and $B_P$, the latter
serves as a rough estimate of $B_{c_2}$ \index{$B_{c_2}$} \cite{Singleton 00a,Symington
01}, see inset of Fig.~\ref{fflo}. Comparing the results with theoretical calculations
derived for a generic quasi-2D metal \cite{Shimahara 94} (solid and dotted lines in
Fig.~\ref{fflo}) using the parameter $B_{c_2}(0) = 35$\,T yielded a fairly good agreement
with the predictions of the FFLO model. In particular, the temperature $T^\ast$ below
which the new state is stabilized was found to meet the theoretical predictions of
$T^\ast = 0.56\,T_c$. However, recent magnetic torque measurements failed to detect any
indication for such a transition \cite{Mola 01}.

\subsection{Magnetic-field-induced superconductivity}
\begin{figure}[t] \center
\sidecaption
\includegraphics[width=.65\textwidth]{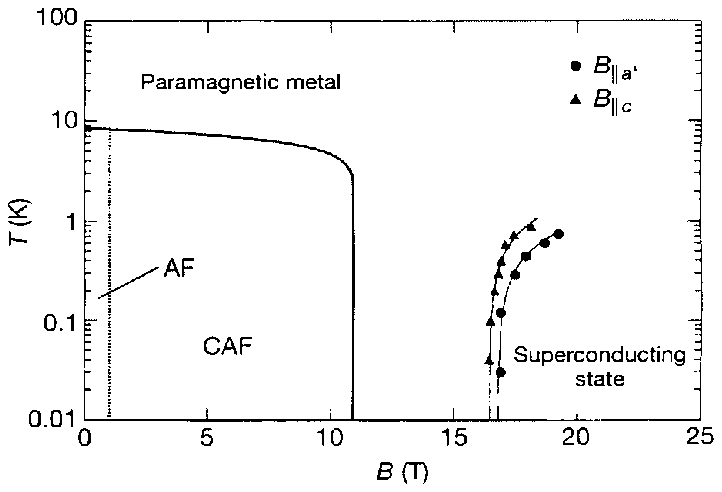}
\caption[]{Temperature vs magnetic field phase diagram for $\lambda$-(BETS)$_2$FeCl$_4$
with the magnetic field aligned parallel to the conducting planes. AF and CAF indicate
the antiferromagnetic and canted antiferromagnetic phase, respectively. Taken from
\cite{Uji 01}.\\} \label{ujipd2}
\end{figure}

The $\lambda$-(BETS)$_2$FeCl$_4$ systems has attracted strong interest recently owing to
the intriguing behavior found when the system is exposed to a magnetic field \cite{Uji
01}. With increasing field aligned parallel to the planes, the low-temperature
antiferromagnetic insulating state becomes suppressed and vanishes above about $10$\,T.
At higher fields the paramagnetic metallic state which governs the $B = 0$\,T behavior
above the N\'{e}el temperature of about $10$\,K is restored. For fields aligned exactly
parallel to the highly conducting planes, a further increase to above 17\,T has been
found to induce a superconducting state at low temperatures, see Fig.~\ref{ujipd2}. For
this field configuration, the orbital effect is strongly suppressed so that
superconductivity is limited only by the Zeeman effect, i.e.\ the Pauli
\index{Pauli-limiting field} limit, resulting in a high upper \index{$B_{c_2}$} critical
field. The field-induced superconducting state which is stable up to 41\,T has been
attributed to the Jaccarino-Peter effect, where the external magnetic field compensates
the exchange field of the aligned Fe$^{3+}$ moments \cite{Balicas 01}. The field-induced
superconductivity is suppressed if the magnetic field is tilted
from the conducting plane.\\
In the concentration range $0.35 \leq x \leq 0.5$ of the series
$\lambda$-(BETS)$_2$Fe$_x$Ga$_{1-x}$Cl$_4$ (cf. Fig.~\ref{ujipd1} in section~\ref{phase
diagrams}), where superconductivity shares a common phase boundary with an
antiferromagnetically ordered insulating state, a field-induced afm
insulator-to-superconductor transition has been observed \cite{Uji 02a}.
It has been suggested that the pairing interaction arises from the magnetic fluctuations
through the paramagnetic Fe moments \cite{Uji 01}.

\subsection{The superconducting state - pairing mechanism and order-parameter symmetry}\label{Gretchenfrage}
Understanding the nature of superconductivity in all its variants continues to pose a
challenging problem of enormous complexity. Elementary superconductors like Al or Zn are
well described by the \index{model} BCS model \cite{Bardeen 57} which envisages pairing
between quasiparticles in a relative zero angular momentum (L=0) and spin-singlet
\index{spin-singlet state} (S=0) state, with an attractive pairing interaction mediated
by the exchange of virtual phonons. In the ground state, all electron pairs (Cooper
pairs) are in the same quantum-mechanical state described by a single macroscopic
wavefunction. Excitations, i.e.\ the creation of unpaired quasiparticles, require an
energy in excess of $2\,\Delta$, where $\Delta$ is the energy gap, known as the
\index{order parameter} order parameter. In the BCS model the gap amplitude
$\Delta$(\boldmath $k$\unboldmath) is uniform over the whole Fermi surface. In addition,
the following quantitative relations are implied in the BCS model: a universal ratio
$\Delta_0 /k_B T_c = 1.76$, where $\Delta_0$ is the energy gap at zero temperature, and
$\Delta C / \gamma T_c = 1.43$ with $\Delta C$ being the jump height in the specific heat
at $T_c$ and $\gamma$ the Sommerfeld coefficient.

In recent decades, novel classes of superconductors have been discovered in complex
materials such as heavy-fermion metals \cite{Steglich 79}, cuprates \cite{Bednorz 86},
ruthenates \cite{Maeno 94} and the present organic materials \cite{Jerome 80}. These
systems have properties which are markedly different from those of the simple elements Al
or Zn. The differences manifest themselves particularly clearly in the role of
electron-electron \index{electron-electron correlations} interactions, which are strong
in the new superconductors but of no relevance in the simple metals. The question at hand
is whether superconductivity in the new materials is of a fundamentally different type to
what is described in the BCS model \index{model} and proved valid to a good approximation
in so many superconductors. As for the present organic materials this question covers the
following aspects:
\begin{enumerate}
  \item[(i)] Are the superconducting carriers "BCS-type" pairs of electrons that form
  below \tc\ (Cooper-pairs)? Or do tightly bound electron pairs preexist already at higher
  temperatures $T > T_c$ which undergo a Bose-Einstein condensation?
  \item[(ii)] Is the relevant pairing mechanism different from a conventional phononic mechanism?
  The exchange of antiferromagnetic spin fluctuations \index{antiferromagnetic spin fluctuations} as one of the models discussed for the
  high-\tc\ cuprates and also for the present organic materials is
  an example for an unconventional pairing interaction.
  \item[(iii)] Is the symmetry of the order parameter \index{order parameter} lower than that of the \index{Fermi surface} Fermi surface?
  It is customary to refer to an unconventional superconducting state as one where the
  order parameter \index{order parameter} breaks at least one of the crystal symmetries.
  \item[(iv)] Is the spin part of the superconducting pair wavefunction an antisymmetric singlet
  or a symmetric triplet state?
\end{enumerate}
Although a local pairing scheme as distinguished from the BCS-type pairing has been
discussed also for the present organic systems \cite{Mazumdar 88,Uemura 91}, the bulk of
experimental observations do not support such a scenario. Rather the formation of
Cooper-pairs below \tc\ implying an energy gap $\Delta$ with a maximum at $T=0$ that
vanishes for $T \geq T_c$ has been widely established, see e.g.\ \cite{Lang 96}.\\ The
discussion of the above issues (ii)$-$(iv) in the context of the organic superconductors
is complicated by the fact that even for the most fundamental properties contradictory
experimental evidences exist. Since some of these discrepancies might simply reflect
extrinsic effects as a consequence of an incomplete sample characterization, we shall
therefore concentrate the discussion on those materials which are best characterized and
where most data are available. These are the $\kappa$-phase (ET)$_2$X and the (TMTSF)$_2$X
salts.

\subsubsection{On the nature of the superconducting state}
The central question for the present organic superconductors concerns the nature of the
pairing interaction. The excitonic mechanism proposed by Little \cite{Little 64,Little 94}
\index{Little's model} for certain quasi-1D organic polymers was of great importance
historically in giving the initial impetus to the development of the field. From today's
perspective, however, it is fair to say that the search for materials with suitable
chemical and physical
properties for such a mechanism to work has been unsuccessful so far. \\
In the discussion of the transport and optical properties in section \ref{Transport and
optical properties} it became clear that there is a substantial coupling of the charge
carriers to both intra- as well as intermolecular phonons. Consequently, some researchers
in the field believe in a conventional electron-phonon coupling mechanism. On the other
hand, the fact that for the present materials - as for the cuprates and heavy-fermion
systems - the superconducting region in the phase diagram lies next to a magnetically
ordered state suggests that magnetic interactions are involved in the pairing mechanism.

\paragraph{Models considering the role of phonons in the pairing mechanism}
One example for a conventional electron-phonon pairing scenario has been discussed by
Yamaji \cite{Yamaji 87}. In this model he considers an attractive interaction mediated by
several high-frequency internal molecular vibrations in addition to one low-frequency
intermolecular phonon.\\
A more general account for the complex role of phonons for superconductivity has been
given by Girlando et al.\ \cite{Girlando 02,Girlando 00a,Girlando 00}. By calculating the
lattice phonons for the $\kappa$-(ET)$_2$I$_3$ and $\beta$-(ET)$_2$I$_3$ superconductors
using the quasi-harmonic-lattice-dynamics method and evaluating the coupling to the
electrons, $\lambda$, they succeeded in reproducing all available experimental data
related to the phonon dynamics, for example the lattice specific heat. They showed that a
lattice mode that couples particularly strongly to the electrons is one in which the
relative displacement of the ET molecule is along the long axis of the molecule, i.e.\
perpendicular to the conducting planes. As an interesting side result of their study on
$\kappa$-(ET)$_2$I$_3$, it is mentioned, that the coupling of acoustic phonons is very
anisotropic and likely to cause gap anisotropies, though of a conventional type. In
addition, it has been shown that the lattice phonons alone cannot account for the $T_c$
values in these compounds. High-frequency intramolecular phonons modulating the on-site
electronic energies have to be taken into account to reproduce the critical temperatures
\cite{Girlando 02}.\\
Recently, a distinct kind of phonon-mediated pairing has been suggested for the \etzx\
salts \cite{Varelogiannis 02}. It is based on the idea that in a system in which Coulomb
correlations are screened to be short range, i.e.\ Hubbard type, the electron-phonon
scattering is dominated by forward processes. This results in an effective small-q
pairing potential. Subsequent self-consistent solutions of the BCS gap equation using a
band structure \index{band structure} based on the effective-dimer approximation, lead to
a gap structure where $d$- and anisotropic $s$-wave states are nearly degenerate.
Furthermore it has been argued that the conflicting experimental results about the gap
symmetry may originate in the decorrelation of superconductivity on various parts of the
Fermi \index{Fermi surface} surface - a consequence of small-q dominated pairing - and
the near degeneracy of $s$- and $d$-wave \index{$d$-wave superconductivity}
superconducting gap states \cite{Varelogiannis 02}.

\paragraph{Models which consider a magnetic interaction}
On the other hand, the rich phase diagram and the anomalous properties of the metallic
state of these materials may suggest that the key elements dominating the physical
behavior are the layered structure and the strong interactions between the electrons
\cite{McKenzie 97}. As a consequence, some researchers even resign from considering any
coupling to the phonons and instead consider mechanisms which are solely based on
two-dimensional electronic interactions. Most of these proposals deal with a
spin-fluctuation-mediated pairing mechanism. The latter is motivated by the close
proximity of the superconducting region in the phase diagram to an antiferromagnetic
insulating state which - in analogy to the high-$T_c$ cuprates - suggests that both
phenomena are closely connected to each other \cite{Kino 96,McKenzie 97,Schmalian 98}.

An approach in this direction has been proposed by Kino and Fukuyama \cite{Kino 96} who
studied the effects of on-site Coulomb interaction and dimer structure in a strictly
two-dimensional system within the Hartree-Fock approximation, see also \cite{Demiralp
97}. In their picture, the antiferromagnetic insulating state of $\kappa$-(ET)$_{2}$X is
a Mott \index{Mott-Hubbard insulating state} insulator. The Mott-Hubbard scenario for the
present organic superconductors implies a half-filled conduction band, together with
strong electron correlations. Because of the approximate square-lattice configuration of
the dimers the authors expect a similar spin-fluctuation mediated superconductivity with
probably $d_{x^2-y^2}$ symmetry as in the cuprates \cite{Kino 96}.\\ Such a possibility
has been studied in detail by Schmalian \cite{Schmalian 98}. Using a two-band Hubbard
model to describe the antibonding orbitals on the ET dimer he succeeded in creating a
superconducting state with $T_c \simeq 10$\,K mediated by short-ranged antiferromagnetic
\index{antiferromagnetic spin fluctuations} spin fluctuations. It has been argued that
despite the frustrating interactions and in-plane anisotropies which distinguish the
organic materials from the high-$T_c$ cuprates, the origin of superconductivity is very
similar for both material classes \cite{Schmalian 98}.\\ A spin-fluctuation-based
superconductivity similar to that of the cuprates has been claimed also by Kondo and
Moriya \cite{Kondo 98,Kondo 99,Kondo 99a}. They investigated the properties of a
half-filled Hubbard model in a fluctuation exchange approximation (FLEX) with a
right-angle isosceles triangular lattice with transfer matrices $-\tau'$ and $-\tau$.
They revealed an energy gap of $d_{x^2-y^2}$ symmetry which upon cooling grows much
faster compared to that expected in the \index{model} BCS model. In addition they showed,
that the appearance of $d$-wave superconductivity \index{$d$-wave superconductivity} near
an antiferromagnetic instability requires a suitable electronic structure, i.e.\ $\tau' /
\tau > 0.3$ \cite{Kondo 98,Kondo 99,Kondo 00}.\\ A spin-fluctuation mediated $d$-wave
superconducting \index{$d$-wave superconductivity} state has been found also by several
other approaches including FLEX, perturbation theory or quantum Monte Carlo simulation
applied to $\kappa$-(ET)$_2$X \cite{Kino 98,Vojta 99,Kuroki 99,Jujo 99} or the
quasi-one-dimensional (TM)$_2$X salts \cite{Kuroki 99,Kino 99}. In \cite{Jujo 99a} and
\cite{Kino 99} an explanation for the pseudogap behavior at elevated temperatures \tst\
has been proposed in terms of strong antiferromagnetic spin-fluctuations.\\
While the starting point of the above models is in the limit of strong correlations,
i.e.\ near the Mott-insulating \index{Mott-Hubbard insulating state} state, a somewhat
different viewpoint is taken in the work by Louati et al.\ \cite{Louati 00}. These
authors studied the effect of spin fluctuations \index{antiferromagnetic spin
fluctuations} in a two-dimensional model in the weak correlation regime by varying the
bandwidth \index{bandwidth} and the nesting \index{nesting} properties of the Fermi
\index{Fermi surface} surface. They argued that spin fluctuations
\index{antiferromagnetic spin fluctuations} are enhanced by the good nesting
\index{nesting} properties which may account for the anomalous NMR relaxation rate
observed at temperatures \tst\ above $T_c$ in the \brom\ salt. Furthermore they found
that spin fluctuations \index{antiferromagnetic spin fluctuations} can induce a
superconducting coupling with $d$-wave symmetry \index{$d$-wave superconductivity} that
lies next to a spin-density-wave \index{density wave} instability \cite{Louati 00}.

The above models address systems at half filling, which is realized in the dimerized
$\kappa$-phase (ET)$_2$X and (TMTSF)$_2$X salts, suggesting spin-fluctuation-mediated
superconductivity with a $d_{x^2-y^2}$ symmetry. A different scenario has been proposed
for the $\theta$ and $\beta''$ structures \cite{Merino 01}. Here the ET molecules are not
dimerized which results in a quarter-filled hole band. In this case, a nearby insulating
phase is believed to be due to a charge ordering, driven by strong inter-site Coulomb
correlations. Applying slave-boson theory to an extended Hubbard model at quarter
filling, superconductivity mediated by charge fluctuations has been found. This results
in a $d_{xy}$ symmetry of the superconducting state \cite{Merino 01} as opposed to the
$d_{x^2-y^2}$ symmetry for the above spin-fluctuation mechanism.

\subsubsection{Experiments probing the superconducting state}
On the experimental side, the determination of the actual pairing mechanism is a most
difficult task as there is no decisive probe to pin down the relevant pairing
interaction. There are, however, some crucial experiments which may help to delineate the
various possibilities. Investigating the mass-isotope effect on \tc\ is such a key
experiment. Another one is to study the phonon system to probe the role of
electron-phonon \index{electron-phonon interaction} interactions. If phonons are involved
in the pairing interaction this would result in renormalization effects in the
temperature dependences of the phonon frequencies and linewidths upon cooling through
$T_c$. Likewise, if a non-phononic mechanism is at work leading to an anisotropic gap
with nodes along certain directions on the Fermi surface, a determination of the
orientation of the gap zeroes by angular-dependent measurements can provide important
information on the pairing mechanism.

For classical phonon-mediated superconductors the gap amplitude $\Delta$(\boldmath
$k$\unboldmath) is assumed to be isotropic or at least to have an isotropic component
combined with a \boldmath $k$\unboldmath-dependent part which obeys all symmetries of the
crystal lattice. In contrast to such a conventional "finite-gap" state, the above
mentioned electronic coupling schemes lead to a pairing state with higher angular
momentum where $L=2$ ($d$-wave) \index{$d$-wave superconductivity} being the most favored
one. In this case the amplitude of the Cooper-pair wave function vanishes at the origin
of the relative coordinate which keeps the constituent quasiparticles of the Cooper pair
apart. Therefore, $L \neq 0$ pairing states are good candidates for materials with strong
on-site Coulomb repulsion. The gap-function of such an $L \neq 0$ state has a \boldmath
$k$\unboldmath-dependence which is given by the spherical harmonics of the same angular
momentum. For those states where the $\Delta$(\boldmath $k$\unboldmath) functions vanish
at certain \boldmath $k$\unboldmath-vectors at the Fermi surface, the quasiparticle
excitation spectrum at low energies is markedly different from that of an isotropic
finite gap state. For the above $d$-wave order parameter \index{order parameter}
$\Delta$(\boldmath $k$\unboldmath)$ = \Delta_0 \left( \cos(k_x a) - \cos(k_y b) \right)$
the zero crossings along the diagonals correspond to line nodes at the Fermi surface.
This should be reflected in all quantities that depend on the number of thermally excited
quasiparticles such as specific heat, NMR relaxation rate, magnetic penetration
\index{penetration depth} depth, etc.\ in the form of simple power-law dependences at
sufficiently low temperatures. In contrast, an exponentially weak $T$-dependence in these
quantities characterizes an isotropic non-vanishing \index{order parameter} order
parameter. Thus, careful measurements of the above thermal properties should, in
principle, provide a handle on the \index{order parameter} order-parameter symmetry, or
at least permit to discriminate whether gap zeroes exist or not. In this context, it
should be noted, however, that the observation of a non-exponential or power-law
$T$-dependence in one of these quantities does not necessarily imply a gap structure with
nodes. As an example, we mention gapless superconductivity via pair breaking which
destroys an exponential behavior \cite{Maki 69,Abrikosov 69}. Likewise, a power-law
$T$-dependence could be of conventional origin, as e.g.\ a $T^3$ dependence in the
specific heat for
$0.2\,T_c < T < T_c$ which has been attributed to strong-coupling effects \cite{Keiber 84,Luethi 85}.\\ 
To our knowledge, the only example where a non-phononic mechanism has been clearly
identified is the heavy-fermion superconductor UPd$_2$Al$_3$. Here, the combination of
tunnel spectroscopy \cite{Jourdan 99} and neutron scattering experiments \cite{Sato 01a}
has provided sound evidence for a magnetic pairing interaction, i.e.\ the exchange of
magnetic excitons.

Although numerous experiments have been devoted to the issue of the order parameter
symmetry \index{order parameter} for the organic materials, no consensus has yet been
achieved. It is fair to say that for the present systems direct evidence for a
non-phononic mechanism such as the one mentioned above does not exist. Also
phase-sensitive experiments as those applied successfully to the high-\tc\ cuprates
\cite{Tsuei 94,van Harlingen 95} have not been
performed so far for the organic materials. \\
In what follows we shall give a discussion of a selection of experimental results on the
pairing mechanism and the symmetry of the \index{order parameter} order parameter. To
begin with we shall focus on the \etzx\ salts.

\subsubsection{Experiments on the pairing mechanism}
\paragraph{Isotope effect and electron-phonon coupling}
The effect of isotope substitution \index{isotope effect, - substitution} has been
studied for various members of the \etzx\ superconductors. Isotopes have been substituted
in the ET molecule by replacing $^1$H by $^2$D in the ethylene \index{ethylene endgroups}
endgroups, by a partial exchange of $^{12}$C by $^{13}$C or $^{32}$S by $^{34}$S atoms in
the inner skeleton of the molecule. In addition, systems have been studied where the
acceptor molecule has been isotopically labelled. First experiments focused on the role
of the electron-molecular-vibration (EMV) \index{electron-molecular-vibration (EMV)
coupling} coupling by substituting $^{13}$C for $^{12}$C in the central double bond of
the ET molecule. The large decrease of $T_c$ of $\Delta T_c / T_c = - 2.5$\,\% found for
the high-temperature variant ($\beta_H$) of $\beta$-(ET)$_2$I$_3$ by the Orsay group
\cite{Auban 93} could not be reproduced by a subsequent study where no systematic
decrease of $T_c$ could be detected \cite{Carlson 92}. Most intensive studies on the
isotope \index{isotope effect, - substitution} effect have been carried out on the
\cuncs\ salt by the Argonne group \cite{Kini 96}. Their investigations include isotope
substitutions \index{isotope effect, - substitution} on both the \index{BEDT-TTF}
BEDT-TTF molecules (all together seven differently labelled BEDT-TTF derivatives) as well
as the anions. In each case, a batch of unlabelled samples has been synthesized under
strictly parallel experimental conditions. These crystals serve as a reference for
comparison. By sampling a large number of crystals, typically eight or more of both
labelled and unlabelled material, a genuine mass-isotope effect on $T_c$ has been found:
upon replacing all eight sulfur atoms by $^{34}$S and the outer-ring-carbon atoms of the
\et\ endgroups by $^{13}$C, which corresponds to a 5\,\% increase of the ET molecule's
mass, a shift of $\Delta T_c = - (0.12 \pm 0.08)$\,K has been observed. Assuming a
BCS-like mass-isotope effect \index{isotope effect, - substitution} $T_c \propto
M^{-\alpha}$ with the whole ET molecule as the relevant mass entity M, this shift
corresponds to $\alpha = 0.26 \pm 0.11$. This experiment provides clear evidence that the
{\em inter}molecular (lattice) phonon modes are strongly involved in the pairing
mechanism. On the other hand, the same group demonstrated the absence of a comparable
isotope effect on $T_c$ for \cuncs\ and \brom\ upon a partial substitution of the central
C=C and also a simultaneous replacement of both the central and ring C=C atoms. The same
holds true for a substitution of the eight sulfur atoms, see \cite{Kini 96} and
references therein. These results indicate that the {\em intra}molecular C=C and C$-$S
bond-stretching vibrational modes of the ET molecule do not provide a substantial
contribution to the Cooper pairing.

\begin{figure}[h]
\sidecaption
\includegraphics[width=.5\textwidth]{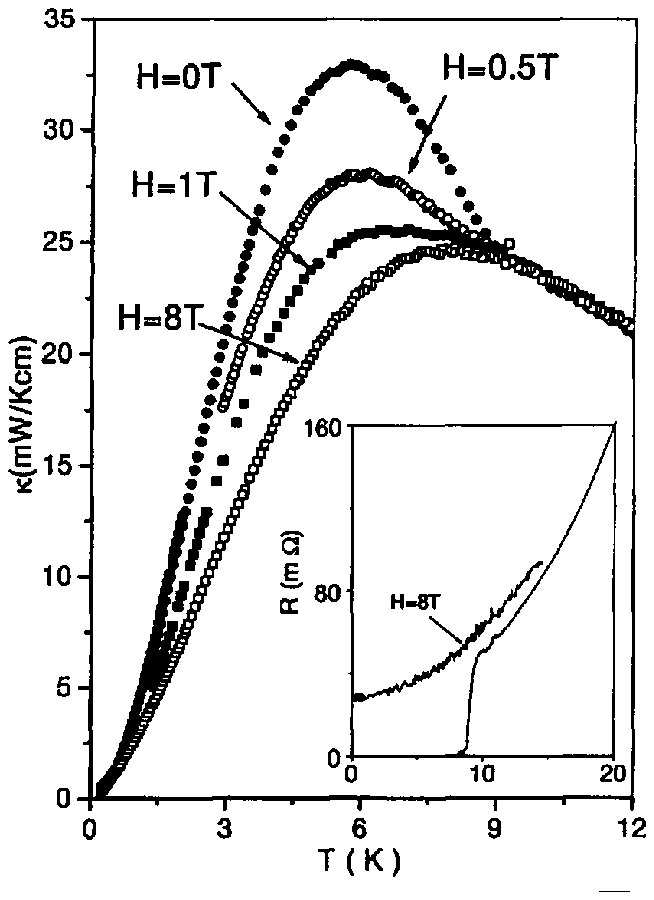}
\caption[]{Temperature dependence of the thermal conductivity of \cuncs\ for $B$ applied
perpendicular to the planes. Inset: temperature dependence of the resistance. Taken from
\cite{Belin 98}.\\}\label{kappa}
\end{figure}
Further indications for a strong electron-phonon coupling \index{electron-phonon coupling}
have been inferred from measurements of the thermal conductivity on \cuncs\ and \brom\
\cite{Belin 98,Izawa 02,Wosnitza 02}. As shown in Fig.~\ref{kappa}, the thermal
conductivity $\kappa(T)$ of \cuncs\ exhibits an upturn at the onset of superconductivity
followed by a pronounced maximum just below \tc. It has been convincingly argued by the
authors that the enhancement of $\kappa(T)$ in the superconducting state is a consequence
of the condensation of electrons into Cooper pairs which strengthens the heat transport
by freezing out the scattering of the main heat carriers, the phonons. By employing the
Wiedemann-Franz law it has been found that just above \tc\ the electronic contribution
amounts to only 5\,\% of the total thermal conductivity \cite{Belin 98}.
Figure~\ref{kappa} also shows $\kappa (T)$ data taken at varying fields applied
perpendicular to the conducting planes. In the normal state, a magnetic field of 8\,T
does not affect the thermal conductivity within the resolution of the experiment but
induces a sizeable decrease in the charge conductivity (see inset) which is an
additional indication of lattice-dominated thermal conductivity in the vicinity of \tc.\\
Temperature-dependent Raman scattering studies of the phonon dynamics of \cuncs\ and
\brom\ substantiate the strong coupling of the superconducting charge carriers to
intermolecular \index{intermolecular electron-phonon coupling} phonons \cite{Eldridge
96,Eldridge 97,Pedron 97,Pedron 99,Faulques 00,Girlando 00}. 
The observed anomalous temperature dependence of the low-frequency phonons around and
below \tc\ were found to be consistent with an isotropic gap $2\,\Delta_0$ close to
2.8\,meV \cite{Faulques 00}. From the reported frequency shifts the electron-phonon
coupling constants $\lambda_i$ have been calculated \index{electron-phonon coupling
constant} yielding a total coupling constant of $\lambda_{tot} = 0.97 \pm 0.11$
\cite{Faulques 00}.

\paragraph{Superconductivity-induced phonon renormalization}
As a consequence of the interaction of charge carriers with the phonon system, the
opening of a gap in the electronic density of states \index{density of states} below \tc\
induces changes in the phonon frequencies and linewidths. These effects were first
observed in the classical superconductors Nb$_3$Sn and Nb \cite{Axe 73,Shapiro 75}. The
results of these studies support the generally accepted picture that superconductivity in
these materials is phonon-mediated.
\begin{figure}[h]
\center
\includegraphics[width=.85\textwidth]{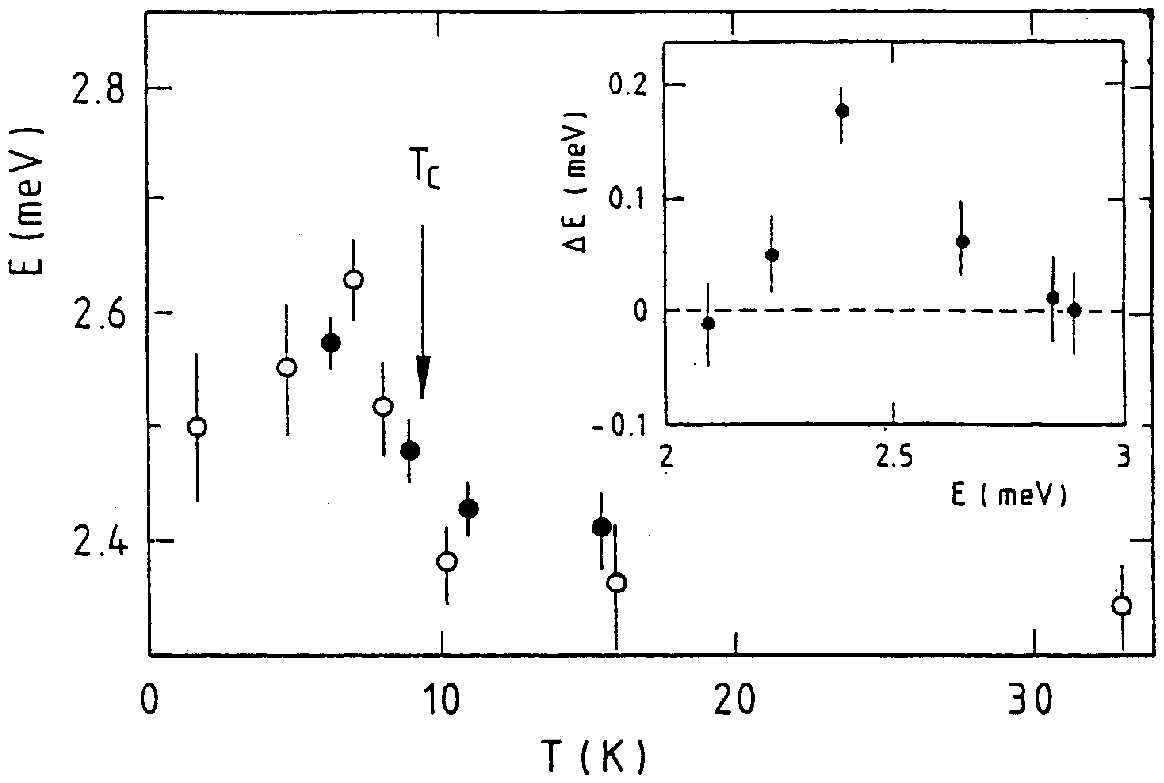}
\caption[]{Temperature dependence of the energy of the transverse acoustic phonon with
wave vector \boldmath $q$ \unboldmath\ $= (-0.225, 0, 0.45)$ derived from inelastic
neutron scattering on two different single crystals (open and closed circles) of
deuterated \cuncs. Inset: observed frequency shifts [$\Delta E = E(T < T_c) - E(T > T_c)$]
of transverse acoustic phonons in the [$- \zeta, 0, 2\,\zeta$] direction. Reproduced from
\cite{Pintschovius 97}.}\label{neutron}
\end{figure}\\
Inelastic neutron-scattering experiments have been performed on single crystals of 
\cuncs\ on both hydrogenated and deuterated crystals \cite{Toyota 97,Pintschovius 97}.
Due to the extraordinarily large incoherent crosssection of the protons, the study of
$\kappa$-(H$_8$-ET)$_2$Cu(NCS)$_2$ allows for a selective investigation of those
vibrational modes that involve the hydrogen atoms at the terminal \index{ethylene
endgroups} ethylene groups. The analysis of measurements above and below $T_c$ suggest a
substantial coupling of these modes to the superconducting charge carriers \cite{Toyota
97}. Figure ~\ref{neutron} shows the results of inelastic neutron-scattering experiments
on deuterated \cuncs\ carried out by Pintschovius et al.\ \cite{Pintschovius 97}. The
data reveal a sudden increase of the frequencies of transverse acoustic phonons upon
cooling through \tc. This phonon hardening was found to be most pronounced for the wave
vector \boldmath $q$\unboldmath\ = ($-0.225, 0, 0.45$) and a phonon energy $2.4$\,meV. As
discussed by Zeyher and Zwicknagl \cite{Zeyher 90}, significant changes are expected only
for those phonons whose energy $\hbar \omega$ is close to the gap value $2\,\Delta$ with a
softening (hardening) for $\hbar \omega < 2\,\Delta$ ($\hbar \omega
> 2\,\Delta$) \cite{Zeyher 90}. The above results thus imply $2\,\Delta \simeq 2.4$\,meV,
i.e.\ $2\,\Delta/k_B T_c \simeq 3.1$, which is close to the BCS \index{ratio} ratio of
3.52. The salient feature of this study is that intermolecular modes strongly couple to
the superconducting charge carriers \index{intermolecular electron-phonon coupling} and
may thus provide a substantial contribution to the pairing interaction \cite{Pintschovius
97}.

\subsubsection{On the order-parameter symmetry in (BEDT-TTF)$_2$X}
\paragraph{Measurements of the gap anisotropy}
A new development in the investigation of the order-parameter symmetry is to
\begin{figure}[t] \sidecaption
\includegraphics[width=.55\textwidth]{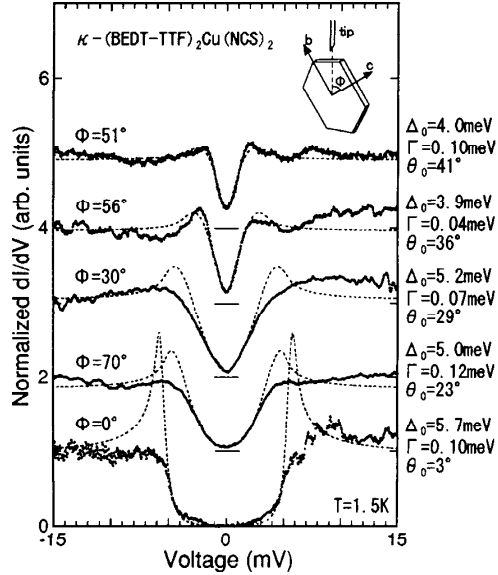}
\caption[]{${\rm d}I/{\rm dV}$-$V$ curves taken at 1.5\,K on the lateral surfaces of
\cuncs\ single crystals. Data have been taken along various tunneling directions at
different angles $\phi$ as defined in the inset. The dashed line represents the
calculated curve based on the $d$-wave gap model with a \boldmath $k$\unboldmath\
dependent tunneling. Taken from \cite{Arai 01}.\\}\label{STM}
\end{figure}
probe the gap anisotropy directly by using orientational-dependent measurements. For the
present organic superconductors, these techniques include mm-wave transmission
\cite{Schrama 99},
STM spectroscopy \cite{Arai 01} and thermal conductivity \cite{Izawa 02} studies.\\
A {\em mm-wave magneto-optical technique} was used to determine the angle dependence of
the high-frequency conductivity of \cuncs\ \cite{Schrama 99,Schrama 01}. The results have
been interpreted to support an anisotropic gap with "X shape", i.e.\ with nodes along the
$b$- and $c$-direction \cite{Schrama 99}, consistent with a $d_{x^2-y^2}$ symmetry of the
order parameter \index{order parameter} as theoretically suggested by Schmalian
\cite{Schmalian 98}. However,
these results have been critically commented upon by other groups \cite{Hill 01,Shibauchi 01}.\\
The superconducting gap structure of the same compound has been investigated using {\em
STM spectroscopy} by Arai et al.\ \cite{Arai 01}. The tunneling curves observed on the
$bc$-plane (parallel to the conducting layers) in the low-energy region could be well
fitted by a $d$-wave \index{$d$-wave superconductivity} gap model. The corresponding
$2\,\Delta_0/k_B T_c$ ratio was found to be $6.7$ which is smaller than a previously
reported value of $9$ \cite{Bando 90} but substantially larger than the BCS \index{ratio}
value of $3.52$. In addition, the in-plane gap anisotropy was investigated, see
Fig.~\ref{STM}. The ${\rm d}I/{\rm dV}$-$V$ curves observed on the lateral surfaces were
found to be also consistent with a $d$-wave \index{$d$-wave superconductivity} gap. For
this configuration a very large $2\,\Delta_0/k_B T_c$ ratio of $8.7 \sim 12.9$ has been
obtained. The analysis of the angular dependence revealed that the direction of the line
nodes of the gap is $\pi/4$ from the $k_b$- and $k_c$-axes, i.e.\ the gap has
$d_{x^2-y^2}$ symmetry \cite{Arai 01}. It has been noted that these orientations of the
gap nodes are at variance with those inferred from the above mm-wave-transmission experiments.\\
The {\em thermal conductivity} has been used as another directional-dependent probe.
\begin{figure}[t]
\sidecaption
\includegraphics[width=.55\textwidth]{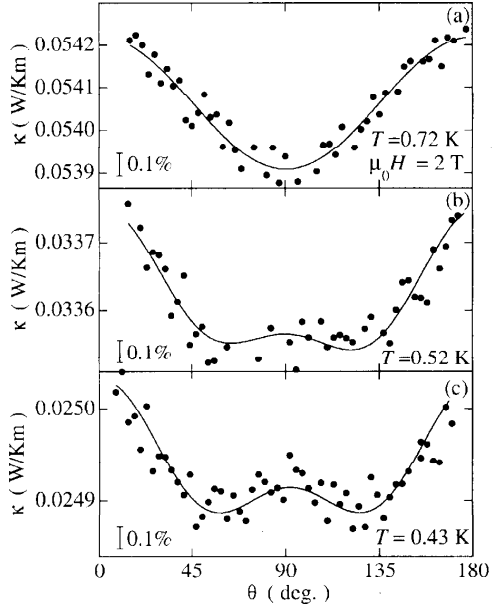}
\caption[]{Angular variation of $\kappa(B,\Theta)$ at 2\,T for different temperatures
where $\Theta$ denotes the angle between a rotating magnetic field in respect to the heat
current flowing along the $b$-axis of the \cuncs\ crystal. The solid lines represent the
results of a fit using the function $\kappa(B,\Theta)=C_0 + C_{2 \Theta} \cos 2 \Theta +
C_{4 \Theta} \cos 4 \Theta$, where $C_0$, $C_{2 \Theta}$ and $C_{4 \Theta}$ are
constants. Taken from \cite{Izawa 02}.\\}\label{fourfold}
\end{figure}
When compared to STM measurements, for example, this quantity has the advantage that it
is free of surface effects. The implications of the symmetry of gap zeroes on the thermal
conductivity in the vortex state have been theoretically investigated by various authors,
see e.g.\ Won and Maki \cite{Won 02}. Measurements have been performed for the \cuncs\
salt in a magnetic field rotating within the 2D superconducting plane.
Figure~\ref{fourfold} shows the angular variation of $\kappa$ at a fixed field of $B =
2$\,T. $\Theta$ denotes the angle between the heat current flowing along the
crystallographic $b$ direction and the magnetic field, i.e.\ $\Theta = 0^\circ$ for $B
\parallel b$. The salient result of this study is the occurrence of a $\kappa(\theta)$
contribution with a fourfold symmetry, $\kappa_{4\theta}$, at low temperatures $T \leq
0.52\,K$ that adds to a predominant term with twofold symmetry. While the latter has been
interpreted as being mainly phononic in origin, it is argued that the former is of a
purely electronic nature and reflects the nodal gap structure \cite{Izawa 02}. Their
analysis revealed that the gap zeroes are oriented along the directions rotated by
$45^\circ$ relative to the $b$- and $c$-axes. It has been pointed out in \cite{Izawa 02}
that this nodal structure is inconsistent with the theories based on antiferromagnetic
spin fluctuation where the nodes are expected to be along the $b$- and $c$-directions.
Based on this observation Izawa et al.\ proposed a $d_{xy}$ symmetry (referring to the
magnetic Brillouin zone, see inset of Fig.~1 in \cite{Izawa 02}) which has been
theoretically suggested for a charge-fluctuation scenario \cite{Scalapino 87,Merino 01}.

\begin{figure}[t]
\sidecaption
\includegraphics[width=.6\textwidth]{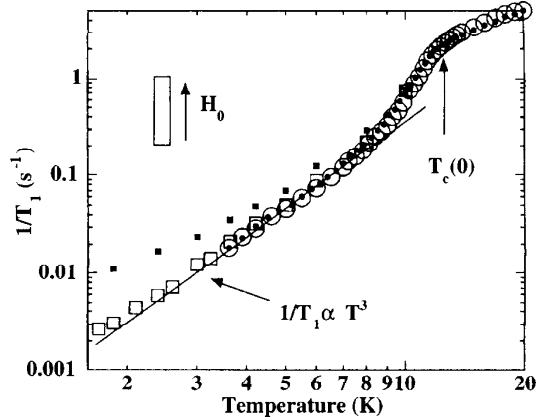}
\caption[]{Spin-lattice-relaxation rate $(T_1)^{-1}$ at fields of 5.6\,T (open circles),
7.8\,T (black circles) and 7\,T (open squares) applied parallel to the conducting planes
of \brom. Black squares correspond to a field of 7\,T with a small misalignment. Taken
from \cite{Mayaffre 95b}.\\}\label{nmr}
\end{figure}

\paragraph{NMR measurements}
A more indirect information on the symmetry of the superconducting order parameter is
provided by temperature-dependent measurements of quantities which depend on the
quasiparticle excitation spectrum. In this context NMR experiments, i.e.\ measurements of
the Knight shift $K_S$ and the spin-lattice relaxation rate $(T_1)^{-1}$ are of
particular interest. The $^{13}$C spin-lattice relaxation rate and Knight shift of \brom\
have been investigated by various groups \cite{de Soto 95,Mayaffre 95b,Kanoda 96} with
similar results. In these experiments single crystalline material was used where both
$^{12}$C atoms in the central carbon double bond of the ET molecule had been replaced by
$^{13}$C. For the investigation of electronic properties, these nuclei are superior since
their coupling to the $\pi$-electron system \index{$\pi$-electrons, -orbital} is much
stronger than that of the protons in the ethylene endgroups \index{ethylene endgroups} of
the ET molecules. The salient results of these studies are: (i) Knight-shift measurements
performed in fields aligned parallel to the conducting planes reveal a spin
susceptibility that tends to zero at low temperatures. Since any contributions from the
pancake vortices \index{pancake vortices} have been excluded for this field configuration,
the above results have been taken as evidence for the spin-singlet \index{spin-singlet
state} character of the pairing state. (ii) The spin-lattice relaxation rate,
$(T_1)^{-1}$, measured in the same parallel field configuration lacks a Hebel-Slichter
peak and shows a power-law $T^n$ behavior at low $T$, with $n$ being close
to 3, see Fig.~\ref{nmr}. 
For the experimental conditions chosen, the authors ascribed the dominant source of
relaxation to the quasiparticle excitations in the superconducting state. Consequently,
the power-law temperature dependence in $(T_1)^{-1}$ has been interpreted as indicating an
anisotropic pairing with nodes in the gap function \cite{de Soto 95,Mayaffre 95b,Kanoda
96}.

\paragraph{Thermal conductivity}
Investigations of the thermal conductivity on quasi-2D organic superconductors at low
temperatures have been performed first on \cuncs\ \cite{Belin 98} and more recently also
on \brom\ \cite{Wosnitza 02}. These studies reveal that the onset of superconductivity is
associated with a sudden increase of $\kappa(T)$ which can be suppressed by a moderate
magnetic field. The enhancement of $\kappa(T)$ at the onset of superconductivity has been
attributed to a strengthening of the phonon heat transport by reducing the scattering due
to the gap formation. Their argument is based on a quantitative analysis of the data
employing the Wiedemann-Franz law. It showed that just above $T_c$ the electronic
contribution amounts to only 5\,\% of the total thermal conductivity. A lattice-dominated
thermal conductivity around $T_c$ is also consistent with the absence of a magnetic field
dependence of $\kappa$ in this temperature range \cite{Belin 98}. As for the question of
the gap symmetry, the data at low temperature have been interpreted as indicating an
excitation spectrum with gap zeroes: an extrapolation of the data for \cuncs\ to $T
\rightarrow 0$ revealed a finite in-$T$ linear term which has been attributed to a
residual electronic contribution \cite{Belin 98}. The latter is expected for an
unconventional superconductor
due to impurity scattering of residual quasiparticles \cite{Graf 96,Norman 96}.\\
On the other hand, recent thermal conductivity measurements on the \brom\ salt showed
that down to the lowest temperatures the phonon scattering length is strongly influenced
by quasiparticle scattering \cite{Wosnitza 02} which renders the analysis of the data on
\cuncs\ \cite{Belin 98} as being questionable.

\paragraph{Magnetic penetration depth}
The quantity which has been studied most intensively for the $\kappa$-(ET)$_2$X
superconductors in connection with the question on the order-parameter symmetry is
the magnetic penetration \index{penetration depth} depth.\\
According to the London theory, the penetration depth $\lambda_L$ in the limit
$T\rightarrow0$ is directly related to the density of superconducting electrons $n_s$ via
\begin{equation}\label{London1}
\lambda_L (0) = \sqrt{\frac{m^\ast c^2}{4 \pi n_s e^2}},
\end{equation}
where $m^\ast$ is the effective mass of the superconducting carriers \cite{Tinkham 96}.
Employing a two-fluid model with $n_e = n_s(T) + n_n(T)$ and $n_e$ the density of
conduction electrons, the temperature dependence of $\lambda_L(T)$ provides information
on the normal-conducting component $n_n(T)$, i.e.\ the quasiparticle excitation spectrum.
Since $n_s(T\rightarrow0) = n_e$, the low-temperature value $\lambda(T\rightarrow0)$ is a
measure of the pair condensate, i.e.\ $\lambda_L^2(0) \propto m^\ast/n_s(0)$. For a
conventional weak-coupling superconductor, the BCS theory \index{model} predicts a
mean-field temperature dependence of $\lambda_L$ around \tc\ and an exponentially small
variation at low temperatures $T \ll T_c$ \cite{Muehlschlegel 59}:
\begin{equation}\label{London2}
\lambda_L(T) \simeq \lambda(0) \left[ 1+ \left( \frac{2 \pi \Delta}{k_B T}
\right)^{\frac{1}{2}} \exp\left({-\frac{\Delta}{k_B T}} \right) \right].
\end{equation}
This holds true also for an anisotropic gap function without nodes, where for $k_B T \ll
\Delta_{\rm min}$ the exponential low-temperature behavior is governed by the minimum
value of the gap $\Delta_{\rm min}$. In contrast, an energy gap which vanishes along
lines or at points at the Fermi surface will result in a power-law dependence of
$\lambda_L(T)$ for $T \ll T_c$.

\begin{figure}[t]
\sidecaption
\includegraphics[width=.6\textwidth]{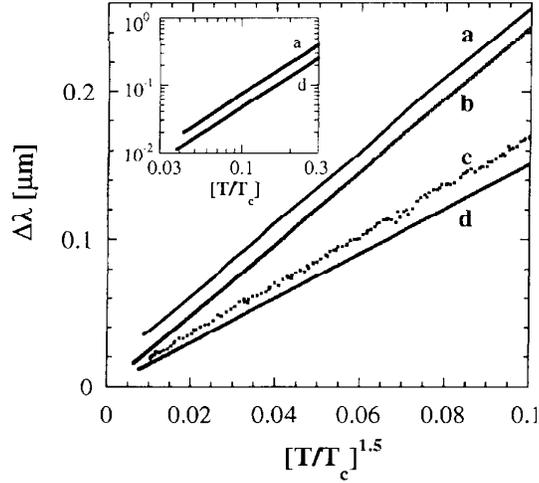}
\caption[]{Changes of the in-plane penetration depth $\Delta \lambda_\parallel (T)$ of
\brom\ [two samples (a),(b)] and \cuncs\ [(c),(d)] plotted versus $(T/T_c)^\frac{3}{2}$.
The data have been offset. Taken from \cite{Carrington 99}.\\}\label{lambda2}
\end{figure}

For the present materials the magnetic penetration depth \index{penetration depth} has
been determined by a variety of different techniques including ac-susceptibility
\cite{Kanoda 90,Pinteric 00}, muon-spin relaxation \cite{Harshman 90,Le 92},
dc-magnetization \cite{Lang 92,Lang 92b}, surface im\-pe\-dan\-ce \cite{Achkir 93,Dressel
94} and a related high-frequency technique using a tunnel diode oscillator
\cite{Carrington 99}. The results of these studies, however, are quite inconsistent
regarding both the temperature dependence as well as the absolute values of
$\lambda_L(T\rightarrow0)$ (see Table~\ref{Tabelle4c} in section~\ref{Superconducting
parameters}) and have led to quite different conclusions as to the symmetry of the
superconducting \index{order parameter}
order parameter.\\
Interestingly enough, these inconsistencies do not only involve results from different
experimental techniques. Contradictory conclusions have been drawn also on the basis of
seemingly identical experiments performed by different groups. This is the case for
surface impedance studies where the penetration depth \index{penetration depth} can be
extracted from the complex conductivity. The latter is derived from the frequency shifts
and variations of the quality factor of the resonator caused by the sample. While the
surface impedance studies using a microwave perturbation technique on \cuncs\ and \brom\
by Klein et al.\ \cite{Klein 91} and Dressel et al.\ \cite{Dressel 94} were found to be
in good agreement with the BCS predictions, other studies by Achkir et al.\ \cite{Achkir
93} on \cuncs\ revealed an in-$T$ linear behavior at low temperatures indicative of an
order parameter \index{order parameter} with zeroes on the Fermi surface. Deviations from
an exponential temperature dependence of $\lambda_L(T)$ for the above two $\kappa$-\etzx\
compounds have been observed also in a more recent experiment using an rf tunnel-diode
oscillator \cite{Carrington 99}. In contrast to the above measurements by Achkir et al.,
however, their data of the in-plane penetration depth \index{penetration depth} rather
follow a $T^\frac{3}{2}$ power law (see Fig.~\ref{lambda2}).
As has been argued by the authors, the data would still be consistent with a quasi-linear
variation of the superfluid density as expected for a $d$-wave \index{$d$-wave
superconductivity} superconductor with impurities or a small residual gap
\cite{Carrington 99}. Alternatively, the authors point out that the exponent $3/2$ may
arise naturally in a model proposed for
short-coherence-length \index{coherence length} superconductors exhibiting a pseudogap \cite{Kosztin 98}.\\
An inconsistency exists also for $\mu$SR experiments performed by different groups. Here
$\lambda_L$ can be determined by measuring the field inhomogeneities in the mixed state,
i.e.\ the spatial variation of the local induction of the vortex lattice. This technique
was first applied to \cuncs\ by  Harshman et al.\ \cite{Harshman 90} who could fit their
data by a BCS temperature dependence. Subsequently, Le et al.\ \cite{Le 92} carried out
similar experiments on the same system as well as on the \brom\ salt and found an in-$T$
linear variation for the in-plane penetration depth \index{penetration depth} at low
temperatures
$\lambda_\parallel (T)\approx 1 + \alpha \cdot (T/T_c)$.\\
A power law temperature dependence of $\lambda_\parallel (T)$ consistent with $d$-wave
superconductivity \index{$d$-wave superconductivity} has been observed also by
ac-susceptibility measurements performed by different groups \cite{Kanoda 90,Pinteric
00,Pinteric 02}. The latter experiments as well as the surface impedance studies are
operating at very small external magnetic fields ($B < B_{c_1}$) attempting to probe the
Meissner state. Possible difficulties in these experiments that may arise from
flux-pinning-related \index{pinning} phenomena near the surface of the superconductor,
i.e.\ an inhomogeneous superconducting state, have been discussed in \cite{Lang 96}.\\ An
alternative way to determine the penetration depth \index{penetration depth} is to make
\begin{figure}[t]
\sidecaption
\includegraphics[width=.6\textwidth]{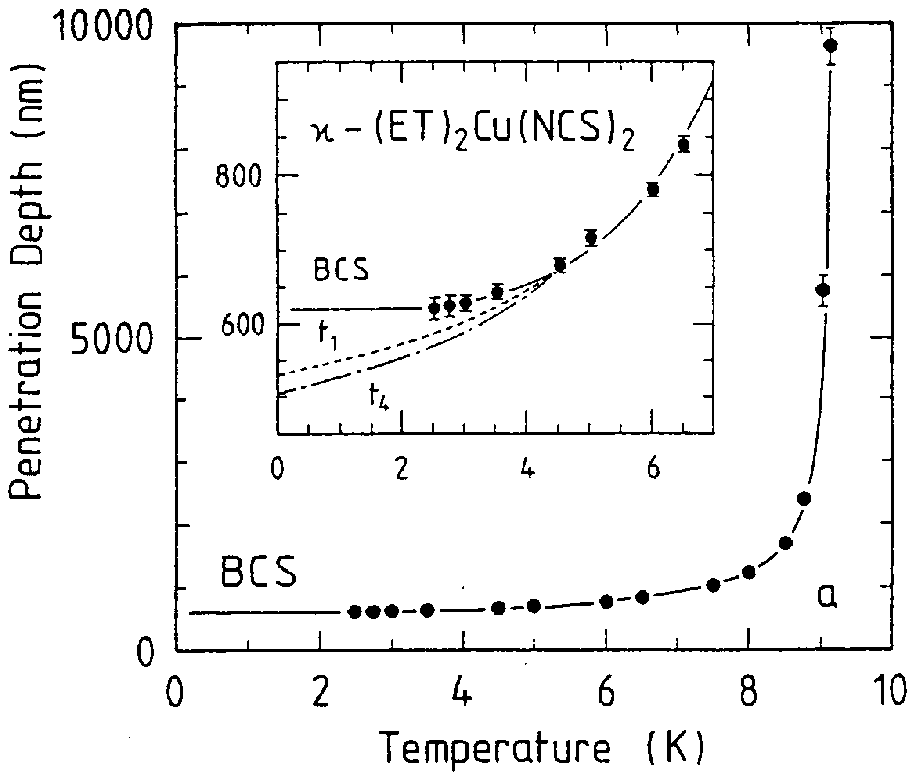}
\caption[]{In-plane penetration depth for single crystalline \cuncs. The solid line
represents a BCS fit. The model calculations labelled t$_1$ and t$_4$ represent those
anisotropic states proposed by Hasegawa et al.\ \cite{Hasegawa 87} and used by Le et al.\
to explain their $\mu$SR data \cite{Le 92} which have the weakest and strongest $T$
dependences, respectively. Taken from \cite{Lang 92}.\\}\label{lambda1}
\end{figure}
use of the reversible mixed-state magnetization, a peculiarity of these strongly
an\-iso\-tro\-pic superconductors with short \index{coherence length} coherence lengths.\\
According to the London model, the field dependence of the magnetization is given by
\begin{equation}\label{London3}
\frac{{\rm d}M}{{\rm d}(\ln H)} = \frac{\phi_0}{32 \pi^2 \lambda^2_{\rm eff}},
\end{equation}
where $\lambda^2_{\rm eff} = \lambda_\parallel^2$ for $B \perp {\rm planes}$ and
$\lambda^2_{\rm eff} = \lambda_\parallel \lambda_\perp$ for $B \parallel {\rm planes}$.
In 3D superconductors, vortex pinning \index{pinning} usually gives rise to an
inhomogeneous distribution of the vortices in the mixed state and thus to an irreversible
behavior of the magnetization upon increasing and decreasing the field. This may cause
substantial uncertainties in determining the penetration depth \index{penetration depth}
from magnetization data. On the contrary, for quasi-2D superconductors with short
coherence \index{coherence length} length the magnetization is entirely reversible over
an extended field range, i.e.\ $B_{c_1} < B_{\rm irr} < B < B_{c_2}$ with $B_{\rm irr}$
being the temperature-dependent irreversibility line \index{irreversibility line} (see
section~\ref{Mixed state}). For \cuncs\ and \brom\ a reversible magnetization has been
observed over an extended range in the $B$-$T$ plane which thus allows for a precise
determination of the in-plane penetration \index{penetration depth} depth \cite{Lang
92,Lang 92b}. The in-plane penetration \index{penetration depth} depths
$\lambda_\parallel(T)$ were determined, see Fig.~\ref{lambda1}, from the slopes ${\rm
d}M/{\rm d}(\ln H)$ of the isotherms taken at different temperatures and using equation
(\ref{London3}). The solid line represents a BCS fit \cite{Muehlschlegel 59} to the data.
For both systems, the data reveal only a weak variation with temperature at low $T$
consistent with an exponential temperature dependence as expected for a finite gap.

\paragraph{Specific heat}
The above variety of contradictory results on the magnetic penetration depth
\index{penetration depth} indicate an extraordinarily high sensitivity of this quantity to
extrinsic effects such as disorder or pinning-related \index{pinning} phenomena. As a
possible source, we mention the disorder \index{disorder} associated with the glass
transition \index{glass-like transition} of the ethylene \index{ethylene endgroups}
endgroups, see section~\ref{glassy phenomena} and \cite{Pinteric 02}.\\ A quantity which
is less sensitive to the above problems but can still provide fundamental information on
the gap structure is the specific heat. In case this integral thermodynamic quantity were
to find a low temperature electronic quasiparticle contribution, $C_{\rm es}$, that varies
exponentially weakly with the temperature, the existence of gap zeroes on the Fermi
surface could be definitely ruled out. On the other hand, the observation of a
non-exponential temperature dependence does not necessarily prove the existence of gap
zeroes as this result might originate in extraneous contributions such as impurity
phases, normal-conducting regions or pair-breaking effects.
\begin{figure}[b]
\center
\includegraphics[width=0.9\textwidth]{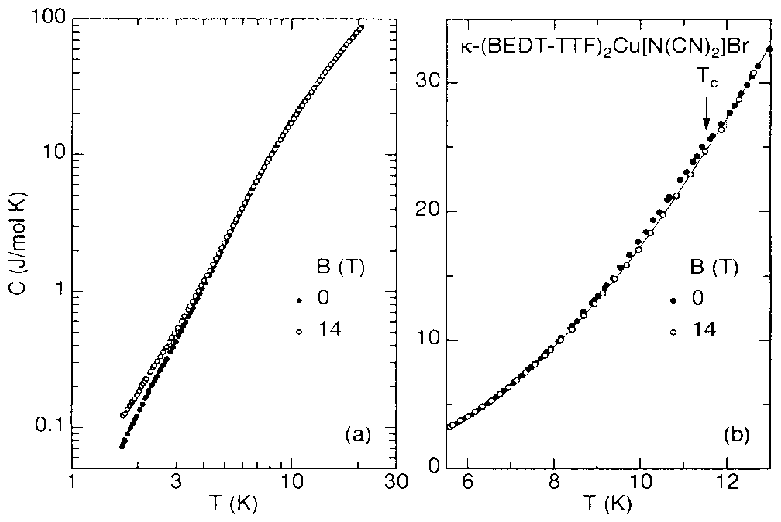}
\caption[]{Temperature dependence of the specific heat of \brom\ in the superconducting
($B=0$) and normal state ($B=14$\,T) over an extended temperature range (left panel) and
in the vicinity of $T_c = 11.5$\,K (right panel). The solid line is a polynomial fit to
the 14\,T data. Taken from \cite{Elsinger 00}.}\label{spezbr}
\end{figure}\\
First specific heat measurements were focussing on the determination of the discontinuity
at $T_c$ which provides information on the coupling strength. From the results of Andraka
et al.\ \cite{Andraka 89} and Graebner et al.\ \cite{Graebner 90} on \cuncs\ yielding a
ratio of $\Delta C/\gamma T_c > 2$ a strong-coupling behavior has been inferred for this
salt. In a series of subsequent experiments, the temperature dependence of the electronic
contribution, \ces, at lower temperatures was at the focus of the investigations. From
experiments on \brom\ Nakazawa et al.\ \cite{Nakazawa 97} reported a quadratic
temperature dependence of \ces\ at low temperatures, which was taken as an indication for
line nodes in the gap. However, recent high-resolution specific heat measurements on the
same compound revealed an exponentially weak low-$T$ electronic contribution to the
specific heat implying a finite energy gap \cite{Elsinger 00}. Moreover it has been shown
in the latter study that the $T^2$ dependence in the \ces\ data by Nakazawa et al.\
\cite{Nakazawa 97} most likely originates in their incorrect determination of the phonon
background \cite{link4g1}. Figures~\ref{spezbr} and \ref{spezbr2} show the results of
specific heat measurements performed by Elsinger et al.\ \cite{Elsinger 00}. The phonon
contribution, which predominates the specific heat near $T_c$ has been determined from
measurements in
\begin{figure}[t]
\sidecaption
\includegraphics[width=0.55\textwidth]{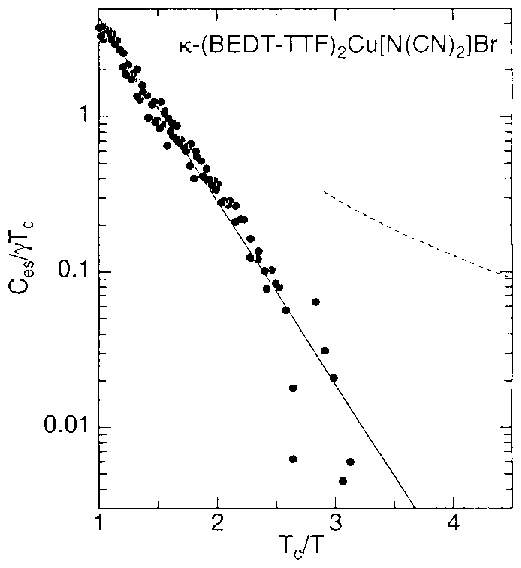}
\caption[]{A semi-logarithmic plot of $C_{es}/(\gamma T_c)$ vs $T_c/T$ as determined from
the data shown in Fig.~\ref{spezbr}. The solid line indicates the exponential variation of
\ces. The dashed line corresponds to \ces\ as determined by Nakazawa et al.\
\cite{Nakazawa 97} (see text and \cite{link4g1}). Taken from \cite{Elsinger
00}.}\label{spezbr2}
\end{figure}
an overcritical field. This standard procedure is valid as long as there are no magnetic
contributions to the specific heat which might change with the field. From the absence of
any measurable field dependence in the data above $T_c$ this assumption appears
justified. Figure~\ref{spezbr2} shows the exponential decrease of \ces\ with decreasing
temperatures. The lack of a finite
\ces\ for $T\rightarrow0$ rules out the existence of zeroes in the energy gap.\\
A similar behavior has been observed also for \cuncs\ \cite{Mueller 02,Wosnitza 02}.
Figure~\ref{spezcuncs} shows the difference $\Delta C(T) = C(T,B=0) - C(T,B=8\,{\rm T}>
B_{c_2})$ used to analyze the specific heat data. The advantage of using this quantity
means that the unknown phonon contribution drops out. As Fig.~\ref{spezcuncs}
demonstrates, $\Delta C(T)$ deviates markedly from the weak-coupling BCS behavior in both
the jump height at \tc\ as well as the overall temperature dependence. However, as was
found also for \brom\ \cite{Elsinger 00}, a much better description of the data is
obtained by using the semi-empirical extension of the BCS formalism \index{model} to
strong-coupling superconductors - the so-called $\alpha$-model \cite{Padamsee 73}. It
contains a single free parameter $\alpha \equiv \Delta_0/k_B T_c$ which scales the BCS
\index{ratio} energy gap $\Delta (T) = (\alpha / \alpha_{\rm BCS}) \cdot \Delta_{\rm
BCS}(T)$ with $\alpha_{\rm BCS}=1.764$. As Fig.~\ref{spezcuncs} clearly demonstrates, the
strong-coupling BCS model \index{model} with $\alpha = 2.8 \pm 0.1$ provides an excellent
description of the data over the entire temperature range investigated \cite{Mueller 02}.
Similar to what has been found for \brom\ (Fig.~\ref{spezbr2}), the data for \cuncs\
(inset of Fig.~\ref{spezcuncs}) are fully consistent with an exponentially small \ces\ at
low temperatures, i.e.\ an energy gap without zeroes at the Fermi surface. The same
behavior has been previously observed for other \etzx\ superconductors \cite{Wosnitza
94,Wanka 98,Wosnitza 02}. The above findings of an exponentially weak specific heat at low
temperatures are clearly incompatible with the existence of gap zeroes as claimed by
\begin{figure}[t]
\center
\includegraphics[width=.75\textwidth]{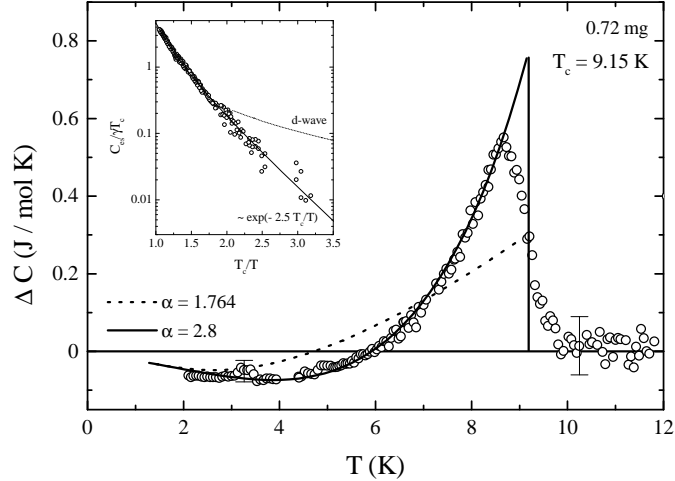}
\caption[]{Specific heat difference $\Delta C=C(0\,{\rm T})- C(8\,{\rm T})$ of a \cuncs\
single crystal of m = 0.72 mg (main panel). The dotted and solid thick lines represent
the BCS curves for weak and strong coupling, respectively. The inset shows the
quasiparticle contribution to the specific heat in the superconducting state as $C_{\rm
es}/\gamma T_c$ vs $T_c/T$ in a semi-logarithmic representation. Here the solid line
represents the strong-coupling BCS behavior while the dotted line indicates a $T^2$
behavior as expected for a $d$-wave order parameter \index{order parameter} \cite{Momono
96}. Taken from \cite{Mueller 02,Mueller 01}.}\label{spezcuncs}
\end{figure}
various of the above mentioned experiments. It has to be shown by future studies whether
or not this controversy can be removed by taking properly into account the influence of
magnetic fields even when applied parallel to the planes and other extraneous effects
such as the cooling-rate-dependent \index{cooling-rate dependence} disorder
\index{disorder} associated with the ethylene groups.

\subsubsection{On the pairing state in (TMTSF)$_2$X}
As for the BEDT-TTF systems, the nature of superconductivity in the \index{Bechgaard
salts} Bechgaard salts (TMTSF)$_2$X is still far from being understood and continues to
be a controversial issue. Early suggestions of a spin-triplet (S=1) \index{spin-triplet
state} state were based on the proximity of superconductivity to a SDW state in the
pressure-temperature phase diagram (cf.\ section \ref{phase diagrams}). This situation
resembles the theoretical expectation for an interacting 1D electron gas \cite{Solyom 79}
where a SDW phase lies next to triplet superconductivity. Further arguments for a
spin-triplet $p$-wave superconductivity \index{spin-triplet state} were derived from the
observation that $T_c$ is extremely sensitive to the introduction of nonmagnetic defects
\cite{Choi 82,Bouffard 82} and substitutional impurities \cite{Coulon 82,Tomic 83a}, see
also \cite{Abrikosov 83,Gorkov 85}. As discussed in section \ref{phase diagrams},
depending on external pressure and magnetic field, the Bechgaard salts \index{Bechgaard
salts} can be tuned to either a superconducting or a SDW ground state which has led to
the proposal that the order parameters \index{order parameter} of both phases are not
independent of each other. The superconducting properties of the Bechgaard salts have
been reviewed by several authors \cite{Jerome 94,Ishiguro 98,Dressel 99}. Since these
articles are comprehensive up to 1998, we shall give only a brief overview on the early
results and concentrate the discussion on the more recent developments including the
possibility of spin-triplet superconductivity.

Among the (TMTSF)$_2$X superconductors, the X = PF$_6$ and ClO$_4$ salts are the most
extensively studied materials although the number of experimental investigations of the
superconducting properties is much less compared to that of the BEDT-TTF salts and their
derivatives. The reason for this is most likely related to the low $T_c$ values of the
former systems which require extensive low-temperature equipment and, in the case of X =
PF$_6$, external pressure of $p \geq 5.8$\,kbar \cite{Lee 00} to stabilize the
superconducting state. For the ambient-pressure superconductor (TMTSF)ClO$_4$ it is the
anion ordering \index{anion ordering} which renders the experimental situation difficult
(see section~\ref{glassy phenomena}). By slowly cooling \index{cooling-rate dependence}
through the anion-ordering temperature $T_{AO}$ at around 24\,K an ambient pressure
superconducting state below $T_c = (1.2 \pm 0.2)$\,K can be stabilized. In this case, the
anions are believed to be well ordered. On the other hand, for samples that have been
cooled rapidly across $T_{AO}$, an insulating state forms below $T_{MI} \simeq 6.1$\,K.\\
Early specific heat measurements on (TMTSF)$_2$ClO$_4$ focusing on the temperature range
close to $T_c$ revealed a discontinuity at $T_c$, $\Delta C / \gamma T_c = 1.67$ which is
in fair agreement with the \index{ratio} BCS value of 1.43 \cite{Garoche 82}. On the
other hand, deviations from a BCS-type of superconductivity have been observed in NMR
measurements by Takigawa et al.\ \cite{Takigawa 87}. These authors reported the absence
of a Hebel-Schlichter peak and a $T^3$ dependence in the spin-lattice relaxation rate.
This has led to the proposal of a $d$-wave \index{$d$-wave superconductivity} order
parameter \index{order parameter} with a gap that vanishes along lines on the Fermi
surface \cite{Takigawa 87}. These results are at variance with more recent thermal
conductivity data on the same salt, showing a rapid decrease of the electronic
contribution to the heat transport below \tc\ which indicates the absence of low-lying
excitations \cite{Belin 97}. Their results provide strong evidence for a nodeless gap
function. However, as pointed out by the authors, this result is not necessarily
associated with an $s$-wave \index{order parameter} \index{$s$-wave superconductivity}
order parameter.\\
By enumerating possible gap functions for quasi-1D systems, Hasegawa and Fukuyama
\cite{Hasegawa 87} showed that besides an anisotropic spin-singlet \index{spin-singlet
state} $d$-wave also a spin-triplet $p$-wave \index{spin-triplet state} state - in both
cases the order parameters \index{order parameter} vanish along lines on the FS - is
possible. The authors suggested the possibility of an antiferromagnetic
spin-fluctuation pairing mechanism for the \index{Bechgaard salts} Bechgaard salts.\\
Arguments in favor of such a spin-fluctuation-mediated superconductivity with $d$-wave
\index{$d$-wave superconductivity} symmetry have been derived from a recent resistivity
study under pressure \cite{Jaccard 01,Wilhelm 01}. According to these results, a minimum
in the resistivity $\rho_a(T)$ at $T_{\rm min}$ marks the onset of AF fluctuations
before, at lower temperatures, an itinerant antiferromagnetic state (SDW) is stabilized.
The width of the region of
critical AF fluctuations in the $T$-$p$ phase diagram (
see Fig.~\ref{bechpd2} in section~\ref{phase diagrams}) is enhanced when the SDW ground
state is approached from the high-pressure side, where the system is close to the SDW/SC
border and largest where $T_c(p)$ reaches its optimum value. The correlation between the
spin-fluctuation regime above the onset of superconductivity and the \tc\ value is taken
as a strong argument for a pairing mechanism mediated by the exchange of these
fluctuations \cite{Jaccard 01,Wilhelm 01}.\\

\begin{figure}[b]
\sidecaption
\includegraphics[width=.55\textwidth]{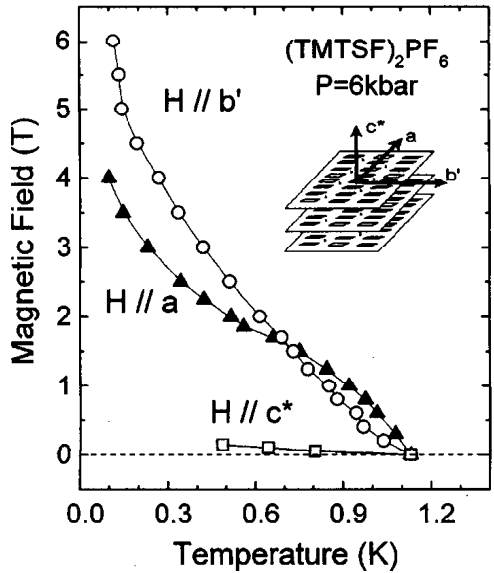}
\caption[]{$B$-$T$ phase diagram of (TMTSF)$_2$PF$_6$ at a pressure of 6\,kbar for
magnetic fields aligned along three perpendicular axes as defined in the inset. Taken
from \cite{Lee 97a}.\\}\label{Bechgaard1}
\end{figure}
In a recent series of papers on the upper \index{$B_{c_2}$} critical fields, the
discussion of a possible realization of a triplet-pairing \index{spin-triplet state}
state in the Bechgaard salts \index{Bechgaard salts} has again been raised \cite{Lee
97a,Lee 00,Lee 02a}. Lee et al.\ examined the upper \index{$B_{c_2}$} critical field
$B_{c_2}(T)$ and its directional dependence in (TMTSF)$_2$PF$_6$ under pressure via
resistivity measurements. The resulting magnetic field-temperature phase diagram is
depicted in Fig.~\ref{Bechgaard1} for fields aligned along the three crystal axes
\cite{Lee 97a}. While the upper-critical-field \index{$B_{c_2}$} curves near $T_c$ were
found to be consistent with earlier results on the same and related compounds, which
indicated a more conventional behavior (see also Fig.~\ref{hc2tmtsf} in
section~\ref{Superconducting parameters}) \cite{Murata 87,Greene 82,Chaikin 83}, the
extension of the measurements to lower temperatures uncovered important new features: (i)
the upper critical fields for $B$ aligned along the $a$- and $b$-directions display
strong positive curvatures {\em without} saturation. This behavior was found to be
independent of the criterion used to extract $B_{c_2}$ from the resistance profiles. (ii)
$B_{c_2}^b$ becomes larger than $B_{c_2}^a$, i.e.\ the anisotropy inverts upon cooling,
and (iii), the critical fields \index{$B_{c_2}$} in both directions exceed the Pauli
limiting field \index{Pauli-limiting field} of $B_{P} = 1.84 \times T_c \simeq 2.2$\,T
\cite{Clogston 62,Chandraskhar 62}. In a subsequent resistivity study under fields along
the $b$-axis and optimum pressure settings \cite{Lee 00}, the onset of superconductivity
was found to persist even up to $9$\,T, which is more than four times $B_{P}$. The
authors discussed various proposals attempting to explain an upturn in the upper critical
field. Among them are strong spin-orbit scattering \cite{Klemm 75,Huang 89}, a
magnetic-field-induced dimensional crossover from 3D to 2D \cite{Lebed 86,Dupuis 93} and
the formation of a spatially inhomogeneous Fulde-Ferrell-Larkin-Ovchinnikov (FFLO) state
\index{Fulde-Ferrell-Larkin-Ovchinnikov (FFLO) state} \cite{Fulde 64,Larkin 65}. It has
been argued by Lee et al.\ that, even with a field-induced dimensional crossover which
greatly enhances the orbital critical field, an additional effect is required to exceed
the pa\-ra\-ma\-gne\-tic limit. This could be either the formation of the FFLO state or
triplet superconductivity. The fact that no indications for a first-order phase
transition into the FFLO state have been observed thus points to the possibility of a
spin-triplet \index{spin-triplet state} superconducting
state.\\
This interpretation has gained further support from recent NMR measurements of the Knight
shift $K_S$ in the superconducting state of pressurized (TMTSF)$_2$PF$_6$ \cite{Lee 02}.
\begin{figure}[t]
\sidecaption
\includegraphics[width=.525\textwidth]{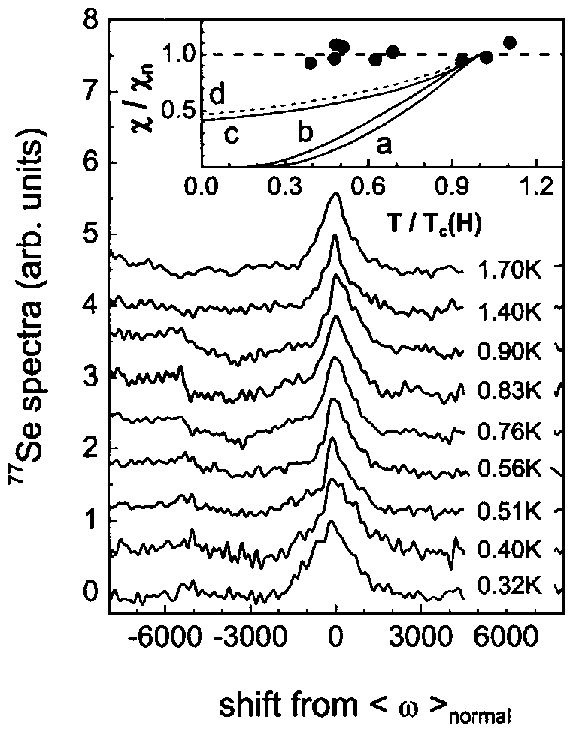}
\caption[]{$^{77}$Se NMR spectra of (TMTSF)$_2$PF$_6$ under 7\,kbar at various
temperatures above and below \tc\ which is 0.81\,K at the applied field of 1.43\,T
aligned along the most conductive $a$ axis. The inset shows normalized \xs\ data compared
with theoretical calculations. Curves a and b are for a singlet state in applied fields
$B/B_{c_2}$ near zero and 0.63, respectively, as calculated by Fulde and Maki \cite{Fulde
65} while curves c and d represent a scenario where the vortex cores induce normal
regions of a fraction equal to $B/B_{c_2}$. Taken from \cite{Lee 02}.\\}\label{Bechgaard2}
\end{figure}
Since $K_S$ is proportional to the spin-susceptibility $\chi_{\rm Spin}$, which is
different for spin singlet ($\chi_{\rm Spin}\rightarrow0$\ for\ $T\rightarrow0$) and
triplet \index{spin-triplet state} ($\chi_{\rm Spin}$ remains unchanged on cooling
through $T_c$) superconductors, measurements of $K_S$ provide a direct probe of the spin
parity of the superconductor. Figure~\ref{Bechgaard2} shows that in contrast to the
expectations for the various scenarios involving singlet superconductivity there is no
significant change in $K_S$ for $B\parallel a$ on cooling into the superconducting state.
These observations together with similar results of a previous study for $B\parallel b$
\cite{Lee 00a} thus argue for a spin-triplet pairing. Besides $K_S$, the above studies
included also measurements on the spin-lattice relaxation rate $1/T_1$ which exhibits a
small peak near \tc. In light of the absence of such a peak in previous reports of zero
field proton-NMR measurements \cite{Takigawa 87} its identification with a
Hebel-Schlichter peak is unclear.

\clearpage

\section{Epilogue}
Studies on superconductivity in organic charge-transfer salts have been carried out for
more than twenty-years. The joint efforts of researchers from the fields of synthetic
chemistry and solid-state physics have created this new class of molecular conductors
which exhibits fascinating properties. Key features are a pronounced anisotropy of the
electronic states spanning the whole range from quasi-1D to anisotropic 3D, a low charge
carrier concentration and a strong coupling of the charge carriers to the lattice degrees
of freedom. The combination of these extraordinary material parameters make the organic
conductors ideal model systems for exploring the interplay of strong electron-electron
and electron-phonon interactions in reduced dimensions. In fact, this material class has
enriched the solid-state physics by a wealth of interesting collective phenomena such as
spin-Peierls or density-wave instabilities, Mott-type metal-insulator transitions, and,
above all, superconductivity. Most remarkably, these systems offer a unique opportunity
to scan through the various ground states either in discrete steps by chemical means, or
even in
a continuous way through changes of the applied pressure.\\

Thanks to the intense interdisciplinary efforts from research groups in chemistry and
physics, our level of understanding of these materials has progressed substantially. The
basis for this development was accomplished by advances in organic and organometallic
chemistry which have provided a rich and varied supply of molecules serving as building
blocks for conductors and superconductors. By employing state-of-the-art preparation
techniques, it is possible nowadays to synthesize crystals of unprecedented high quality
which permit highly accurate measurements of their electronic parameters. In particular,
these clean materials enable direct experimental access to the electronic band structure
with a high level of accuracy. For the 2D materials, the various experiments agree in
that the normal-state electronic properties are well described by quasiparticles within a
Fermi-liquid approach. These quasiparticles reveal a considerable mass renormalization
due to both electron-electron as well as electron-phonon interactions. This contrasts
with the behavior found in the quasi-1D salts, where non-Fermi-liquid properties have
been reported for transport and optical properties. Indeed, these results indicate the
separation of spin and charge degrees of freedom - one
of the hallmarks of a Luttinger liquid.\\

The phase diagrams of the quasi-1D and 2D materials have been mapped out in great detail
utilizing a wide spectrum of experimental techniques - some of which have been applied
under extreme conditions such as high magnetic fields and high pressure. These studies
confirmed, on the one hand, the universal character of the phase diagrams for both
families and revealed, on the other hand, intriguing new details of the interplay of the
various phases. Despite such detailed information, in particular on the phase boundary
between superconductivity and magnetism, there is still no generally accepted picture on
the nature of the superconducting state. The fact that both phases share a common phase
boundary have led some researchers to believe that both phenomena have the same origin.
In such a scenario, the attractive pairing interaction would have to be provided by the
exchange of antiferromagnetic spin fluctuations. In fact, some of the experimental data
seem to indicate an anisotropic pairing state compatible with such a magnetic pairing
interaction. While for the quasi-2D salts a spin-singlet state with $d$-wave symmetry is
the most favored one among the anisotropic states proposed, a number of experiments on the
quasi-1D materials even suggest triplet superconductivity. However, for the quasi-2D
salts, the notion of an anisotropic superconducting state is in clear conflict with
results of other experiments - in particular specific heat measurements - which
demonstrate the existence of a
finite superconducting energy gap on the whole Fermi surface.\\
It is remarkable that despite the continuous experimental efforts to unravel the
fundamental features of the superconducting state, the above controversy still persists.
These efforts include the application of standard experimental techniques, some of which
have been driven to an extraordinarily high resolution, as well as new, more sophisticated
methods such as angular-dependent investigations of the gap functions.\\ An important
lesson to be learnt from this controversy concerns the role of disorder, in particular
the \textit{intrinsic}-type of disorder which is inherent to many of the materials even
when prepared under ideal conditions. As was recognized early on for the quasi-1D
materials but has now proved to be true also for the quasi-2D systems, \textit{intrinsic}
disorder can be of crucial importance and should not be overlooked in exploring and
discussing superconducting-state properties. This is of concern for all those materials
where, by the symmetry, certain structural elements can adopt more than one orientation.
Since these states are almost degenerate in energy, a cooling-rate dependent frozen
\textit{intrinsic} disorder will result at low temperatures which may have a severe
influence on both the electronic and also the elastic properties of the sample. It is to
be hoped that part of
the above controversy can be removed by taking these disorder effects into account properly.\\

In viewing the field of organic conductors and superconductors as a whole, it is clear
that these materials have provided a considerable contribution to our understanding of
strongly interacting electrons in low dimensions, though important questions still remain
open. This concerns, in particular, the nature of the metallic state above $T_c$ and the
role of electron-electron and electron-phonon interactions in the pairing mechanism. In
light of the progress we have made in recent years, one can be optimistic that more
systematic investigations on the relationship between molecular properties and crystal
structure on the one hand, and the collective features of the bulk, on the other, will
not only provide an answer to these questions but also will be a guidance for the
synthesis of new materials with hopefully even more fascinating properties.

\clearpage



\begin{thebibliography}{}
\bibitem{link4g1}In \cite{Nakazawa 97} the lattice specific
heat of \brom\ was estimated by measuring a second, deuterated sample after quench cooling
to suppress superconductivity. The data analysis in \cite{Nakazawa 97} is based on the
assumption that by this procedure the lattice specific heat remains unaffected and is
identical to that of the hydrogenated superconducting compound. In \cite{Elsinger 00} it
has been shown, however, that the so-derived phonon background differs substantially from
that determined in an overcritical field. The reason for this might be related to the
glass-like transition at \tg\ observed in this system \cite{Mueller 02a}, see
section~\ref{glassy phenomena}.
\end{thebibliography}

\begin{theindex}

  \item activation energy, 33, 34
  \item anion ordering, 32, 80
  \item anisotropic order parameter (see order parameter)
  \item anisotropy parameter, 43, 44, 58
  \item antiferromagnetic order, 30, 37, 41
  \item antiferromagnetic spin fluctuations, 20, 25--27, 63, 66

  \indexspace

  \item band filling, 13, 14
  \item band structure, 5, 13, 50, 65
  \item bandwidth, 5, 16, 38, 66
  \item BCS
  \subitem model, 63, 66, 75, 79
  \subitem ratio, 52, 54, 71, 72, 79, 80
  \subitem relation, 50
  \item Bechgaard salts, 6--8, 31, 80, 81
  \item BEDT-TSF, 5, 41, 42
  \item BEDT-TTF, 5--7, 41, 42, 51, 68
  \item BETS, 5, 12, 20, 41, 54
  \item $B_{c_1}$, 52, 55, 56
  \item $B_{c_2}$, 43, 52--56, 58, 60, 62, 81
  \item $B_{c_{th}}$, 45, 52, 55--57

  \indexspace

  \item carrier concentration, 5, 21, 42, 45
  \item charge transfer, 5, 13, 14
  \item coherence length, 43, 45, 53, 54, 56, 57, 76, 77
  \item cooling-rate dependence, 33--35, 40, 80
  \item crystal growth, 7

  \indexspace

  \item density of states, 15, 24, 27, 38, 50, 55, 70
  \item density wave, 6, 20, 25, 30, 37, 40, 50, 66
  \item dimerization, 7--9, 13, 14, 16, 38
  \item Dingle temperature, 15
  \item disorder, 7, 9, 20, 31--33, 35, 40, 77, 80
  \item $d$-wave superconductivity, 65--67, 72, 75, 76, 81

  \indexspace

  \item eclipsed conformation, 9
  \item effective masses, 16, 17, 22, 43, 54
  \item electron-electron interaction/correlations, 16, 17, 21, 37, 63
  \item electron-intermolecular-phonon interaction/coupling (see also electron-phonon interaction),
  23, 24, 70, 71
  \item electron-molecular-vibration (EMV) coupling (see also electron-phonon interaction), 23, 24, 68
  \item electron-phonon
  \subitem coupling constant, 24, 50, 70
  \subitem interaction/coupling, 16, 17, 22, 23, 37, 67, 69
  \item ET, 5, 8, 9, 20
  \item ethylene conformation, 9, 34, 40
  \item ethylene endgroups, 9, 20, 28, 33, 34, 40, 68, 70, 73, 77

  \indexspace

  \item Fermi liquid, 13, 14, 19, 20, 22, 27
  \item Fermi surface, 13, 14, 16, 21, 22, 25, 27, 29, 31, 37, 39, 40
        51, 54, 63, 65, 66
  \item Fulde-Ferrell-Larkin-Ovchinnikov (FFLO) state, 55, 60, 82

  \indexspace

  \item Ginzburg number, 45
  \item Ginzburg-Landau coherence length (see coherence length)
  \item glass-like transition, 28, 33, 39, 77

  \indexspace

  \item half-filled band (see band filling)
  \item H-bonding, 9, 51
  \item highest occupied molecular orbital (HOMO), 13, 14, 23, 24, 50

  \indexspace

  \item impurities, 31
  \item irreversibility line, 57--59, 77
  \item isotope effect, - substitution, 34, 50, 51, 68

  \indexspace

  \item Josephson
  \subitem coupling, 57
  \subitem effect, 43
  \subitem vortices, 57, 60

  \indexspace

  \item Little's model, 3, 64
  \item lock-in transition, 58
  \item London penetration depth (see penetration depth)
  \item lower critical field (see $B_{c_1}$)
  \item lowest unoccupied molecular orbital (LUMO), 13
  \item Luttinger liquid, 19, 22

  \indexspace

  \item many-body effects, 16
  \item mean-field transition temperature, 45, 46, 53
  \item Mott-Hubbard insulating state, 31, 37, 65, 66

  \indexspace

  \item nesting, 13, 25, 27, 31, 37, 38, 51, 66
  \item non-centrosymmetric anion, 7, 31

  \indexspace

  \item order parameter, 42, 53, 55, 57, 60, 63, 64, 67, 68, 72, 75,
        79--81
  \item overlap/transfer integral, 13, 18, 21, 22, 24, 36, 50, 54

  \indexspace

  \item packing, 5, 8, 9, 51
  \item pancake vortices, 57, 58, 74
  \item Pauli-limiting field $B_P$, 52--55, 60, 62, 81
  \item penetration depth, 35, 43, 53, 56, 57, 67, 74--77
  \item pinning, 56, 58--60, 76, 77
  \item $\pi$-electrons, -orbital, 5, 6, 10, 13, 37, 50, 73

  \indexspace

  \item resistance maximum, 19, 20, 31

  \indexspace

  \item scaling behavior, 46
  \item spin fluctuations (see antiferromagnetic spin fluctuations)
  \item spin-singlet state, 52, 55, 60, 63, 74, 81
  \item spin-triplet state, 53, 80--82
  \item staggered conformation, 9
  \item superconducting fluctuations, 44, 45, 53
  \item $s$-wave superconductivity, 81
  \item $\sigma$-electrons, -orbital, 5, 13

  \indexspace

  \item TCNQ, 5
  \item terminal ethylene groups (see ethylene endgroups)
  \item thermodynamic critical field (see $B_{c_{th}}$)
  \item TMTSF, 5, 6, 42
  \item TMTTF, 5
  \item TTF, 5, 6

  \indexspace
  \item upper critical field (see $B_{c_2}$)

  \indexspace

  \item van der Waals radius, 7, 8

\end{theindex}

\end{document}